\documentclass[aps,amsfonts]{iopart}
\usepackage{amsmath}
\usepackage{graphicx}
\usepackage{amssymb,amsfonts,bm}

\begin{document}

\topical[L.Barack]
{Gravitational self force in extreme mass-ratio inspirals}


\author{Leor Barack\dag}

\address{\dag\ School of Mathematics, University of Southampton, Southampton,
SO17 1BJ, United Kingdom}

\begin{abstract}
This review is concerned with the gravitational self-force acting on a mass
particle in orbit around a large black hole. Renewed interest in this old
problem is driven by the prospects of detecting gravitational waves from
strongly gravitating binaries with extreme mass ratios.
We begin here with a summary of recent advances in the theory of
gravitational self-interaction in curved spacetime, and proceed to survey
some of the ideas and computational strategies devised for implementing this
theory in the case of a particle orbiting a Kerr black hole. We review in detail
two of these methods: (i) the standard {\it mode-sum method}, in which the
metric perturbation is regularized mode-by-mode in a multipole decomposition,
and (ii) {\em $m$-mode regularization}, whereby individual azimuthal modes of the
metric perturbation are regularized in 2+1 dimensions. We discuss several practical
issues that arise, including the choice of gauge, the numerical representation of the
particle singularity, and how high-frequency contributions near the particle
are dealt with in frequency-domain calculations. As an example of a full
end-to-end implementation of the mode-sum method, we discuss the computation
of the gravitational self-force for eccentric geodesic orbits in Schwarzschild,
using a direct integration of the Lorenz-gauge perturbation equations in the
time domain. With the computational framework now in place, researchers have
recently turned to explore the physical consequences of the gravitational
self force; we will describe some preliminary results in this area.
An appendix to this review presents, for the first time, a detailed derivation
of the ``regularization parameters'' necessary for implementing the mode-sum
method in Kerr spacetime.
\end{abstract}

\pacs{04.30.Db,04.30.-w,04.25.Nx,04.70.Bw}

\ead{leor@soton.ac.uk}

\date{\today}



\section{Introduction}
\label{sec:intro}

\paragraph{Context and motivation.}

Within Newtonian theory, the gravitational two-body problem, in its basic form,
is readily solvable. An isolated system of two gravitationally-bound point masses
admits two conserved integrals---the energy and angular momentum---and
the resulting motion is precisely periodic. The corresponding
{\it general-relativistic} problem is radically---and notoriously---more difficult.
In General Relativity (GR), the orbits in a bound binary are never periodic:
Gravitational radiation continually removes energy and angular momentum
from the system, and the radiation back-reaction gradually drives the
two objects tighter together until they eventually merge. Furthermore,
in GR one cannot usually work consistently with pointlike mass particles
(as we discuss later), and the simplest and most universal
two-body problem becomes that of a binary black hole system.

In general, the description of the nonlinear radiative dynamics of a
black hole binary entails a full Numerical-Relativistic (NR) treatment.
There has been a much celebrated advance in NR research in the past
few years \cite{Pretorius:2005gq,Campanelli:2005dd,Baker:2005vv},
with NR codes now capable of tracking
the complicated nonlinear evolution of a binary black hole spacetime during
the final stages of the merger. However, NR methods become less efficient
when the two black holes are far apart, or when one of the components is much
heavier than the other. Each of these two regimes features two greatly different
lengthscales, which is not easily accommodated in a NR framework \cite{Gonzalez:2008bi}.
Fortunately, the occurrence of two separate lengthscales also restores a sense of
``pointlikeness'' (at least approximately), which allows for simpler, perturbative
treatments. In the first regime---at sufficiently large separations---the
dynamics is best analyzed using post-Newtonian (PN) methods \cite{Blanchet:2002av},
wherein GR corrections to the Newtonian dynamics (accounting for radiation
reaction, the objects' internal structure, etc.)~are incorporated into
the equations of motion order by order in the binary separation.

The second, so-called {\it extreme mass-ratio} regime---which will concern us in
this review---is most naturally explored
within the framework of black hole perturbation theory. Here the ``zeroth-order''
configuration is that of a test particle (the lighter black hole) moving along
a geodesic of the fixed background spacetime of the large black hole. This can then
serve as a basis for a systematic expansion, wherein corrections due to
the finite mass of the small object (and due possibly also to its internal structure)
are included order by order in the small mass ratio. At first order in the mass
ratio, the gravitational field of the small object is a linear perturbation
of the background black hole geometry. The back reaction from this perturbation
gives rise to a gravitational {\it self force} (SF) which gradually diverts the
small object from its geodesic motion. In this picture, it is the SF that is
responsible (in particular) for the radiative decay of the orbit. In principle, knowledge
of the SF (along with the metric perturbation) forms a complete picture of the orbital
dynamics at linear order in the small mass ratio. One therefore hopes that a
calculation of the SF would facilitate a faithful, if only approximate
description of the orbital dynamics in binary systems with small mass ratios.

The determination of the SF in curved spacetime is an old problem in
mathematical relativity, made very relevant following recent
developments in gravitational-wave research. Interest in this problem was
renewed in the mid 1990s, when it was first proposed that the planned
space-based gravitational wave detector LISA (the Laser Interferometer
Space Antenna \cite{LISA}) could observe signals from the inspiral of compact
objects (white dwarfs, neutron stars, or stellar-mass black holes) into
massive black holes in galactic nuclei. It is now believed that LISA should
be able to detect tens to thousands such events \cite{Gair:2004iv},
out to cosmological distances ($z\sim 1$) \cite{Gair:2008bx}---
depending on the astrophysical rates (inspirals per galaxy per year),
which are still highly uncertain.\footnote{
In fact, if inspiral rates are near the higher end of their estimated range,
the unresolvable stochastic background from thousands weak inspirals
could dominate the noise budget of LISA at its most sensitive frequency
band. \cite{Barack:2004wc}} Dubbed {\em EMRI}s (Extreme Mass Ratio Inspirals), these
potential sources became key targets for LISA due to their unique facility
as precision probes of strong-field gravity. A typical LISA EMRI will
spend the last few years of inspiral in a very tight orbit around the
massive hole, emitting some $10^5$--$10^6$ gravitational wavecycles wholly
within the LISA frequency band. These complicated waveforms carry
extremely accurate information about the physical parameters of the
inspiral system, as well as a detailed map of the spacetime geometry
around the massive hole. The potential scientific implications are broad
and far-reaching, ranging from astrophysics to cosmology and to fundamental
theory \cite{Barack:2003fp,Collins:2004ex,Barack:2006pq,AmaroSeoane:2007aw,
Gair:2007kr,Prince:2009uc,Schutz:2009tz}.

The odds are that even the strongest detected EMRI signals will be buried
deeply inside LISA's instrumental noise. However, since an EMRI signal is detectable
over a long time, it can be extracted with significant signal-to-noise ratio (of
order $\sim 100$ for the strongest sources) using {\em matched filtering} \cite{Barack:2003fp}.
This, however, requires precise theoretical templates of EMRI waveforms across
the relevant parameter space of inspirals---indeed, the detection statistics
cited above assumes that such templates were at hand by the time LISA flies.
It is likely that other, non-template-based data analysis techniques
could be used to detect some of the brightest EMRIs \cite{Gair:2008ec,Babak:2009ua}.
Nonetheless, it remains
true that an accurate parameter extraction, crucial for exploiting the full scientific
value of the EMRI signal, will rely on (or be restricted by) the availability
of accurate and faithful theoretical templates of the inspiral waveforms.

The theoretical challenge is derived from the astrophysical specifications
of the LISA-relevant EMRIs. With its peak sensitivity at a few mHz, LISA will
observe inspirals into galactic (likely Kerr) black holes with masses in the
range $\sim 5\times 10^5$--$5\times 10^7 M_{\odot}$; the inspiraling
objects (compact stars or stellar-mass black holes) may have masses in the range
$0.5$--$50M_{\odot}$, giving mass ratios of $10^{-4}$--$10^{-8}$---well within
the ``extreme mass ratio'' domain of the two-body problem.
Inspirals are not expected to have any preferred
orientation with respect to the central hole's spin direction, and orbits may
remain quite eccentric throughout the inspiral \cite{Barack:2003fp}. It is not likely that
interaction with a possible accretion disk around the massive hole would play
a dominant role in the orbital dynamics \cite{Barausse:2007dy}. In a typical
LISA-band EMRI---a $10M_{\odot}/10^6M_{\odot}$ system---gravitational radiation
reaction drives the orbital decay over a timescale of months. More importantly,
it affects the phasing of the inspiral orbit over mere hours. It is therefore
clear that a useful model of the long-term orbital phase evolution in a
LISA-relevant system ought to incorporate properly the effect of radiation
reaction.

The above translates, at first approximation, to a very clean problem
in black hole perturbation theory: A point mass is set in a generic (eccentric,
inclined) strong-field orbit around a Kerr black hole of a much larger mass,
and one wishes to calculate the gravitational waveforms emitted as radiation
reaction drives the gradual inspiral up until the eventual plunge through
the event horizon. While it remains important to quantify the effect of
higher-order corrections to this picture (e.g., due to the spin of the inspiralling
object), one expects that, thanks to the extreme mass ratio, the above
simple setup should provide a good model for astrophysical inspirals.
Indeed, from the perspective of GR theorists working in the field, much
of the appeal of the SF problem comes from its unusual dual nature as
both an elementary theoretical problem in GR, and an exciting problem
in contemporary astrophysics.

\paragraph{Scope and relation to other reviews.}

Our main aim here is to review the challenges and main developments
in the program to calculate the gravitational SF in Kerr spacetime.
Our discussion will be focused mostly on work concerned with the evaluation
of the SF along a specified, non-evolving orbit (normally taken to be a
geodesic of the background spacetime); we will not consider here the
important question of how orbits evolve under the effect of the SF.
The strategic approach envisaged here is one in which the complete analysis
of the orbital evolution is carried out in two separate, consequent steps.
In the first, preparatory step, one calculates the SF across the entire relevant phase
space (i.e., obtain the value of the SF as a function of location and
velocity, perhaps through interpolation of numerical results). In the
second step one then uses the SF information to calculate the inspiral
orbits of particles with given initial conditions. Here we will be concerned
only with the first, most crucial step.

In parallel to the work on SFs described in this review, there is
already a substantial research effort aimed to formulate a reliable scheme
for the orbital evolution, assuming the SF has been calculated \cite{Pound:2007th,
Hinderer:2008dm}. This work incorporates techniques from multiple-scale
perturbation theory. Another parallel effort aims to obtain an approximate model
of the orbital evolution and emitted waveforms {\em without} resorting to the
loacl SF \cite{Hughes:2005qb,Drasco:2005is,Drasco:2005kz,Sago:2005fn,Ganz:2007rf,
Mino:2007ft}. This work is largely based on a strategy proposed by
Mino \cite{Mino:2003yg,Mino:2005qj,Sago:2005gd,Mino:2006em}, in which a time-average
measure of the rate of change of the geodesic ``constants of motion'' is calculated
from a certain ``radiative'' solution of the perturbation equations (to be described
later in this review), which accounts for the long-term radiative aspects of
the dynamics but neglects some of the conservative effects. This ``inspiral
without SF'' approach is reviewed by Mino in Ref.\ \cite{Mino:2005qj}, and also
by Tanaka in Ref.\ \cite{Tanaka:2005ue}. It is well
possible that this method will prove sufficiently accurate for LISA applications.
However, ultimately, its performance and accuracy could only be assessed
against actual calculations of the full SF.

The fundamental formulation of the SF in curved spacetime is described in
a comprehensive Living Review article by Poisson \cite{Poisson:2003nc}.
This is an excellent treatise which is both self-contained and pedagogical,
and it makes an essential reading for anyone wishing to introduce oneself
to the field. (The less endurant reader would find a slightly abridged version
of this review in Ref.\ \cite{Poisson:CQG}; there is also a concise introduction
in \cite{Poisson:2004gg}.) It would not be useful to review here at any great
detail the basic SF theory already covered in \cite{Poisson:2003nc}. Instead,
we merely give a succinct summary of the essential formal results, and move on
to describe, in the rest of our review, how the general formalism is implemented
in actual calculations of the gravitational SF in Kerr. In this respect, our
review starts where Poisson's review ends. We will, however, briefly survey
a few recent developments in SF theory not covered in Ref.\ \cite{Poisson:2003nc}
(last updated in 2004).

A good 2005 snapshot of the activity surrounding SF calculations is offered by
a collection of review articles published in a special issue of Classical and
Quantum Gravity \cite{CQGspecial}. Drasco's review from 2006 \cite{Drasco:2006ws}
explores the utility of the various computation methods within the context of the
LISA EMRI problem. The website of the 12th Capra Meeting on Radiation Reaction
in Relativity (Bloomington IN, 2009) \cite{Capra12} is an excellent resource of information
on current research in the field, with links to all talks given in the meeting.
The website also includes links to the websites of previous meetings in the series (1998--2008).
There are several good reviews of EMRI astrophysics and the science potential
of EMRI detections, including ones by Hopman \cite{Hopman:2006pv}, Amaro-Seoane
{\it et al.}~\cite{AmaroSeoane:2007aw}, and Miller {\it et al.}~\cite{Miller:2009wv}.
Finally, an abridged version of some of the material in our current review is
to appear in Ref.\ \cite{OrleansBook}.

\paragraph{Structure.}
The overall structure of our presentation will follow the logical route of
development in SF research: From basic theory (Sec.\ \ref{sec:theory}) through
to practical calculation schemes (Secs.\ \ref{sec:methods}--\ref{sec:modesum}) and
to numerical implementation in Schwarzschild (Secs.\ \ref{sec:numerics}--\ref{sec:Lorenz})
and Kerr (Sec.\ \ref{sec:puncture}), concluding with a discussion of some physical
consequences (Sec.\ \ref{sec:effects}).

We begin in Sec.\ \ref{sec:theory} with a brief introduction to the general
theory of the gravitational SF in curved spacetime, highlighting from the outset
the role of gauge dependence in this theory. We then summarize (rather than reproduce)
the main theoretical developments that over the past twelve years helped establish
a robust formal framework for SF calculations.
Section \ref{sec:methods} is a brief survey of the main strategies that have
been proposed for implementing this formal framework in actual calculations,
particularly for orbits in Kerr (or Schwarzschild) spacetimes. This section
also provides a quick-reference catalogue (Tables \ref{tab:1}--\ref{tab:3})
of actual computations of the SF carried out so far.
Section \ref{sec:modesum} is a self-contained introduction to the {\em
mode-sum method}, one of the leading techniques for SF calculations.
The basic idea is presented through an elementary example, followed by
a formulation of the method as applied to generic orbits in Kerr. In
an accompanying appendix we provide, for the first time, a full derivation
of the regularization parameters necessary for implementing the mode sum
scheme in Kerr. (The values of these parameters where published in the past
\cite{Barack:2002mh} without a detailed derivation.)

An essential preliminary step in almost all calculations of the SF involves
the numerical integration of the relevant perturbation equations sourced by
the point particle. In Sec.\ \ref{sec:numerics} we discuss the practicalities
of such calculations, and review some of the numerical strategies that were
proposed for dealing with the particle singularity in both the frequency and
time domains. Section \ref{sec:Lorenz} then focuses on a particular implementation
strategy, based on a direct time-domain integration of the Lorenz-gauge metric
perturbation equations. This approach recently led to a first computation of
the gravitational SF for eccentric orbits in Schwarzschild, and we show some
results from this calculation. In Sec.\ \ref{sec:puncture} we discuss higher-dimensional
alternatives to standard mode-sum, focusing on the recently proposed $m$-mode
regularization, which may offer a more efficient treatment in the Kerr case.
Section \ref{sec:effects} reviews initial work aimed to understand and
quantify the physical, gauge-invariant effects of the gravitational SF.
Finally, in Sec.\ \ref{sec:conc} we reflect on recent advances and comment
on future directions.

\paragraph{Notation.}
We follow here the notation conventions of Misner, Thorne and Wheeler
\cite{MTW}. Hence, the metric signature is $({-}{+}{+}{+})$, the connection
coefficients and Riemann tensor are
$\Gamma^{\lambda}_{\mu\nu}=\frac{1}{2}g^{\lambda\sigma}(g_{\sigma\mu,\nu}
+g_{\sigma\nu,\mu}-g_{\mu\nu,\sigma}$) and
$R^{\alpha}{}_{\!\lambda\mu\nu}=\Gamma^{\alpha}_{\lambda\nu,\mu}
-\Gamma^{\alpha}_{\lambda\mu,\nu}+\Gamma^{\alpha}_{\sigma\mu}\Gamma^{\sigma}_{\lambda\nu}
-\Gamma^{\alpha}_{\sigma\nu}\Gamma^{\sigma}_{\lambda\mu}$, the Ricci tensor
and scalar are $R_{\alpha\beta}=R^{\mu}{}_{\!\alpha\mu\beta}$ and
$R=R_{\alpha}{}^{\!\alpha}$, and the Einstein equations read
$G_{\alpha\beta}=R_{\alpha\beta}-\frac{1}{2}g_{\alpha\beta}R=8\pi T_{\alpha\beta}$.
We use Greek letters for spacetime indices, and adopt standard geometrized
units (with $c=G=1$) throughout.

\section{Essential theory}
\label{sec:theory}
\subsection{Gravitational forces}

The concept of a gravitational force is, of course, fundamentally strange
to GR: In a purely gravitational system one expects no
acceleration and hence no ``forces'' in the ordinary sense. Since the
gravitational SF is an example of a force of a purely gravitational origin,
we begin by explaining what is generally meant by ``gravitational forces''
in our context.

Consider a smooth region of spacetime with metric $g_{\alpha\beta}$
(which may be, for example, the stationary vacuum exterior of a Kerr
black hole), and a smooth weak gravitational perturbation $h_{\alpha\beta}$
of that spacetime (which may represent, for example, an incident gravitational
wave). Now consider a test particle of mass $\mu$ which is moving freely in the
perturbed spacetime.
Neglecting SF
effects, the particle's trajectory will be a geodesic of the perturbed
spacetime, described, in a given coordinate system $x^{\alpha}$, by
\begin{equation} \label{eq1:10}
\frac{d^2x^{\alpha}}{d\tau'^2}+\Gamma'^{\alpha}_{\mu\nu}
\frac{dx^{\mu}}{d\tau'}\frac{dx^{\nu}}{d\tau'}=0,
\end{equation}
where $\tau ^{\prime}$ is an affine parameter along the trajectory and
$\Gamma'^{\alpha}_{\mu\nu}$ are the connection coefficients associated with
the perturbed metric $g+h$.

In some occasions, however, it is practically useful to reinterpret
the particle's motion in terms of a trajectory in the background
spacetime $g$. Under this interpretation, the trajectory (in $g$) is no
longer geodesic; rather, the particle experiences ``an external gravitational
force'', which is exerted by the perturbation $h$. This (fictitious) force
is defined through Newton's second law as
\begin{equation} \label{eq1:20}
F_{\rm grav}^{\alpha}=\mu\left(
\frac{d^2x^{\alpha}}{d\tau^2}+\Gamma_{\mu\nu}^{\alpha}\frac{dx^{\mu}}
{d\tau}\frac{dx^{\nu}}{d\tau}\right),
\end{equation}
where $\tau$ is an affine parameter in the background metric $g$, and
$\Gamma_{\alpha\beta}^{\mu}$ are the connection coefficients associated with
$g$. Note that, in this non-covariant description, $h$ and $\Gamma$ are
treated as tensor fields in $g$, and similarly the force $F_{\rm grav}^{\alpha}$ and
four-velocity $u^{\alpha}\equiv dx^{\alpha}/d\tau$ are defined as vectors
in $g$. The tensorial indices of all these quantities are raised and lowered
using the background metric $g$.

To express $F_{\rm grav}^{\alpha}$ in terms of the linear perturbation
field $h_{\alpha\beta}$, we use $d/d\tau=(d\tau'/d\tau)d/d\tau'$ in Eq.\
(\ref{eq1:20}) and introduce $\Delta\Gamma _{\mu\nu }^{\alpha}\equiv
\Gamma'^{\alpha}_{\mu\nu}-\Gamma _{\mu\nu}^{\alpha}$. By virtue of
Eq.\ (\ref{eq1:10}) this gives
\begin{equation}\label{eq1:30}
F_{{\rm grav}}^{\alpha }=-\mu\Delta \Gamma _{\mu \nu }^{\alpha }u^{\mu
}u^{\nu }+\zeta u^{\alpha },
\end{equation}
with $\zeta\equiv \mu(d\tau/d\tau')(d^2\tau'/d\tau^2)$.
From its definition
in Eq.\ (\ref{eq1:20}), the force $F_{{\rm {grav}}}^{\alpha }$ must be
perpendicular to the four velocity $u^{\alpha}$. Hence, projecting
$F_{{\rm {grav}}}^{\alpha }$ orthogonally to $u^{\alpha}$ keeps it unchanged,
and we may use this fact to dispose of the term $\zeta u^{\alpha}$ in Eq.\
(\ref{eq1:30}):
\begin{equation}\label{eq1:40}
F_{\rm grav}^{\alpha}=-\mu(\delta_{\lambda}^{\alpha}+u^{\alpha}u_{\lambda })
\Delta\Gamma_{\mu \nu}^{\lambda}u^{\mu}u^{\nu}.
\end{equation}
Finally, expressing $\Delta\Gamma$ in terms of $h$ (keeping only terms
that are linear in $h$) we obtain
\begin{equation}\label{eq1:50}
F_{\rm grav}^{\alpha}=-\frac{1}{2}\mu(g^{\alpha\lambda }
+u^{\alpha}u^{\lambda})\left(\nabla_{\nu}h_{\lambda\mu}+
\nabla_{\mu}h_{\lambda\nu}-\nabla_{\lambda}h_{\mu \nu}\right)
u^{\mu}u^{\nu}
\equiv \mu \nabla^{\alpha\beta\gamma}h_{\beta\gamma},
\end{equation}
where $\nabla_{\alpha}$ denotes covariant differentiation with respect to the
background metric $g$. The differential operator $\nabla^{\alpha\beta\gamma}$
determines the gravitational force exerted by any given external perturbation;
it is given explicitly by
\begin{equation}\label{eq1:60}
\nabla^{\alpha\beta\gamma}=
\frac{1}{2}\left(g^{\alpha\delta}u^{\beta}-2g^{\alpha\beta}u^{\delta}
-u^{\alpha}u^{\beta}u^{\delta}\right)u^{\gamma}\nabla_{\delta}.
\end{equation}
Later we will often work with the trace-reversed metric perturbation,
\begin{equation}\label{eq1:70}
\bar h_{\alpha\beta}=h_{\alpha\beta}-\frac{1}{2}g_{\alpha\beta}
g^{\mu\nu}h_{\mu\nu}.
\end{equation}
In terms of $\bar h_{\alpha\beta}$, Eq.\ (\ref{eq1:50}) becomes
\begin{equation}\label{eq1:80}
F_{\rm grav}^{\alpha}=\mu \bar\nabla^{\alpha\beta\gamma}\bar h_{\beta\gamma},
\end{equation}
with
\begin{equation}\label{eq1:90}
\bar\nabla^{\alpha\beta\gamma}=
\frac{1}{4}\left(
2g^{\alpha\delta}u^{\beta}u^{\gamma}
-4g^{\alpha\beta}u^{\gamma}u^{\delta}
-2u^{\alpha}u^{\beta}u^{\gamma}u^{\delta}
+u^{\alpha}g^{\beta\gamma}u^{\delta}
+g^{\alpha\delta}g^{\beta\gamma}\right)
\nabla_{\delta}.
\end{equation}

Formally, the above interpretation of the motion requires a suitable
procedure for mapping the physical trajectory from the full spacetime
$g+h$ onto the background spacetime $g$. In the above discussion we
adopted the following procedure: First, we assume that $g+h$ and $g$
were covered with two coordinate meshes that are ``similar'' in the sense
that, at the limit $h\to 0$, any given physical event would attain the same
coordinate values in both spacetimes. Then, we identify each event $x^{\alpha}$
in $g+h$ with an event having the same coordinate value $x^{\alpha}$ in $g$.
This, in particular, results in a projection of the trajectory in $g+h$ onto
one in $g$.

Of course, there is not just one way of specifying the
coordinate systems in the two spacetimes, which leads to an ambiguity in
the projected trajectory: Different coordinate choices would, in general,
give rise to different projected trajectories in $g$, showing different
accelerations and hence interpreted as being under the influence of
different gravitational forces. This, of course, is nothing but an example
of the usual gauge ambiguity intrinsic to perturbation theory in GR.
The gravitational force, just like the metric perturbation itself, is
{\em gauge dependent}.

It is straightforward to write down a gauge transformation law for
$F_{\rm grav}^{\alpha}$. Consider a small gauge displacement
\begin{equation}\label{eq1:100}
x^{\mu}\to x^{\mu}-\xi^{\mu},
\end{equation}
where the magnitude of $\xi^{\mu}$ is assumed to scale like that of the
external perturbation $h_{\alpha \beta}$. Under $\xi^{\mu}$, the perturbation
transforms as $h_{\alpha\beta}\to h_{\alpha\beta}+\delta_{(\xi)}h_{\alpha\beta}$,
where
\begin{equation}\label{eq1:110}
\delta_{(\xi)} h_{\alpha\beta}=\nabla_{\alpha}\xi_{\beta}
+\nabla_{\beta}\xi_{\alpha}.
\end{equation}
From Eq.\ (\ref{eq1:50}), this will induce a change in the gravitational
force, given by
\begin{equation}\label{eq1:120}
\delta_{(\xi)} F_{\rm grav}^{\alpha}=
\mu \nabla^{\alpha\beta\gamma}\delta_{(\xi)} h_{\beta\gamma}.
\end{equation}
(Terms arising from the gauge transformation of $\nabla^{\alpha\beta\gamma}$
are quadratic in the magnitude of the perturbation and we neglect them here.)
Substituting from Eq.\ (\ref{eq1:110}) in Eq.\ (\ref{eq1:120}) and using the
commutation relation $\nabla_{\mu}\nabla_{\nu}\xi_{\alpha}-
\nabla_{\nu}\nabla_{\mu}\xi_{\alpha}=\xi_{\lambda}R^{\lambda}{}_{\alpha\nu\mu}$,
one readily arrives at
\begin{equation} \label{eq1:130}
\delta_{(\xi)} F_{\rm grav}^{\alpha}=
-\mu\left[\left(g^{\alpha\lambda}+u^{\alpha}u^{\lambda}\right)
\ddot{\xi}_{\lambda}+{R^{\alpha}}_{\mu\lambda\nu}u^{\mu}
\xi^{\lambda}u^{\nu}\right],
\end{equation}
where overdots denote covariant derivatives with respect to the affine
parameter $\tau$ along the background trajectory. This is the general
gauge transformation formula for external gravitation forces.\footnote{
One might notice that Eq.\ (\ref{eq1:130}) reproduces the geodesic deviation
equation if one sets the left-hand side to zero and reinterprets $\xi^{\alpha}$
as the displacement vector connecting two adjacent geodesics in $g$. Indeed,
a vanishing $\delta_{(\xi)} F_{\rm grav}^{\alpha}$ implies that the two adjacent
trajectories (the original projected trajectory and its gauge-transformed
counterpart) have the same acceleration in $g$, in which case the displacement
vector between them is known to satisfy the geodesic deviation equation.}

\setcounter{footnote}{0}

\subsection{Gravitational self force}

To devise a theory of the gravitational SF, one might be tempted to simply
interpret it as an example of a gravitational force of the type discussed
above, with the source of the metric perturbation now being the particle itself.
This naive interpretation would be problematic, for several reasons.
First, the physical perturbation due to the particle (a retarded solution of
the linearized Einstein equations, $h_{\alpha\beta}^{\rm ret}$) is singular at
the location of the particle, and the statement that the particle follows a
geodesic of $g+h^{\rm ret}$ is therefore physically meaningless.
Obviously, trying to apply Eq.\ (\ref{eq1:50}) [or (\ref{eq1:80})] with the
external perturbation replaced with the self-perturbation
$h_{\alpha\beta}^{\rm ret}$ would yield a singular, and hence meaningless result.

Second (and relatedly), since we are now considering the self-gravity of
the particle (it is no longer a test particle), we must make a mathematical
sense of its being ``pointlike''. This is not a trivial matter to address in
curved spacetime. Mathematically, the usual delta-function representation of
a point particle stress-energy is known to be inconsistent with the nonlinearity
of the full Einstein equations \cite{Geroch,Wald}. A familiar physical manifestation
of this is that one cannot squeeze a finite amount of mass to a point without
creating a black hole (of a finite size). The mathematical consistency of a
delta-function source is restored in the linear theory, but it remains a challenging
task to understand how the notion of a point particle might emerge (rather than
be pre-assumed) from a suitable limiting procedure.

Third, the SF is conceptually different from the external forces discussed
above, in that the latter are, in truth, just fictitious forces resulting
from our insistence to artificially split the physical spacetime into a background
and a perturbation. The SF, in contrast, must be viewed as a genuine
physical effect (even if a delicate one, as the SF too is gauge dependent---see
below). There indeed exists an interpretation of the motion---we will discuss
it later---wherein the particle moves freely on a geodesic of a certain smooth,
perturbed spacetime, subject to no SF. However, in this description the
smooth geometry is {\em not} the physical spacetime of the background + particle
system (the metric of this geometry is not a retarded solution of the
linearized Einstein equations), and so this effective spacetime cannot be
said to represent ``physical reality'' any better than the SF itself.

Thanks to work commencing around 1997 and continuing to these days,
we now have a rather satisfactory theory of the gravitational SF in
curved spacetime, which, in particular, addresses all of the above issues.
Initial work was strongly inspired by the classical analyses of the analogous
electromagnetic self-force problem in flat (Dirac, 1938 \cite{Dirac})
and curved (DeWitt and Brehme, 1960 \cite{DeWitt:1960fc}) spacetimes.
DeWitt and Brehme's work, especially, provided much of the mathematical
framework---that of covariant bi-tensors in curved spacetime---needed
for analyzing the gravitational problem too. In 1997 two independent
groups published three independent derivations of the gravitational SF.
Quinn and Wald \cite{Quinn:1996am} used an axiomatic approach, in which
the physical self-acceleration is deduced, essentially, by comparing the
(divergent) self-field of the particle in question with that of a particle
in a suitably constructed tangent flat space. Independently, Mino, Sasaki
and Tanaka (MST) \cite{Mino:1996nk} obtained the same result using a local
energy-momentum conservation argument---a direct application of DeWitt
and Brehme's method to the gravitational case. Both these methods
pre-assume a notion of a point mass particle, and neither seeks to
make a consistent sense of this notion. However, MST's
paper also reported a second, independent derivation of the SF, using
an approach which, for the first time, offered a fully GR-consistent
treatment of the problem.

This approach---a new implementation of the old idea of {\it matched
asymptotic expansions}---relies on the assumption that there can be
identified two separate lengthscales in the problem: One associated with
the particle's mass $\mu$, and another, much larger, associated with
the typical radius of curvature of the geometry in which the particle
is moving. In the strong-field EMRI problem, the second lengthscale
is provided by the mass of the central black hole, $M\gg\mu$.
MST's construction further assumes that the ``particle'' is actually
a Schwarzschild black hole of mass $\mu$. The two separate scales in the
setup define a ``near zone'', $r\ll M$, and a ``far zone'', $r\gg m$
(where $r$ is a suitable measure of distance from the small hole).
In the near zone, the geometry is approximately that of the small
Schwarzschild hole, with small tidal-type corrections from the background
geometry. As we zoom away from the small object and enter the far zone,
the effect of the small object's detailed structure becomes gradually
less important, and at the far zone limit the geometry becomes that of
the background spacetime, weakly perturbed by what is now a distant
``point particle''---it is indeed the far-zone limit through which
a notion of point mass can be defined in a consistent way.
In situations where $M\gg m$ one would have a ``buffer zone'' where
$m\ll r\ll M$ and both ``near zone'' and ``far zone'' descriptions of
the geometry are valid. MST showed that matching the near zone and far
zone metrics (expressed as asymptotic expansions in $r/M$ and
$m/r$, respectively) constrains the motion of the particle (from a far-zone
point of view) and thus yields an expression for the SF. This expression agreed
with those obtained by Quinn and Wald and by using DeWitt and Brehme's method;
it was later coined the {\em MiSaTaQuWa formula}, an acronym based on the
names of the five contributing authors.
A self-contained and pedagogical review of the MiSaTaQuWa formula, including
an elegant reproduction of previous derivations, can be found in Poisson's
\cite{Poisson:2003nc}.

More recently, Gralla and Wald \cite{Gralla:2008fg} (see also \cite{Gralla:2009uf}
for a more concise presentation) developed a new procedure for deriving the
gravitational SF, which offers improved mathematical rigor as well as a
generalization of the MiSaTaQuWa formula. Rather than relying on two
separate asymptotic expansions of the metric as in MST's original method,
Gralla and Wald introduce a single one-parameter family of metrics, which,
through two different limiting procedures, can produce both near and
far zone metrics in a natural way. This allows to define more robustly the
criteria for existence of the two zones, and enables a more elegant
buffer-zone matching. The analysis proves that in the far-zone limit the
particle is described
precisely by the usual delta-function distribution, and that at the very
limit $\mu\to 0$ this particle moves on a geodesic of the background.
Furthermore, the analysis relaxes all assumptions about the
nature of the small object: It no longer need to be a Schwarzschild
black hole, but can assume the form of any sufficiently small black hole
or a blob of ordinary matter. This allows, in particular, for a spin-force
term to appear in the resulting, generalized version of the MiSaTaQuWa formula.

The main end product of the above theoretical developments is a
firmly-established general formula for the SF in a class of background
spacetimes including Kerr. It should be stressed that the SF formula
stems in a deterministic way from nothing else than the Einstein
equations with the usual conservation laws; it does not rely---and
{\em should} not rely, as a matter of principle---on any form of
ambiguous ``regularization'' or ``subtraction of infinities''. This
idea is implemented explicitly in the matched asymptotic expansions
approach, and even more so in the Gralla--Wald analysis.

\subsection{MiSaTaQuWa equation}

We now state the MiSaTaQuWa formula \cite{Quinn:1996am,Mino:1996nk,
Poisson:2003nc}---Eq.\ (\ref{eq1:500}) below. (For simplicity we ignore
spin-force terms and focus on the self-interaction part of the formula.)
This will serve as a starting point for the rest of this review.

Consider a timelike geodesic $\Gamma$ in a background spacetime with
metric $g_{\alpha\beta}$. For concreteness, let us think of $\Gamma$ as
a test particle orbit outside a Kerr black hole, so $g_{\alpha\beta}$ is
the Kerr metric. Let $\tau$ be the proper time along $\Gamma$, and let
$x^{\alpha}=z^{\alpha}(\tau)$ describe $\Gamma$ in some smooth coordinate
system and $u^{\alpha}\equiv dz^{\alpha}/d\tau$ be the four velocity of
the test particle. Denote by $h_{\alpha\beta}^{\rm ret}$ the physical,
retarded metric perturbation from a particle of mass $\mu$ whose worldline
is $\Gamma$.
Assume $h_{\alpha\beta}^{\rm ret}$ is given in the {\em Lorenz} gauge:
\begin{equation} \label{eq1:140}
\nabla^{\beta}\bar h^{\rm ret}_{\alpha\beta}=0,
\end{equation}
where $\bar h^{\rm ret}_{\alpha\beta}$, recall, is the trace-reversed
version of $h_{\alpha\beta}^{\rm ret}$ [see Eq.\ (\ref{eq1:70})].
Remember that throughout our discussion indices are raised and lowered
using the background metric $g_{\alpha\beta}$, and covariant derivatives
are taken with respect to that metric.

\begin{figure}[htb]
\input{epsf}
\centerline{\includegraphics[totalheight=4.5cm]{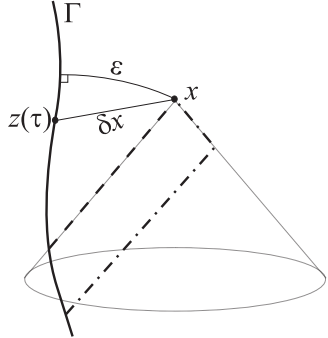}}
\caption{
An illustration of the setup described in the text. $z(\tau)$ is a
point on the timelike worldline $\Gamma$ (thick solid line) and $x$ is a
field point close to $z$, shown with a portion of its past light cone.
$\epsilon$ is the spatial geodesic distance from $x$ to $\Gamma$ and
$\delta x^{\alpha}\equiv x^{\alpha}-z^{\alpha}$. The metric perturbation at
$x$ consists of a {\bf direct} and a {\bf tail} contributions,
illustrated by the thick dashed and dash-dot lines, respectively.
[{\it Graphics reproduced from Ref.\ }\cite{OrleansBook}.]
}
\label{fig1}       
\end{figure}

At any specetime point $x$, the retarded perturbation can be written
as a sum of two pieces,
\begin{equation} \label{eq1:150}
\bar h^{\rm ret}_{\alpha\beta}=\bar h^{\rm dir}_{\alpha\beta}+
\bar h^{\rm tail}_{\alpha\beta},
\end{equation}
the former being the ``direct'' contribution
coming from the intersection of the past light-cone of $x$ with $\Gamma$,
and the latter being the ``tail'' contribution arising from the part
of $\Gamma$ {\it inside} this light cone (see Fig.\ \ref{fig1}).
The occurrence of a tail term is a well-known feature of the wave equation
in 3+1D curved spacetime, and it can be interpreted physically as arising
from the effect of waves being scattered off spacetime curvature (``failure
of the Huygens principle'').
Both $\bar h^{\rm ret}_{\alpha\beta}$ and $\bar h^{\rm dir}_{\alpha\beta}$
obviously diverge when evaluated on $\Gamma$; however, $\bar h^{\rm tail}_{\alpha\beta}$
is continuous and differentiable everywhere, including on the worldline.
Notably, though, the tail field is not a smooth function on the worldline,
and is not a vacuum solution of the linearized Einstein equations.

The MiSaTaQuWa formula states that the gravitational SF at a given point
$z$ along $\Gamma$ results simply from the back reaction of the tail field:
\begin{equation} \label{eq1:500}
F^{\alpha}_{\rm self}(z)=\mu \bar\nabla^{\alpha\beta\gamma}
\bar h^{\rm tail}_{\beta\gamma}(z).
\end{equation}
Here $\bar\nabla^{\alpha\beta\gamma}$ is the usual ``force operator'', given
in equation (\ref{eq1:90}), which, recall, is dependent upon the four-velocity
$u^{\alpha}$ and the background metric $g_{\alpha\beta}$ at point z.\footnote{
In the original MiSaTaQuWa formulation, the SF is not expressed directly
in terms of the gradient of $\bar h^{\rm tail}_{\alpha\beta}$, but rather
as an integral over the gradient of the relevant retarded Green's function
along the portion of the worldline to the past of $z$ [cf.\ Eq.\ (1.9.6) of
Poisson \cite{Poisson:2003nc}].
The commutation of the derivative and the worldline integral
produces local terms at $z$, which, however, vanish upon contraction with
$\bar\nabla^{\alpha\beta\gamma}$. This leads to the equivalent formulation
shown here in Eq.\ (\ref{eq1:500}).}

\subsection{Detweiler-Whiting reformulation}

In a 2003 paper \cite{Detweiler:2002mi} (see also a preliminary discussion in
\cite{Detweiler:2000gt}) Detweiler and Whiting (DW) proposed an alternative
formulation, which offers an interesting re-interpretation of the perturbed motion.
DW replaced the direct/tail decomposition of the retarded perturbation
[Eq.\ (\ref{eq1:150})] with a new decomposition,
\begin{equation} \label{eq1:160}
\bar h^{\rm ret}_{\alpha\beta}=\bar h^{\rm S}_{\alpha\beta}+
\bar h^{\rm R}_{\alpha\beta},
\end{equation}
where the $R$-field $\bar h^{\rm R}_{\alpha\beta}$ (R for Regular), unlike
the tail field, is a certain {\em smooth}, vacuum solution of the perturbation
equations, which, nonetheless, gives rise to the same physical SF as the tail
field:
\begin{equation} \label{eq1:600}
F^{\alpha}_{\rm self}(z)=\mu\bar\nabla^{\alpha\beta\gamma}
\bar h^{\rm R}_{\beta\gamma}(z).
\end{equation}
The $S$-field $\bar h^{\rm S}_{\alpha\beta}$ (S for Singular), which
precisely mimics the singular behavior of the retarded field near the
particle, exerts no SF and does not affect the motion of the particle.
The precise formal prescription for constructing the R and S fields is
described nicely in Poisson's review \cite{Poisson:2003nc}.

DW's discovery that the MiSaTaQuWa SF can also be expressed as the
back-reaction force from a smooth vacuum perturbation leads to an
interesting re-interpretation of the gravitational SF effect: The particle
effectively moves {\em freely} along a geodesic of a smooth perturbed
spacetime with metric $g_{\alpha\beta}+h_{\alpha\beta}^R$. In this
alternative picture---more in the spirit of GR's equivalence principle---the
notion of a SF becomes artificial (and obsolete) in much the the same way that
the notion of an external gravitational force is artificial.

It should be understood, however, that the $R$-field does not represent
the actual physical perturbation from the particle (the physical perturbation
field is of course $\bar h^{\rm ret}_{\alpha\beta}$). The $R$-field has peculiar
causal properties which make it problematic as a candidate for what we may call
a ``physical field'': The value of the $R$-field at an event $x$ depends not
only on events in the the causal past of $x$ but also on events {\em outside}
the light-cone of $x$ \cite{Poisson:2003nc}. Rather than an entity of physical
substance, the $R$-field should be viewed as an {\em effective} field that allows
us to describe the dynamics in terms of geodesic motion.

The two descriptions of the perturbed motion---self-accelerated motion in $g$
vs.\ geodesic motion in $g+h^R$---are alternative (equivalent) interpretations
of the same (genuine) physical effect. The two points of view are not contradictory
but rather they are complementary in their perspective on the problem.
Workers in the field often find it useful to invoke both descriptions
alternately in order to get a fuller picture of the physics in question.
Sago {\it et al.}~\cite{Sago:2008id} recently demonstrated the equivalence
of the two approaches with an explicit calculation of a certain gauge-invariant
physical SF effect in a particular example; we shall return to discuss
this work in Sec.\ \ref{sec:effects}.

\subsection{Singular field}

Equations (\ref{eq1:500}) and (\ref{eq1:600}) prescribe the correct
regularization of the gravitational SF and form the fundamental basis
for all modern SF calculations. In later sections we will discuss practical
methods for implementing these formulas. We will then need some more
information on the properties of the direct (or singular) field,
which we now give.


The fields $\bar h^{\rm dir}_{\alpha\beta}$ and $\bar h^{\rm S}_{\alpha\beta}$
share the same leading-order singular structure near the particle's worldline
$\Gamma$, but they differ in their sub-dominant singular behavior.
More precisely (referring again to Fig.\ \ref{fig1}), consider a particular point
$z$ on $\Gamma$ and a nearby off-$\Gamma$ field point $x$.
The form of both the direct and $S$ fields is then given, in the Lorenz gauge,
by \cite{Mino:2001mq,Barack:2002bt,Nakano:2003he}
\begin{equation}\label{eq1:170}
\bar h^{S,{\rm dir}}_{\alpha\beta}(x;z)
=\frac{4\mu\, \hat u_{\alpha}(x;z)\hat u_{\beta}(x;z)}{\epsilon(x;z)}
+\frac{\mu\, w_{\alpha\beta}^{\rm S,dir}(x;z)}{\epsilon(x;z)}
+c_{\alpha\beta}^{\rm S,dir},
\end{equation}
where $\hat u_{\beta}$ is the four-velocity vector parallelly-propagated
from $z$ to $x$, $\epsilon$ is the spatial geodesic distance from $x$ to $\Gamma$
(i.e., the proper length of the short normal geodesic section connecting
$x$ to $\Gamma$),
$w^{\rm S,dir}_{\alpha\beta}$ are smooth functions of $x$ (and $z$) which
vanish at $x\to z$ at least {\em quadratically} in the coordinate differences
$x^{\alpha}-z^{\alpha}$, and $c_{\alpha\beta}^{\rm S,dir}$ are constants
(dependent on $z$ but not on $x$).
The S and direct fields differ only in the explicit form of $w_{\alpha\beta}$
and (possibly) in the value of $c_{\alpha\beta}$, but neither of these will
be important for us in this review.

It is, however, important to emphasize that the singular form (\ref{eq1:170})
is generally gauge dependent. The expression given here is specific to the
Lorenz gauge, and is not guaranteed to keep its form if one makes
other gauge choices for the perturbation. Indeed, it has been demonstrated
\cite{Barack:2001ph} that some common gauge choices can ``distort'' the
local rest-frame isotropy of the Lorenz-gauge singularity manifest
(at leading order) in Eq.\ (\ref{eq1:170}).

\subsection{Gauge dependence}

Earlier we emphasized some important differences between the concepts of
a general (external) gravitational force and that of the SF. The two
notions, however, share a basic common feature: They are both defined
via a mapping of the physical trajectory from a ``perturbed'' spacetime
to a ``background'' spacetime. In both cases, such a mapping procedure gives
rise to a gauge ambiguity. A thorough analysis of the gauge dependence of the
gravitational SF was presented in Ref.\ \cite{Barack:2001ph}, and a gauge
transformation law for the SF was derived. Consider again the infinitesimal
gauge transformation of Eq.\ (\ref{eq1:100}), where now the gauge displacement
vector $\xi^{\alpha}$ is assumed to scale like the particle's mass $\mu$.
The change this induces on the physical SF was found \cite{Barack:2001ph}
to be given by
\begin{equation} \label{eq1:180}
\delta_{(\xi)}F^{\alpha}_{\rm self} =
-\mu\left[(g^{\alpha\lambda}+u^{\alpha}u^{\lambda})\ddot{\xi}_{\lambda}
+R^{\alpha}_{\ \mu\lambda\nu}u^{\mu}\xi^{\lambda}u^{\nu}
\right].
\end{equation}
This has the same form as the gauge transformation law for the external
gravitational force [recall Eq.\ (\ref{eq1:130})], which is not surprising
given the similar geometrical origin of the gauge freedom in both cases.

We note, however, that the derivation of the transformation rule (\ref{eq1:180})
in Ref.\ \cite{Barack:2001ph} (which pre-dates the DW analysis) could not---and
did not---follow the procedure we implemented above in deriving Eq.\ (\ref{eq1:130}),
simply replacing the external perturbation $h_{\alpha\beta}$ with the
physical perturbation from the particle, $h^{\rm ret}_{\alpha\beta}$.
That is because $h^{\rm ret}_{\alpha\beta}$, unlike the smooth external field
$h_{\alpha\beta}$, is singular at the particle, and so expressions
such as (\ref{eq1:110}) or (\ref{eq1:120}) (with $h_{\alpha\beta}\to
h^{\rm ret}_{\alpha\beta}$) would make no sense when evaluated along the
particle's worldline. DW's later $R$-field interpretation suggests an alternative
derivation of Eq.\ (\ref{eq1:180}): Since the $R$-field can be viewed, effectively,
as a smooth external perturbation (even if one with peculiar causal properties),
and since the SF is the force exerted by this
perturbation, the derivation leading to Eq.\ (\ref{eq1:130}) can be repeated in
full with $h_{\alpha\beta}\to h^{\rm R}_{\alpha\beta}$, and Eq.\ (\ref{eq1:180})
follows immediately. In effect, the $R$-field interpretation allows us here to treat
the SF on an equal footing with the gravitational force from an external perturbation.

The gauge dependence of the SF by no means implies that there is something
``unphysical'' about it---the SF is as physical as the metric perturbation
itself, which is also gauge dependent. The gauge dependence does mean,
however, that one needs to exercise some care in decoding the physical content
of the SF. One cannot expect to be able to describe the physical effect of the
SF based on the value of the SF alone (to put this to extreme: one can always
make a gauge choice that nullifies the SF anywhere along the orbit!).
Instead, a meaningful description of the physical effect must involve
both the SF and the gauge information associated with it (in the form of
the metric perturbation, for example).

Another crucial point to have in mind is that the MiSaTaQuWa formula
(\ref{eq1:500}) is guaranteed to hold true only if $\bar h^{\rm ret}
_{\alpha\beta}$ satisfies the Lorenz-gauge condition (\ref{eq1:140}).
Strictly speaking, the SF in a given non-Lorenz gauge only makes
sense if, for $\xi^{\alpha}$ relating that gauge to the Lorenz gauge,
the expression on the right-hand side of Eq.\ (\ref{eq1:180}) is well defined.
This is not at all an obvious condition; some simple counter-examples are
analyzed in \cite{Barack:2001ph}. In gauges for which the right-hand side
of (\ref{eq1:180}) does not have a well defined (i.e., finite and direction independent)
particle limit, one might still devise a useful notion of the SF by averaging over
angular directions, or by taking a directional limit in a consistent fashion.
Such possibilities are discussed in Refs.\ \cite{Barack:2001ph,Gralla:2008fg}.

\subsection{Equations of motion}

Given the SF, the particle's equation of motion becomes
\begin{equation} \label{eq1:190}
\mu u^{\beta}\nabla_{\beta}u^{\alpha}=F^{\alpha}_{\rm self},
\end{equation}
where on the left-hand side we have the usual four-acceleration (times $\mu$)
along the background trajectory. This equation, along with the MiSaTaQuWa equation
(\ref{eq1:500}), describe the dynamics of the particle given the metric
perturbation (and assuming one has a way of extracting the tail piece out
of the full perturbation). To close the system of equations we need to know
how the metric perturbation is determined from the particle's trajectory.
This, of course, is provided by the linearized Einstein equation, which
takes the Lorenz-gauge form
\begin{equation}\label{eq1:200}
\nabla^{\gamma}\nabla_{\gamma} \bar h^{\rm ret}_{\alpha\beta}
+2R^{\mu}{}_{\alpha}{}^{\nu}{}_{\beta}\bar h^{\rm ret}_{\mu\nu}
=-16\pi \mu \int_{-\infty}^{\infty}
(-g)^{-1/2}\,\delta^4[x^{\mu}-z^{\mu}(\tau)]u_{\alpha}u_{\beta}\,d\tau,
\end{equation}
where $g$ is the determinant of $g_{\alpha\beta}$ and $z^{\mu}(\tau)$,
recall, describes the particle's worldline. The source on the right-hand side
is the usual distributional representation of the point particle's energy-momentum.
The field equation (\ref{eq1:200}) is to be supplemented by the gauge condition
(\ref{eq1:140}) and by suitable boundary conditions.

The set of equations (\ref{eq1:500}), (\ref{eq1:190}), (\ref{eq1:200})
and (\ref{eq1:140}) (together with a method to obtain $\bar h^{\rm tail}$ out of
$\bar h^{\rm ret}$) should in principle determine the dynamics of the orbit
at linear order in perturbation theory. However, as pointed out by
Garlla and Wald recently \cite{Gralla:2008fg}, the field equation (\ref{eq1:200})
is only consistent with the Lorenz gauge condition (\ref{eq1:140}) if
the particle is moving strictly along a geodesic---which would then be
inconsistent with the equation of motion (\ref{eq1:190}). To resolve this
inconsistency while allowing for orbital evolution, Gralla and Wald suggested
a ``Lorenz-gauge relaxation'' approach, wherein one relaxes the gauge
condition (\ref{eq1:140}) and considers solutions of the set
(\ref{eq1:500},\ref{eq1:190},\ref{eq1:200}). One then expects that, in situations
where the orbit is very nearly geodesic (as is usually the case with LISA-relevant
astrophysical inspirals), such solutions would give a faithful, albeit approximate
description of the actual orbit. This approach is yet to implemented and
tested in actual calculations of the orbital evolution.

The proposal to use self-consistent solutions of the set (\ref{eq1:500},
\ref{eq1:190},\ref{eq1:200}) for modeling the slow orbital evolution at linear
order in perturbation theory was put forward by Gralla and Wald in
\cite{Gralla:2008fg}. A different mathematical framework, based on techniques
from multi-scale perturbation theory, was developed by Hinderer and Flanagan in
\cite{Hinderer:2008dm}. Pound and Poisson \cite{Pound:2007ti,Pound:2007th}
performed first actual calculations of the orbital evolution under the full
SF effect, using multi-scale analysis, within a weak-field, PN framework.
Thus far there are no calculations of the full orbital evolution in
strong-field scenarios.

In the rest of this review we will not consider any further the question of
orbital evolution, but rather focus on the calculation of the SF [via Eqs.\
(\ref{eq1:500}) or (\ref{eq1:600})] along a pre-determined, non-evolving orbit.
We consider this a preliminary step, whose output (e.g., the value of the
SF at sufficiently many points across the relevant phase space) can later be
incorporated into whichever evolution scheme one chooses to apply.

\section{Overview of implementation frameworks and calculations to date}
\label{sec:methods}

Starting in the late 1990s, work began to translate MiSaTaQuWa's SF formalism
into practical working schemes and to implement it in actual calculations.
While the ``holy grail'' of this program has from the outset been---and still
remains---the calculation of the gravitational SF for generic orbits around
Kerr black holes, much of the initial effort has focused on the simpler toy
problem of the scalar-field SF, and on simple classes of orbits (radial, circular)
in Schwarzschild spacetime. The last few years, however, saw first
calculations of the gravitational (and electromagnetic) SFs for generic
orbits in Schwarzschild---and initial work on Kerr is now underway.
In this section we give a broad overview of the methods that have been
suggested for implementing MiSaTaQuWa's formalism, and we survey the
actual calculations performed so far.

Our starting point are the formal expressions (\ref{eq1:500}) and (\ref{eq1:600})
for the gravitational SF. To facilitate the following discussion, we first use
Eqs.\ (\ref{eq1:150}) and (\ref{eq1:160}) to recast these expressions in the
more practical form
\begin{equation} \label{eq3:10}
F^{\alpha}_{\rm self}(z)=
\mu \lim_{x\to z}\left[\bar\nabla^{\alpha\beta\gamma}_x\bar h^{\rm ret}_{\beta\gamma}(x)
-\bar\nabla_x^{\alpha\beta\gamma}\bar h^{\rm dir/S}_{\beta\gamma}(x)\right],
\end{equation}
referring once again to Fig.\ \ref{fig1} for notation. This describes a
``regularization'' procedure, which one can perform using either
$\bar h_{\beta\gamma}^{\rm dir}$ or $\bar h_{\beta\gamma}^{\rm S}$---both
producing the same final value for the SF.
The limit procedure is necessary here because the individual fields
$\bar h^{\rm ret}_{\beta\gamma}$ and $\bar h^{\rm dir/S}_{\beta\gamma}$,
unlike their difference $\bar h^{\rm tail/R}_{\beta\gamma}$, are each
singular at $x\to z$, and so are their derivatives. Since the operator
$\bar\nabla^{\alpha\beta\gamma}$ is only defined along the particle's
worldline [as it involves the four-velocity---recall Eq.\ (\ref{eq1:90})],
in Eq.\ (\ref{eq3:10}) we needed to introduce an extension of this operator
off the worldline---denoted $\bar\nabla^{\alpha\beta\gamma}_x$.
For all $x$ in the neighborhood of a given worldline point $z$, the operator
$\bar\nabla^{\alpha\beta\gamma}_x$ is given by Eq.\ (\ref{eq1:90}), where
$g^{\alpha\beta}$ and $u^{\alpha}$ take the same values they have at $z$,
and $\nabla_{\delta}$ is the standard covariant derivative at point $x$.
Here, and throughout this review, we use this ``fixed contravariant components''
extension of $\bar\nabla^{\alpha\beta\gamma}$ exclusively, although other
natural extensions are possible
\cite{Barack:2002bt}. Of course, the final value of the SF in Eq.\ (\ref{eq3:10}),
after the limit $x\to z$ is taken, is not sensitive to the choice of extension.
Notice that the definition of $\bar\nabla^{\alpha\beta\gamma}_x$ is, of course,
coordinate dependent, and it becomes well-posed only in reference to a specific
coordinate system. Also note that the operator $\bar\nabla^{\alpha\beta\gamma}_x$
is defined with respect to a given point $z$.

The most basic technical challenge one faces in preparing to implement the
MiSaTaQuWa formula is the so-called {\it subtraction problem}: How does one go
about extracting the tail (or R) piece from the full retarded perturbation
in practice? Equation (\ref{eq3:10}) suggests applying the subtraction
$\bar h^{\rm ret}_{\alpha\beta}-\bar h^{\rm dir/S}_{\alpha\beta}$,
using approximate analytic expressions for the direct/S fields,
such as the one in Eq.\ (\ref{eq1:170}).
However, this involves the removal of one divergent
quantity from another, which is not easily tractable in actual numerical
calculations. Several strategies have been proposed to address this
problem, and in the main part of this review we shall describe a
few of them in some detail. Here we proceed with a brief overview of the main
avenues of approach to this problem.

\paragraph{Quasi-local/matched expansions calculations.}
\cite{Anderson:2003qa,Anderson:2004eg,Anderson:2005gb,Anderson:2005ds,
Ottewill:2007mz,Ottewill:2008uu} This approach tackles the calculation
of the tail contribution directly, by analytically evaluating the Hadamard
expansion of the relevant Green's function. Such calculations capture the ``near''
part of the tail, which, one might hope, represents the dominant contribution
in problems of interest. Quasi-local calculations can be supplemented
by a numerical computation of the ``far'' part of the tail, a strategy
referred to as ``matched expansions'' (not to be confused with matched
asymptotic expansions). The applicability of this idea was demonstrated
very recently \cite{Casals:2009zh} with a full calculation in Nariai
spacetime (a simple toy spacetime featuring many of the characteristics
of Schwarzschild).

\paragraph{Weak-field analysis.}
\cite{DeWitt*2,Pfenning:2000zf,Pound:2005fs,Pound:2007ti,Pound:2007th}
The tail formula can be evaluated analytically
for certain weak-field configurations, within a Newtonian or a post-Newtonian
(PN) framework. Such work has provided important insight into the nature and
properties of the SF. PN techniques have also been implemented
in combination with the mode-sum method discussed below
\cite{Nakano:2000ne,Nakano:2001kw,Nakano:2003he,Hikida:2003pi,Hikida:2004jw}.

\paragraph{Radiation-gauge regularization.} \cite{Keidl:2006wk}
This approach proposes a reformulation of the MiSaTaQuWa regularization, in which
one reconstructs the R-part of the metric perturbation in a radiation gauge (rather
than in the Lorenz gauge) from a suitably regularized Newman--Penrose curvature
scalar ($\psi_0$ or $\psi_4$). The main advantage of this method is that it reduces
the numerical component of the calculation to a
solution of a single scalar-like (Teukolsky's) equation. However, some of the technical
complexity is relegated to the metric reconstruction step. This technique
was implemented so far only for a particle held static in Schwarzschild,
but more interesting cases are currently being studied \cite{:2007mn,Friedman}.
See the discussion in Sec.\ \ref{sec:3.1} for more details.

\paragraph{Mode-sum method.}
\cite{Barack:1999wf,Lousto:1999za,Barack:2001bw,Mino:2001mq,Barack:2001gx,Barack:2002mha,
Barack:2002bt,Barack:2002mh,Detweiler:2002gi,Hikida:2003pi,Hikida:2004jw,Haas:2006ne}
An approach whereby one evaluates the tail contribution mode by mode in
a multipole expansion. The subtraction ``ret$-$dir'' (or ``ret$-$S'') in Eq.\
(\ref{eq3:10}) is performed mode by mode, avoiding the
need to deal with divergent quantities. The method exploits the
separability of the field equations in Kerr into multipole harmonics.
The mode-sum method has provided the framework for the bulk of work on
SF calculations over the last decade. We will discuss it in detail in
Sec.\ \ref{sec:modesum}.

\paragraph{Puncture methods.}
\cite{Barack:2007jh,Barack:2007we,Vega:2007mc,:2007mn,Lousto:2008mb}
A set of recently-proposed methods custom-built for time-domain numerical
implementation in 2+1 or 3+1 dimensions. Common to these methods is
the idea to utilize as a variable for the numerical time-evolution
a ``punctured'' field, constructed from the full (retarded) field by removing
a suitable singular piece, given analytically. The piece removed approximates
the correct S-field sufficiently well that the resulting ``residual''
field is guaranteed to yield the correct MiSaTaQuWa SF.  In the
2+1D version of this approach the regularization is done
mode by mode in the azimuthal ($m$-mode) expansion of the full field.
This procedure offers significant simplification; we shall review
it in detail in Sec.\ \ref{sec:puncture}.

\subsection{SF calculations to date}

As we have mentioned already, the program to calculate the SF for black
hole orbits has been progressing gradually, through the study of a
set of simplified model problems.
Some of the necessary computational techniques were first tested
within the simpler framework of a scalar-field toy model before being
applied to the electromagnetic (EM) and gravitational problems.
Authors have considered special classes of orbits (static, radial, circular)
before attempting more generic cases, and much of the work so far has
focused on Schwarzschild orbits. The state of the art in the field are
numerical codes to compute the scalar, EM and gravitational SFs along
any given (geodesic) orbit outside a Schwarzschild black hole. It is reasonable
to expect that attention will now be increasingly drawn to the Kerr problem.

The information in Tables \ref{tab:1}--\ref{tab:3} is meant to provide a
quick reference to work done so far. It covers actual evaluations of the
local SF that are based on the MiSaTaQuWa formulation (or the analogous
scalar-field and EM formulations of Refs.\ \cite{Quinn:2000wa} and
\cite{DeWitt:1960fc,Hobbs:1968,Quinn:1996am}, respectively), either
directly or through one of the aforementioned implementation methods.
We have included weak-field and PN implementations, but have not included
work based on the radiative field approach. Some of the techniques
referred to under ``strategy'' will be discussed in the following sections.
\begin{table}[thb]
\begin{flushright}
\begin{tabular}{lll}
\hline\hline\noalign{\smallskip}
\multicolumn{3}{c}{\bf Scalar-field self force} \\
\hline\hline\noalign{\smallskip}
Case  &  Author(s) & Strategy   \\
\noalign{\smallskip}\hline \hline
Newtonian potential
& Pfenning \& Poisson \cite{Pfenning:2000zf}
& direct, analytic  \\
\hline
spherical mass shell:
& Burko {\it et al.}\ \cite{Burko:2000yx}
&  mode sum, analytic \\
static particle && \\
\hline
isotropic cosmology:
&  Burko {\it et al.}\  \cite{Burko:2002ge}
&  direct, analytic \\
 static particle && \\
\hline
isotropic cosmology:
& Haas \& Poisson \cite{Haas:2004kw}
&  direct, analytic  \\
slow motion && \\
\hline
Nariai spacetime:
& Casals  {\it et al.}\ \cite{Casals:2009zh}
&  matched expansions \\
static particle && \\
\hline
Schwarzschild: & Burko \cite{Burko:1999zy} &  mode sum, analytic \\ \cline{2-3}
static particle & Wiseman \cite{Wiseman:2000rm} & direct, analytic\\
\hline
Schwarzschild:
& Barack \& Burko \cite{Barack:2000zq}&  mode sum, numerical    \\
 radial geodesics & &(1+1D evolution)  \\
\hline
Schwarzschild:
& Nakano {\it et al.}\  \cite{Nakano:2001kw}
&  post-Newtonian, analytic \\
circular geodesics & Hikida {\it et al.}\ \cite{Hikida:2004jw} & \\
\cline{2-3}
\mbox{}
& Burko \cite{Burko:2000xx} &  mode sum, numerical  \\
\mbox{} & Detweiler {\it et al.}\
\cite{Detweiler:2002gi,DiazRivera:2004ik}& (frequency domain) \\
\cline{2-3}
\mbox{}
& Vega \& Detweiler \cite{Vega:2007mc}
&  puncture, numerical  \\
&& (1+1D evolution) \\
\cline{2-3}
\mbox{}
& vega {\it et al.}\  \cite{Vega:2009qb}
&  puncture, numerical  \\
&& (3+1D evolution) \\
\hline
Schwarzschild:
& Haas \cite{Haas:2007kz} &  mode sum, numerical  \\
eccentric geodesics && (1+1D evolution) \\
\hline
Kerr--Newman:
&  Burko \& Liu \cite{Burko:2001kr} &  mode sum, analytic \\
static particle && \\
\hline
Kerr: circular-
& Warburton \& Barack  \cite{BWprep}
& mode sum, numerical   \\
equatorial geodesics && (frequency domain)\\
\hline\hline
\end{tabular}
\caption{Calculations of the {\em scalar-field} SF as a toy model for the
gravitational problem. In this table, as well as in Tables \ref{tab:2} and \ref{tab:3},
``direct'' implies explicit evaluation of the tail contribution to the SF.
``Mode sum'' and ``puncture'' refer to the two implementation schemes
described, respectively, in Sec.\ \ref{sec:modesum} and \ref{sec:puncture}
of this review. The method of ``matched expansions'' (different from ``matched
asymptotic expansions'') is described briefly in Secs.\ \ref{sec:methods}.}
\label{tab:1}
\end{flushright}
\end{table}
%
\begin{table}[thb]
\begin{flushright}
\begin{tabular}{lll}
\hline\hline\noalign{\smallskip}
\multicolumn{3}{c}{\bf Electromagnetic self force} \\
\hline\hline\noalign{\smallskip}
Case  &  Author(s) & Strategy   \\
\noalign{\smallskip}\hline \hline
Newtonian potential
& Pfenning \& Poisson \cite{Pfenning:2000zf}
& direct, analytic\\
\hline
Isotropic cosmology:
& Haas \& Poisson \cite{Haas:2004kw}
&  direct, analytic \\
slow motion && \\
\hline
Schwarzschild: & Burko \cite{Burko:1999zy}
&  mode sum, analytic\\
static particle && \\
\cline{2-3}
 & Keidl {\it et al.}\ \cite{Keidl:2006wk}
&  radiation-gauge \\
& & regularization, analytic \\
\hline
Schwarzschild:  & Haas \cite{Haas:Capra}
        &  mode sum, numerical\\
eccentric geodesics & &  (1+1D evolution) \\
\noalign{\smallskip}\hline\hline
\end{tabular}
\caption{Recent calculations of the {\em Electromagnetic} self-force.}
\label{tab:2}
\end{flushright}
\end{table}
%
\begin{table}[thb]
\begin{flushright}
\begin{tabular}{lll}
\hline\hline\noalign{\smallskip}
\multicolumn{3}{c}{\bf Gravitational self force} \\
\hline\hline\noalign{\smallskip}
Case  &  Author(s) & Strategy   \\
\noalign{\smallskip}\hline \hline
Newtonian potential
& Pfenning \& Poisson \cite{Pfenning:2000zf}
& direct, analytic\\
\hline
Schwarzschild:  & Barack \& Lousto \cite{Barack:2002ku}
        & mode sum, 1+1D evolution \\
radial geodesics && in Regge--Wheeler gauge \\
\hline
Schwarzschild: & Keidl {\it et al.}\ \cite{Keidl:2006wk}
&  radiation-gauge \\
static particle & & regularization, analytic  \\
\hline
Schwarzschild: & Barack \& Sago \cite{Barack:2007tm}
        & mode sum, 1+1D evolution \\
circular geodesics && in Lorenz gauge \\
\cline{2-3}
  & Detweiler \cite{Detweiler:2008ft}
        & mode sum, frequency-domain \\
 && in Regge--Wheeler gauge \\
\hline
Schwarzschild: & Barack \& Sago \cite{Barack:2009ey,BSprep}
        & mode sum, 1+1D evolution\\
eccentric geodesics &&  in Lorenz gauge \\
\noalign{\smallskip}\hline\hline
\end{tabular}
\caption{Calculations of the {\em Gravitational} self-force.}
\label{tab:3}
\end{flushright}
\end{table}
%

\section{Mode-sum method}
\label{sec:modesum}

Let us write the subtraction formula (\ref{eq3:10}) using the more
compact notation
\begin{equation} \label{eq4:10}
F^{\alpha}_{\rm self}(z)=\lim_{x\to z}
\left[F_{\rm ret}^{\alpha}(x)-F^{\alpha}_{\rm S}(x)\right],
\end{equation}
where we have introduced the fields ($\propto\mu^2$)
\begin{equation} \label{eq4:20}
F_{\rm ret}^{\alpha}(x)=\mu \bar\nabla_x^{\alpha\beta\gamma}
\bar h^{\rm ret}_{\beta\gamma}(x), \quad\quad
F^{\alpha}_{\rm S}(x)=\mu \bar\nabla_x^{\alpha\beta\gamma}
                        \bar h^{\rm S}_{\beta\gamma}(x).
\end{equation}
For concreteness and simplicity we adopt here the S-field subtraction,
noting that the entire discussion in this section would not be altered
upon replacing S$\to$direct and R$\to$tail.\footnote{The only exception is that
statements referring to the smoothness of the R field would need to be
formulated more carefully to reflect the irregularity in the higher derivatives
of the tail field. This irregularity, however, would have little practical impact
on the discussion in this section.}
The fields $F_{\rm ret}^{\alpha}$ and $F^{\alpha}_{\rm S}$ inherit the extension
ambiguity of $\bar\nabla_x^{\alpha\beta\gamma}$,
but here we shall always use the (coordinate dependent) ``fixed'' extension
described above. Both $F_{\rm ret}^{\alpha}$ and $F^{\alpha}_{\rm S}$,
of course, diverge at the particle, but their difference is a {\em smooth}
(analytic) function of $x$ even at the particle.

In the mode-sum method we formally decompose each vectorial component of both
$F_{\rm dir}^{\alpha}$ and $F^{\alpha}_{\rm S}$ into spherical harmonics.
These harmonics are defined in the Kerr/Schwarzschild
background based on the Boyer-Lindquist/Schwarzschild coordinates
$(t,r,\theta,\varphi)$ in the standard way, i.e., through a projection
onto an orthogonal basis of angular functions defined on surfaces
of constant $t$ and $r$.  Let us denote by $F_{\rm ret}^{\alpha l}(x)$ and
$F^{\alpha l}_{\rm S}(x)$ the $l$-mode contributions to
$F_{\rm ret}^{\alpha}$ and $F^{\alpha}_{\rm S}$, respectively (summed over $m$).
A key observation is that each of these $l$-mode fields is finite
even at the particle. This suggests a natural regularization procedure,
which, essentially, amounts to performing the subtraction
$F_{\rm ret}^{\alpha}-F^{\alpha}_{\rm S}$ {\em mode by mode}.
The idea is best developed through an elementary example, as follows.

\subsection{An elementary example}\label{sec:2.1}

Consider a pointlike particle of mass $\mu$ at rest in flat space.
The location of the particle is $\vec{x}=\vec{x}_{\rm p}$ in a given
Cartesian system. In this simple static configuration the perturbed
Einstein equations (\ref{eq1:200}) read
\begin{equation} \label{eq4:30}
\nabla^2\bar h^{\rm ret}_{tt}=-16\pi \mu\, \delta^3(\vec{x}-\vec{x}_{\rm p}),
\end{equation}
where $\nabla^2$ is the 3D Laplacian, and with all other components of
$\bar h^{\rm ret}_{\alpha\beta}$ vanishing.\footnote{Here the label `ret'
is less appropriate, but we shall retain it to adhere to our general notation.
As in the general case, $\bar h^{\rm ret}_{\alpha\beta}$ represents the unique
physical perturbation---here the unique regular static solution of Eq.\
(\ref{eq4:30}).} The static perturbation $\bar h^{\rm ret}_{\alpha\beta}$
automatically satisfies the Lorenz-gauge condition (\ref{eq1:140}). Of course,
in this simple case we can immediately write down the exact physical
(Coulomb-like) static solution, $\bar h^{\rm ret}_{tt}=4\mu/|\vec{x}-\vec{x}_{\rm p}|$,
and we also trivially have $F^{\alpha}_{\rm self}=0$.
However, for the sake of our discussion, let us proceed by considering the
multipole expansion of the perturbation.

\setcounter{footnote}{0}

\begin{figure}[htb]
\input{epsf}
\centerline{\includegraphics[totalheight=4.5cm]{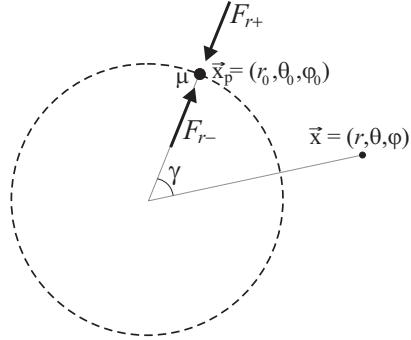}}
\caption{
An illustration of the simple setup described in the text:
A particle of mass $\mu$ in flat space is at rest at
$\vec x_{\rm p}=(r_0,\theta_0,\varphi_0)$. The gravitational field of the particle
is decomposed into spherical harmonics, each contributing a finite
amount to the full radial force acting on the particle: either $F^l_{r+}$
or $F^l_{r-}$, depending on whether the force is calculated from
$r\to r^+$ or $r\to r^-$. $\vec x=(r,\theta,\varphi)$ is an arbitrary field
point used in the construction described in the text.
}
\label{fig2}       
\end{figure}

To this end, introduce polar coordinates $(r,\theta,\varphi)$, such that
our particle is located at $\vec{x}_{\rm p}=(r_0\ne 0,\theta_0,\varphi_0)$,
and expand the physical solution $\bar h_{tt}$ in spherical harmonics
on the spheres $r=$const, in the form
\begin{equation}\label{eq4:120}
\bar h^{\rm ret}_{tt}=\sum_{l=0}^{\infty}\bar h_{tt}^l(r,\theta),\quad
{\rm where}
\quad \bar h_{tt}^{l}(r,\theta)= \sum_{m=-l}^{l}
\tilde h_{tt}^{lm}(r)\,Y_{lm}(\theta,\varphi).
\end{equation}
This expansion separates the field equation (\ref{eq4:30}) into radial
an angular parts, the former reading (for each $l,m$)
\begin{equation} \label{eq4:130}
\tilde h^{lm}_{tt,rr}+\frac{2}{r}\tilde h^{lm}_{tt,r}-\frac{l(l+1)}{r^2}\,
\tilde h_{tt}^{lm}=
\frac{-16\pi\mu}{r_0^2}\,Y_{lm}^*(\theta_0,\varphi_0)\,\delta(r-r_0),
\end{equation}
where an asterisk denotes complex conjugation and a comma denotes partial
differentiation. The unique physical $l,m$-mode solution, continuous everywhere
and regular at both $r=0$ and $r\to\infty$, reads
\begin{equation} \label{eq4:140}
\tilde h_{tt}^{lm}(r)=\frac{16\pi\mu}{(2l+1)r_0}\, Y_{lm}^*(\theta_0,\varphi_0)
\times\left\{
\begin{array}{ll}
(r/r_0)^{-l-1}, & r\geq r_0, \\
(r/r_0)^l     , & r\leq r_0,
\end{array}
\right.
\end{equation}
giving
\begin{equation} \label{eq4:145}
\bar h_{tt}^l(r,\theta)=\frac{4\mu}{r_0}\, P_l(\cos\gamma)
\times\left\{
\begin{array}{ll}
(r/r_0)^{-l-1}, & r\geq r_0, \\
(r/r_0)^l     , & r\leq r_0,
\end{array}
\right.
\end{equation}
where $P_l$ is the Legendre polynomial and $\gamma$ is the
angle subtended by the two radius-vectors to $\vec x$ and $\vec x_{\rm p}$
(see Fig.\ \ref{fig2}).

We now construct the force field $F_{\rm ret}^{\alpha}$ as it is defined in
Eq.\ (\ref{eq4:20}). We find $F_{\rm ret}^t=0$, and the spatial components are
$F_{\rm ret}^{i}=\mu\bar\nabla^{itt}\bar h_{tt}=(\mu/4)\bar h_{tt}^{\;\, ,i}$.
Focus now on the $r$ component. The $l$ mode of $F_{\rm ret}^r$
is given simply as $F_{\rm ret}^{rl}=(\mu/4)\bar h^l_{tt,r}$.
Using Eq.\ (\ref{eq4:145}) and evaluating $F^{rl}_{\rm ret}$ at the particle
(taking $\gamma\to 0$ followed by $r\to r_0^{\pm}$), we obtain
\begin{equation} \label{eq4:150}
F^{rl}_{{\rm ret}\pm}(\vec x_{\rm p})=
\mp L\frac{\mu^2}{r_0^2}-\frac{\mu^2}{2r_0^2}, \quad {\rm where}
\quad L\equiv l+\frac{1}{2}.
\end{equation}
Here the subscripts $\pm$ indicate the two (different) values obtained
by taking the particle limit from ``outside'' ($r\to r^+$) and ``inside''
($r\to r^-$).

Let us note the following features manifest in the above simple analysis:
\begin{itemize}
\item The individual $l$ modes of the metric perturbation, $\bar h^l_{\alpha\beta}$,
are each continuous at the particle's location, although their derivatives are
discontinuous there.
\item The individual $l$ modes $F^{\alpha l}_{\rm ret}$, have {\em finite} one-sided
values at the particle.
\item At large $l$, each of the one-sided values of $F^{\alpha l}_{\rm ret}$ at the
particle is dominated by a term $\propto l$. (The mode sum obviously diverges
at the particle, reflecting the divergence of the full force
$F_{\rm ret}^\alpha$ there.)
\end{itemize}

It turns out (as we shall see later) that all above features are quite
generic, and they carry over intact to the much more general problem
of a particle moving in Kerr spacetime. Specifically, we find that,
at any point along the particle's trajectory, the (one-sided values of
the) modes $F^{\alpha l}_{\rm ret}$ always admit the large-$l$ form
\begin{equation} \label{eq4:160}
F^{\alpha l}_{{\rm ret}\pm}=\pm L A^{\alpha} +B^{\alpha}+ C^{\alpha}/L + O(L^{-2}).
\end{equation}
In our elementary problem the power series in $1/L$ truncates at
the $L^0$ term, but in general the series can be infinite.
The $l$-independent coefficients $A_{\alpha}$, $B_{\alpha}$ and $C_{\alpha}$,
whose values depend on the background geometry as well on the particle's location and
four-velocity, are characteristic of the local structure of the particle singularity
at large $l$. These coefficients, called {\it Regularization Parameters},
play a crucial role in the mode-sum regularization procedure, as we
describe next.

\subsection{The mode-sum formula}

Consider a mass particle moving on a geodesic trajectory in Kerr,
and suppose we are interested in the value of the SF at a point $z$
along the trajectory, with Boyer-Lindquist coordinates ($t_0,r_0,\theta_0,
\varphi_0$). Starting with Eq.\ (\ref{eq4:10}), let us formally
expand $F_{\rm ret}^{\alpha}(x)$ and $F^{\alpha}_{\rm S}(x)$ in spherical
harmonics on the surfaces $t,r$=const. Here we ignore the vectorial nature of
$F_{\rm ret}^{\alpha}$ and $F^{\alpha}_{\rm S}$ and, for mathematical
simplicity, treat each of their Boyer-Lindquist components as a scalar
function (see \cite{Haas:2006ne} for a more sophisticated, covariant treatment).
Denoting the respective $l$-mode contributions (summed over $m$) by
$F_{\rm ret}^{\alpha l}(x)$ and $F^{\alpha l}_{\rm S}(x)$, we write
\begin{equation} \label{eq4:170}
F^{\alpha}_{\rm self}(z)=\lim_{x\to z}
\sum_{l=0}^{\infty}\left[F_{\rm ret}^{\alpha l}(x)-F^{\alpha l}_{\rm S}(x)\right].
\end{equation}
We remind that the form of $F_{\rm ret}^{\alpha l}$ and $F^{\alpha l}_{\rm S}$
will depend on the specific off-worldline extension chosen for
$\bar\nabla^{\alpha\beta\gamma}$; however, this ambiguity disappears upon taking
the limit $x\to z$, and the final SF is, of course, insensitive to the choice of
extension.

Since $F_{\rm ret}^{\alpha}(x)-F^{\alpha}_{\rm S}(x)$ is a smooth function for all
$x$, the mode sum in Eq.\ (\ref{eq4:170}) is guaranteed to converge
exponentially for all $x$ (this is a general mathematical property of
the multipole expansion). In particular, the sum converges
uniformly at $x=z$, and we are allowed to change the order of
limit and summation. We expect, however, that the particle limit
of the individual terms $F_{\rm ret}^{\alpha l}$ and $F^{\alpha l}_{\rm S}$ is
only defined in a one-sided sense, as in the elementary example studied
above. Hence, we write
\begin{equation} \label{eq4:175}
F^{\alpha}_{\rm self}(z)=
\sum_{l=0}^{\infty}\left[F_{{\rm ret}\pm}^{\alpha l}(z)
-F^{\alpha l}_{{\rm S}\pm}(z)\right],
\end{equation}
where $\pm$ indicates the values obtained by first taking the limits
$t\to t_0$, $\theta\to\theta_0$ and $\varphi\to\varphi_0$, and then
taking $r\to r_0^{\pm}$. Of course, the difference
$F_{{\rm ret},\pm}^{\alpha l}(z)-F^{\alpha l}_{{\rm S},\pm}(z)$ does {\em not}
depend on the direction from which the radial limit is taken: As
the difference $F^{\alpha l}_{\rm ret}(x)-F^{\alpha l}_{\rm S}(x)$ is the $l$ mode
of a smooth function, it is itself smooth for all $x$.

Furthermore, since the mode sum in Eq.\ (\ref{eq4:175}) converges
faster than any power of $1/L$, we expect $F^{\alpha l}_{\rm ret}$ and $F^{\alpha l}_{\rm S}$
to share the same large-$l$ power expansion (\ref{eq4:160}), with the
same expansion coefficients. This motivates us to re-express
Eq.\ (\ref{eq4:175}) in the form
\begin{eqnarray} \label{eq4:180}
F^{\alpha}_{\rm self}(z)&=&
\sum_{l=0}^{\infty}\left[F_{{\rm ret}\pm}^{\alpha l}(z)\mp LA^{\alpha}-B^{\alpha}
-C^{\alpha}/L\right] \nonumber\\
&&-\sum_{l=0}^{\infty}\left[F_{{\rm S}\pm}^{\alpha l}(z)\mp
LA^{\alpha}-B^{\alpha}-C^{\alpha}/L\right].
\end{eqnarray}
Here, each of the terms in square brackets falls off at least
as $\sim l^{-2}$ as
$l\to\infty$, and so each of the two sums converges at least as $\sim 1/l$.
We hence arrive at the following mode-sum reformulation of the MiSaTaQuWa
equation:
\begin{equation} \label{eq4:200}
F^{\alpha}_{\rm self}(z)=\sum_{l=0}^{\infty}
\left[F_{{\rm ret}\pm}^{\alpha l}(z)\mp LA^{\alpha}-B^{\alpha}-C^{\alpha}\right]
-D^{\alpha},
\end{equation}
with
\begin{equation} \label{eq4:210}
D^{\alpha}\equiv \sum_{l=0}^{\infty}
\left[F_{{\rm S}\pm}^{\alpha l}(z)\mp LA^{\alpha}-B^{\alpha}-C^{\alpha}\right].
\end{equation}

The {\it mode-sum formula} (\ref{eq4:200}), first proposed in Refs.\
\cite{Barack:1999wf} (scalar-field case) and \cite{Barack:2001bw}
(gravitational case) provides a practical way of calculating the SF,
once the regularization parameters $A^{\alpha}$, $B^{\alpha}$,
$C^{\alpha}$ and $D^{\alpha}$ are known. It is {\em practical} because
(i) it involves no subtraction of divergent quantities, and (ii) it
builds naturally on the fact that in black hole perturbation theory
one usually calculates the metric perturbation mode by mode in a multipole
decomposition.

The values of the regularization parameters are obtained analytically via a local
analysis of the singular (or direct) field near the particle. Early calculations
of these parameters \cite{Barack:1999wf,Barack:2000eh,Barack:2001bw},
which were restricted to specific orbits, were based on a local analysis of
the Green's function near coincidence ($x\to z$) at large $l$. Later, after
the local form of the direct field had been derived explicitly to sufficient
accuracy \cite{Mino:2001mq}, workers have been able to derive general
expressions for the regularization parameters, valid for generic (geodesic)
orbits---first in Schwarzschild \cite{Barack:2001gx,Barack:2002mha,Barack:2002bt},
then in Kerr \cite{Barack:2002mh}. Later improvements
(in the scalar case) are due to Ref.\ \cite{Detweiler:2002gi}, where higher-order
terms in the $1/L$ expansion were derived analytically in order to accelerate
the convergence of the mode sum; and Ref.\ \cite{Haas:2006ne}, where the
regularization parameters were redefined as scalar quantities using a
covariant-form projection of the singular field onto a null tetrad based
on the worldline.

With the regularization parameters given in analytic form, the calculation
of the SF within the mode-sum scheme follows this procedure:
\begin{enumerate}
\item For a given geodesic orbit, calculate sufficiently many multipole modes
of the physical, retarded metric perturbation in the Lorenz gauge.
(Here ``multipole modes'' may refer to tensor harmonics, Fourier-tensor
harmonics, or spheroidal harmonics, depending on the case considered and on
the perturbation framework used for tackling the field equations.)
How this might be (and is being) done in practice---usually using numerical
methods---will be discussed in Sec.\ \ref{sec:numerics}.
\item Construct the $l$-modes $F_{{\rm ret}\pm}^{\alpha l}$ at the particle.
An example of how this is done in practice will be provided in Sec.\ \ref{sec:Lorenz}.
\item Use the modes $F_{{\rm ret}\pm}^{\alpha l}(z)$ as input for the mode-sum
formula (\ref{eq4:200}).
\end{enumerate}

The mode-sum procedure has the very useful
ability to automatically test itself: If either the values of the regularization
parameters $A^{\alpha}$, $B^{\alpha}$ and $C^{\alpha}$, or the values
of the $l$-modes $F_{{\rm ret}\pm}^{\alpha l}$ (usually computed numerically)
are wrongly calculated, the $l$-mode sum in Eq.\ (\ref{eq4:200}) will very
likely fail to converge. Conversely, if one finds that the calculated $l$-term
in the sum has the expected $\sim 1/L^2$ fall-off at large $L$, that provides
an excellent validation test for the entire calculation. So far, the analytic values
of the parameters $A^{\alpha}$, $B^{\alpha}$ and $C^{\alpha}$ have been
successfully ``tested'' (in the above sense) against numerical calculations
for generic orbits in Schwarzschild, and also (in the scalar case) for circular
equatorial orbits in Kerr.

In what follows we state the values of the regularization parameters in the
gravitational case, for generic geodesics in Kerr, as derived in Ref.\
\cite{Barack:2002mh}. In the Appendix we provide a full derivation of these
parameters (not included in  Ref.\ \cite{Barack:2002mh}).

In passing, we remind that the mode-sum formula (in the gravitational case)
is formulated in the Lorenz gauge, just like the MiSaTaQuWa formula on which
it relies. The values of the regularization parameters; the form of the
large-$l$ expansion in Eq.\ (\ref{eq4:160}); or even the very definiteness
of the SF---all of these are gauge dependent.
It has been shown, however, that the regularization parameters remain invariant
under gauge transformations (from the Lorenz gauge) that are sufficiently
regular \cite{Barack:2001ph}.

\subsection{Regularization parameters for generic orbits in Kerr}\label{subsec:RP}

We state here the values of the regularization parameters for the gravitational
SF at an arbitrary point $z$ along an arbitrary geodesic in Kerr spacetime.
We assume the point $z$ has Boyer--Lindquist coordinates
$(t_0,r_0,\theta_0,\varphi_0)$. The regularization parameters $C^{\alpha}$ and
$D^{\alpha}$  are always zero:
\begin{equation} \label{C}
C^{\alpha}=D^{\alpha}=0.
\end{equation}
The Boyer--Lindquist components of the regularization parameter $A^{\alpha}$
at $z$ are given by
\begin{eqnarray} \label{A}
A^{r} &=&-\frac{\mu^{2}}{V}\left(\frac{\sin^{2}\theta
_{0}}{g_{rr}g_{\theta \theta }g_{\varphi \varphi }}\right) ^{1/2}\left(
V+u_{r}^{2}/g_{rr}\right) ^{1/2},  \nonumber   \\
A^{t} &=&-(u_{r}/u_{t})A^{r},\quad \quad A^{\theta}
=A^{\varphi }=0,
\end{eqnarray}
where
\begin{equation} \label{V}
V\equiv 1+u_{\theta }^{2}/g_{\theta\theta}+u_{\varphi}^{2}/g_{\varphi\varphi}.
\end{equation}
Here $u^{\alpha}$ is the particle's four-velocity and $g_{\alpha\beta}$
is the background Kerr metric---both at point $z$.

The value of the parameter $B^{\alpha}$ is more complicated.
It can be expressed in a compact form as
\begin{equation}
B^{\alpha }=\mu^2(2\pi)^{-1}P^{\alpha}_{\ abcd}I^{abcd},  \label{B}
\end{equation}
where hereafter roman indices ($a,b,c,...$) run over the two
Boyer--Lindquist angular coordinates $\theta,\varphi$.
The coefficients $P^{\alpha}_{\ abcd}$ are
given by
\begin{eqnarray}\label{eqA60}
P^{\alpha}_{\ abcd} &=&
\frac{1}{2}\left[P^{\alpha}_{\ d}(3P_{abc}+2P_{ab}P_{c})-P^{\alpha\lambda}
(2P_{\lambda ab}+P_{ab\lambda})P_{cd}\right] \nonumber\\
&&+ (3P^{\alpha}_{\ a}P_{be}-P^{\alpha}_{\ e}P_{ab})C_{cd}^{e},
\end{eqnarray}
where
\begin{equation} \label{eqA70}
P_{\alpha}\equiv u^{\lambda}u^{\rho}g_{\lambda\rho,\alpha},
\quad\quad
P_{\alpha \beta} \equiv g_{\alpha\beta}+u_{\alpha}u_{\beta},
\quad\quad
P_{\alpha\beta\gamma} \equiv
\Gamma_{\alpha \beta}^{\lambda}P_{\lambda\gamma},
\end{equation}
and
\begin{equation}  \label{CC}
C_{\varphi \varphi }^{\theta }=\frac{1}{2}\sin \theta _{0}\cos \theta
_{0},\quad\quad C_{\theta \varphi }^{\varphi }=C_{\varphi \theta }^{\varphi }=-%
\frac{1}{2}\cot \theta _{0},
\end{equation}
with all other coefficients $C_{\alpha\beta}^{\gamma}$ vanishing, and with
$\Gamma_{\alpha\beta}^{\lambda}$ being the background connection coefficients at $z$.
The quantities $I^{abcd}$ are
\begin{equation} \label{Iabcd}
I^{abcd}=(\sin \theta _{0})^{-N}\int_{0}^{2\pi }G(\chi)^{-5/2}
(\sin\chi)^{N}(\cos\chi)^{4-N}\,d\chi ,
\end{equation}
where
\begin{equation}
G(\chi)=P_{\theta\theta}\cos^{2}\chi
+2P_{\theta{\varphi}}\sin\chi\cos\chi/\sin\theta_0
+P_{{\varphi}{\varphi}}\sin^{2}\chi/\sin^{2}\theta_0,
\label{G}
\end{equation}
and $N\equiv N(abcd)$ is the number of times the index $\varphi$ occurs
in the combination $(a,b,c,d)$, namely
\begin{equation}
N=\delta _{\varphi }^{a}+\delta _{\varphi }^{b}+\delta _{\varphi
}^{c}+\delta _{\varphi }^{d}.  \label{N}
\end{equation}


We may write $I^{abcd}$ explicitly in terms of standard complete
Elliptic integrals. Introducing the short-hand notation
\begin{equation}
\alpha \equiv \sin^2\theta_0\, P_{\theta \theta }/P_{{\varphi}{\varphi}}-1,
\quad\quad
\beta \equiv 2\sin\theta_0\, P_{\theta{\varphi}}/P_{{\varphi}\varphi},
\label{eq75}
\end{equation}
we have
\begin{equation}
I^{abcd}=\frac{(\sin \theta _{0})^{-N}}{d}\left[ QI_{K}^{(N)}\hat{K}%
(w)+I_{E}^{(N)}\hat{E}(w)\right] ,  \label{eqIabcd}
\end{equation}
where
\begin{equation}
Q=\alpha +2-(\alpha ^{2}+\beta ^{2})^{1/2},  \label{Q}
\end{equation}
\begin{equation}
d=3P_{{\varphi}{\varphi}}^{5/2}(\sin\theta_0)^{-5}(\alpha ^{2}+\beta
^{2})^{2}(4\alpha +4-\beta ^{2})^{3/2}(Q/2)^{1/2},  \label{d}
\end{equation}
$\hat{K}(w)\equiv \int_0^{\pi/2}(1-w\sin^2 x)^{-1/2}dx$ and 
$\hat{E}(w)\equiv \int_0^{\pi/2}(1-w\sin^2 x)^{1/2}dx$  are complete 
Elliptic integrals of the 1st and 2nd kinds, respectively, and the argument is
\begin{equation}
w=\frac{2(\alpha ^{2}+\beta ^{2})^{1/2}}{\alpha +2+(\alpha ^{2}+\beta
^{2})^{1/2}}.  \label{w}
\end{equation}
The ten coefficients $I_{K}^{(N)},I_{E}^{(N)}$ in Eq.\ (\ref{eqIabcd}) are given by
\label{eq125}
\begin{eqnarray}
I_{K}^{(0)} &=&4\left[ 12\alpha ^{3}+\alpha ^{2}(8-3\beta ^{2})-4\alpha
\beta ^{2}+\beta ^{2}(\beta ^{2}-8)\right] ,  \nonumber \\
I_{E}^{(0)} &=&-16[8\alpha ^{3}+\alpha ^{2}(4-7\beta ^{2})+\alpha \beta
^{2}(\beta ^{2}-4)-\beta ^{2}(\beta ^{2}+4)],
\end{eqnarray}
\begin{eqnarray}
I_{K}^{(1)} &=&8\beta \left[ 9\alpha ^{2}-2\alpha (\beta ^{2}-4)+\beta
^{2}\right] ,  \nonumber \\
I_{E}^{(1)} &=&-4\beta [12\alpha ^{3}-\alpha ^{2}(\beta ^{2}-52)+\alpha
(32-12\beta ^{2})+\beta ^{2}(3\beta ^{2}+4)],
\end{eqnarray}
\begin{eqnarray}
I_{K}^{(2)} &=&-4\left[ 8\alpha ^{3}-\alpha ^{2}(\beta ^{2}-8)-8\alpha \beta
^{2}+\beta ^{2}(3\beta ^{2}-8)\right] ,  \nonumber \\
I_{E}^{(2)} &=&8[4\alpha ^{4}+\alpha ^{3}(\beta ^{2}+12)-2\alpha ^{2}(\beta
^{2}-4)  \nonumber \\
&&+3\alpha \beta ^{2}(\beta ^{2}-4)+2\beta ^{2}(3\beta ^{2}-4)],
\end{eqnarray}
\begin{eqnarray}
I_{K}^{(3)} &=&8\beta \left[ \alpha ^{3}-7\alpha ^{2}+\alpha (3\beta
^{2}-8)+\beta ^{2}\right] ,  \nonumber \\
I_{E}^{(3)} &=&-4\beta [8\alpha ^{4}-4\alpha ^{3}+\alpha ^{2}(15\beta
^{2}-44)+4\alpha (5\beta ^{2}-8)  \nonumber \\
&&+\beta ^{2}(3\beta ^{2}+4)],
\end{eqnarray}
\begin{eqnarray}
I_{K}^{(4)} &=&-4[4\alpha ^{4}-4\alpha ^{3}+\alpha ^{2}(7\beta
^{2}-8)+12\alpha \beta ^{2}-\beta ^{2}(\beta ^{2}-8)],  \nonumber \\
I_{E}^{(4)} &=&16[4\alpha ^{5}+4\alpha ^{4}+\alpha ^{3}(7\beta
^{2}-4)+\alpha ^{2}(11\beta ^{2}-4)  \nonumber \\
&&+(2\alpha+1)\beta^2(\beta ^{2}+4)].
\end{eqnarray}

The sharp-eyed reader may observe that the expression for
$P^{\alpha}_{\ abcd}$ in Eq.\ (\ref{eqA60}) differs somewhat from that
given in Ref.\ \cite{Barack:2002mh}. The reason for this discrepancy
is that in Ref.\ \cite{Barack:2002mh} we have made a different choice
of extension for $\bar\nabla^{\alpha\beta\gamma}_x$ [one in which the metric
functions $g^{\alpha\beta}$ in Eq.\ (\ref{eq1:90}) take their actual
value at $x$, and $u^{\alpha}$ is parallely-propagated from the worldline
to $x$ along a normal geodesic]. This affects the value of the parameter
$B^{\alpha}$ only. We have opted here to give the parameter values
corresponding to the ``fixed'' extension because that extension
is more easily implemented in actual numerical calculations.
To obtain the value of $B^{\alpha}$ in the extension
of Ref.\ \cite{Barack:2002mh}, all one needs to do is replace the expression
in Eq.\ (\ref{eqA60}) with
\begin{equation}\label{eqA600}
P^{\alpha}_{\ abcd}[PP] =
\frac{1}{2}\left[ 3P^{\alpha}_{\ d}P_{abc}-(2P^{\alpha}_{\ ab}+P_{ab}^{\ \ \alpha })
P_{cd}\right]
+(3P^{\alpha}_{\ a}P_{be}-P^{\alpha}_{\ e}P_{ab})C_{cd}^{e},
\end{equation}
where `PP' denotes the extension of Ref.\ \cite{Barack:2002mh} (which
involves a Parallel Propagation of the four-velocity).
Interestingly, the regularization parameters for the {\it scalar-field} SF
are precisely the same as the gravitational-field ones, with the parameter $B^{\alpha}$
calculated using $P^{\alpha}_{\ abcd}[PP]$ (and with the obvious replacement
of the mass $\mu$ with the scalar charge) \cite{Barack:2002mh}.

\section{Numerical implementation strategies}
\label{sec:numerics}

There are two broad (somewhat related) issues that need to be addressed
in preparing to implement the mode-sum formula in practice (these
issues arise, in some form, whether one uses the mode sum scheme or any
of the other implementation methods based on the MiSaTaQuWa formalism).
The first practical issue has often been referred to as the {\it gauge problem}:
The mode-sum method (as the MiSaTaQuWa formalism underpinning it) is formulated
in terms of the metric perturbation in the Lorenz gauge, while standard
methods in black hole perturbation theory are formulated in other gauges.
The second practical issue concerns the numerical treatment of the
point-particle singularity; the problem takes a different form in
the frequency domain and in the time domain, and we shall discuss these
two frameworks separately below. We start, however, with a discussion
of the gauge problem and the methods developed to address it.

\subsection{Overcoming the gauge problem} \label{sec:3.1}

The calculation of the gravitational SF requires direct information about
the local metric perturbation near (and at) the particle. More specifically,
one needs to be able to construct the metric perturbation, along with its
derivatives, in the Lorenz gauge, at the particle's location. The mode-sum
approach allows us to do so without encountering infinities by considering
individual multipoles of the perturbation, but the problem remains how
to obtain these multipoles in the desired Lorenz gauge.
Unfortunately, standard formulations of black hole perturbations employ
other gauges, favored for their algebraic simplicity. Such gauges are simple
because they reflect faithfully the global symmetries of the underlying
black hole spacetimes---unlike the Lorenz gauge, which is suitable
for describing the locally-isotropic particle singularity, but
complies less well with the global symmetry of the background.

A common gauge choice for perturbation studies in Schwarzschild is the one
introduced many years ago by Regge and Wheeler \cite{RW}, and further developed
by Zerilli \cite{Zerilli} and Moncrief \cite{Moncrief}.
In this gauge, certain projections of the metric perturbation onto a tensor-harmonic
basis are taken to vanish, which results in a significant
simplification of the perturbation equations. Another such useful ``algebraic''
gauge is the {\em radiation gauge} introduced by Chrzanowski \cite{Chrz},
in which one sets to zero the projection of the perturbation along a principle
null direction of the background black hole geometry. Perturbations of the
{\em Kerr} geometry have been studied almost exclusively using the powerful
formalism by Teukolsky \cite{Teukolsky}, in which the perturbation is formulated
in terms of the Newman--Penrose gauge-invariant scalars, rather than the metric.
A reconstruction procedure
for the metric perturbation out of the Teukolsky variables (in vacuum) was
prescribed by Chrzanowski \cite{Chrz} (with later supplements by Wald \cite{WaldReconst}
and Ori \cite{OriReconst}), but only in the radiation gauge. It is not known how
to similarly reconstruct the metric in the Lorenz gauge.

Several strategies have been proposed for dealing with this gauge-related
difficulty. Some involve a deviation from the original MiSaTaQuWa notion of SF,
while others seek to tackle the calculation of the Lorenz-gauge perturbation
directly. Here is a survey of the main strategies.

\subparagraph{Self force in a ``hybrid'' gauge.}

Equation (\ref{eq1:180}) in Sec.\ \ref{sec:theory} describes the gauge
transformation of the SF. Let us refer here to a certain  gauge as ``regular''
if the transformation from the Lorenz gauge yields a
well-defined SF in that gauge [this requires that
the expression in square brackets in Eq.\ (\ref{eq1:180}) admits a definite
and finite particle limit]. It has been shown \cite{Barack:2001ph}
that the mode-sum formula maintains its form (\ref{eq4:200}), with the
same regularization parameters, for any such regular gauges.
Namely, for the SF in a specific regular gauge `reg' we have the mode-sum
formula
\begin{equation} \label{eq5:10}
F^{\alpha}_{\rm self}[{\rm reg}]=\sum_{l=0}^{\infty}
\left[F_{{\rm ret}\pm}^{\alpha l}[{\rm reg}]\mp LA^{\alpha}-B^{\alpha}\right],
\end{equation}
where the force modes $F_{{\rm ret}\pm}^{\alpha l}[{\rm reg}]$ are these
constructed from the retarded metric perturbation in the `reg' gauge
(we have already set $C^{\alpha}=D^{\alpha}=0$ here).

Now recall the radiative inspiral problem which motivates us here:
Although the momentary SF is gauge-dependent, the long-term radiative
evolution of the orbit (as expressed, for example, through the drift of
the constants of motion) has gauge-invariant characteristics that
should be accessible from the SF in whatever regular gauge.
And so, insofar as the physical inspiral problem is concerned, one might
have hoped to circumvent the gauge problem by simply evaluating the
SF in any gauge which is both regular and practical, using Eq.\ (\ref{eq5:10}).
Unfortunately, while the Lorenz gauge itself is ``regular but not practical'',
both the Regge-Wheeler gauge and the radiation gauge are generally ``practical
but not regular'', as demonstrated in Ref.\ \cite{Barack:2001ph}.

However, a practical solution now suggests itself: Devise a gauge which
is regular in the above sense, and yet practical in that it relates to one
of the ``practical'' gauges---say, the radiation gauge---through a simple,
explicit gauge transformation (unlike the Lorenz gauge itself).
Heuristically, one may picture such a
``hybrid'' gauge (also referred to as ``intermediate'' gauge \cite{Barack:2002mh})
as one in which the metric perturbation retains its isotropic Lorenz-like
form near the particle, while away from the particle it deforms so as to
resemble the radiation-gauge perturbation.
The SF in such a hybrid gauge would have the mode-sum formula
\begin{equation} \label{eq5:20}
F^{\alpha}_{\rm self}[{\rm hyb}]=\sum_{l=0}^{\infty}
\left[F_{\pm}^{\alpha l}[{\rm rad}]\mp LA^{\alpha}-B^{\alpha}\right]
-\delta D^{\alpha},
\end{equation}
where $F_{\pm}^{\alpha l}[{\rm rad}]$ are the $l$ modes of the full force
in the radiation gauge, and the ``counter term'' $\delta D^{\alpha}$ is the
difference $F_{\pm}^{\alpha l}[{\rm rad}]-F_{\pm}^{\alpha l}[{\rm hyb}]$,
summed over $l$ and evaluated at the particle. With a suitable choice of the
hybrid gauge, the term $\delta D^{\alpha}$ can be calculated analytically,
and Eq.\ (\ref{eq5:20}) then prescribes a practical way of constructing the
SF in a useful, regular gauge, out of the numerically calculated modes of
the perturbation in the radiation gauge.

Different variants of this idea were studied by several authors
\cite{MN1998,Barack:2001ph,Sago:2002fe,Barack:2002mh,Nakano:2003he},
but it has not been implemented in full so far.

\subparagraph{Generalized SF and gauge invariants.}

Another idea (set out in Ref.\ \cite{Barack:2001ph} and further developed
in Ref.\ \cite{Gralla:2008fg}) involves the generalization of the SF notion
through the introduction of a suitable averaging over angular directions.
In some gauges which are not strictly regular in the aforementioned sense, the
SF could still be defined in a directional sense. Such is the case in which
the expression in square brackets in Eq.\ (\ref{eq1:180}) has a finite yet
direction-dependent particle limit (upon transforming from the Lorenz gauge),
and the resulting ``directional'' SF is bounded for any chosen direction.
(This seems to be the situation in the Regge-Wheeler gauge,
but not in the radiation gauge---in the latter, the metric perturbation
from a point particle develops a one-dimensional string-like singularity
\cite{Barack:2001ph}.) In such cases, a suitable averaging over angular
directions introduces a well-defined notion of an ``average'' SF, which
generalized the original MiSaTaQuWa SF (it represents a generalization
since, obviously, the average SF coincides with the standard MiSaTaQuWa
SF for all regular gauges, including Lorenz's). The notion of an average SF
could be a useful one if it can be used in a simple way to construct
gauge-invariant quantities which describe the radiative motion. This
is yet to be demonstrated.

A related method invokes the directional SF itself as an agent for
constructing the desired gauge invariants. In this approach, one defines
(for example) a ``Regge-Wheeler'' SF  by taking a particular directional limit
consistently throughout the calculation, and then using the value of this
SF to construct the gauge invariants.  This approach has been applied
successfully, in combination with the mode-sum method, by Detweiler and
others \cite{Detweiler:2005kq,Detweiler:2008ft}.

\subparagraph{Radiation-gauge regularization.}

Friedman  and collaborators \cite{Keidl:2006wk,:2007mn,Friedman} proposed the following
construction: Starting with the Lorenz-gauge $S$ field, construct the
associated gauge-invariant Newman--Penrose scalar $\psi_0^{\rm S}$, and
decompose it into spin-weighted spheroidal
harmonics. Then obtain the harmonics of the retarded field $\psi_0^{\rm ret}$
by solving the Teukolsky equation with suitable boundary conditions, and
(for each harmonic) define the $R$ part through
$\psi_0^{\rm R}\equiv \psi_0^{\rm ret}-\psi_0^{\rm S}$.
If $\psi_0^{\rm S}$ is known precisely, then $\psi_0^{\rm R}$ is
a vacuum solution of the Teukolsky equation. To this solution, then,
apply Chrzanowski's reconstruction procedure to obtain a smooth
radiation-gauge metric perturbation $h^{\rm R}_{\alpha\beta}[{\rm rad}]$,
and use that to construct a ``radiation gauge'' SF (via, e.g., the mode-sum
method). The relation between this definition of the radiation-gauge SF and
the one obtained by applying the gauge transformation formula (\ref{eq1:180})
to the standard Lorenz-gauge MiSaTaQuWa SF is yet to be investigated.

In reality, the $S$ field is usually known only approximately, resulting
in that $\psi_0^{\rm R}$ retains some non-smoothness. How to apply
Chrzanowski's reconstruction to non-smooth potentials is a matter of
appreciable technical challenge. Also, the above procedure cannot
account for the contribution to the SF from the two non-radiative modes,
$l=0,1$, which then need to be treated separately, using other methods.
Nonetheless, the technique offers a natural way of circumventing the
gauge problem, and has much potential promise.

\subparagraph{Direct Lorenz-gauge implementation.}

In 2005, Barack and Lousto \cite{Barack:2005nr} succeeded in solving the
full set of Lorenz-gauge perturbation equations in Schwarzschild, using
numerical evolution in 1+1D. (An alternative formulation was later developed
and implemented by Berndtson \cite{Berndtson:2009hp}.) This development
opened the door for a direct implementation of the mode-sum formula in the
Lorenz gauge. It later facilitated the first calculations of the gravitational
SF for bound orbits in Schwarzschild \cite{Barack:2007tm,BSprep,Barack:2009ey}.

In Sec.\ \ref{sec:Lorenz} we shall review this approach in some detail.
Here we just point to a few of its advantages:
(i)
This direct approach obviously circumvents the gauge problem.
The entire calculation is done within the Lorenz gauge, and the mode-sum
formula can be implemented directly, in its original form.
(ii)
The Lorenz-gauge perturbation equations take a fully hyperbolic form,
making them particularly suitable for numerical implementation
in the time domain. Conveniently, the supplementary gauge conditions
(which take the form of elliptic ``constraint'' equations) can be made
to hold automatically, as we explain in Sec.\ \ref{sec:Lorenz}.
(iii)
In this approach one solves directly for the metric perturbation components,
without having to resort to complicated reconstruction procedures.
This is an important advantage because metric reconstruction involves
differentiation of the field variables, which inevitably results in loss
of numerical accuracy.
(iv)
Working with the Lorenz-gauge perturbation components as field variables is
also advantageous in that these behave more regularly near point particles
than do Teukolsky's or Moncrief's variables. This has a simple manifestation,
for example, within the 1+1D treatment in Schwarzschild: The individual multipole
modes of the Lorenz-gauge perturbation are always {\em continuous} at the particle,
just like in the simple example of Sec.\ \ref{sec:2.1}; on the other hand,
the multipole modes of Teukolsky's or Moncrief's
variables are {\em discontinuous} at the particle, and so are, in general, the
modes of the metric perturbation in the Regge-Wheeler gauge. Obviously, the
continuity of the Lorenz-gauge modes makes them easier to deal with as
numerical variables.

\subsection{Numerical representation of the point particle}

Common to all numerical implementation methods is the basic preliminary
task of solving the field equations (in whatever formulation) for the
full (retarded) perturbation from a point particle in a specified orbit.
This immediately brings about the practical issue of the numerical
representation of the particle singularity. The particulars of the
challenge depend on the methodological framework: In time-domain methods
one faces the problem of dealing with the irregularity of the field
variables near the worldline; in frequency-domain (spectral) treatments,
such irregularity manifests itself in a problematic high-frequency
behavior. We now survey some of the relevant methods.

\subsubsection{Particle representation in the time domain}\label{sec:3.2.1}

\subparagraph{Extended-body representations.}

In the context of fully nonlinear Numerical Relativity, the
problem of a binary black hole with a small mass ratio remains a difficult
challenge, because of the large span of lengthscales intrinsic in this problem.
(Current NR technology can handle mass ratios as small as $1\!:\!10$
\cite{Gonzalez:2008bi}---still nothing near the $\sim 1\!:\!10^4$--$10^8$ ratios
needed for LISA EMRI applications.) Bishop {\it et al.}~\cite{Bishop:2003bs}
attempted a NR treatment in which the particle is modeled by a quasi-rigid
widely-extended body whose `center' follows a geodesic. Comparison with
perturbation results did not show sufficient accuracy, and the method
requires further development.

An extended-body approach has also been implemented in perturbative studies.
Khanna {\it et al.}~\cite{LopezAleman:2003ik,Khanna:2003qv,Burko:2006ua} solved
the Teukolsky equation in the time domain (i.e., in 2+1D, for each azimuthal
$m$ mode) with a source `particle' represented by a narrow Gaussian distribution.
This crude technique was much improved recently by Sundararajan
{\it et al.}~\cite{Sundararajan:2007jg,Sundararajan:2008zm} using
a ``finite impulse representation'', whereby the source is
modeled by a series of spikes whose relative magnitude is carefully
controlled so as to assure that the source has integral properties similar
to that of a delta function. Such methods were demonstrated to
reproduce wave-zone solutions with great accuracy (indeed, that is
what they are designed to do), but they are likely to remain less useful
for computing the accurate local perturbation near the particle as required
in SF calculations.

Extended-body representations suffer from the inevitable tradeoff between
smoothness and localization: One can only smoothen the solution by making
it less localized, and one can better localize it only by making it less smooth.
In what follows we concentrate on methods in which the source particle is
{\em precisely} localized on the orbit: The energy-momentum
of the particle in represented by a delta-function source term [as in
Eq.\ (\ref{eq1:200})], and the delta distribution is
treated analytically within the numerical scheme, in an exact manner.

\subparagraph{Delta-function representation in 1+1D.}

In full 3+1D spacetime, the full (retarded) metric perturbation obviously
diverges on the particle (at any given time). The divergence is
asymptotically Coulomb-like in the Lorenz gauge (and can take a different
form in other gauges). In spherically-symmetric spacetimes one can decompose
the perturbation into tensor harmonics and solve a separated version of the
field equations in 1+1D (time+radius) for each harmonic separately.
In the particle problem this becomes beneficial not only thanks to the
obvious dimensional reduction, but also because it mitigates the
problematics introduced by the particle's singularity: The angular
integration involved in constructing the individual $l$ modes effectively
``smears'' the Coulomb-like singularity across the surface of a 2-sphere,
and the resulting $l$ modes are finite even at the location of the particle.
Furthermore, in the case of the metric perturbation in the Lorenz gauge,
the individual $l$ modes are also continuous at the particle [cf.\ Eq.\
(\ref{eq4:145}) in Sec.\ \ref{sec:modesum}].
The corresponding $l$ modes of the Teukolsky or Moncrief gauge-invariant
variables are generally {\it not} continuous at the particle, and generally
neither are the modes of the metric perturbation in non-Lorenz gauges.

The boundedness of the $l$ modes is, of course, a crucial feature
of the $l$-mode regularization scheme, as we have already discussed.
That same feature also greatly simplifies the 1+1D numerical treatment of the particle.
Lousto and Price \cite{Lousto:1997wf} formulated a general method for incorporating
a delta-function source in a finite-difference treatment of the field equations
in 1+1D. In this method, the
finite-difference approximation at a numerical grid cell (in $t$--$r$ space)
traversed by the particle's worldline is obtained, essentially, by integrating
the field equation ``by hand'' over the grid cell at the required accuracy.
The original (1D) delta function present in the source term thereby integrates
out to contribute a finite term at each time step. The original Lousto--Price scheme
(formulated with a 1st-order global numerical convergence) was later improved by Martel
and Poisson \cite{Martel:2001yf} (2nd-order convergence) and Lousto \cite{Lousto:2005ip}
(4th-order convergence). This simple but powerful idea is at the core of many
of the 1+1D finite-difference implementations presented in the last
few years \cite{Martel:2003jj,Haas:2007kz,Barack:2007tm}, including the work
discussed in Sec.\ \ref{sec:Lorenz} below.

Despite such advances, 1+1D particles remain numerically expensive to handle,
because the nonsmoothness associated with them introduces a large scale
variance in the solutions: The $l$-mode field gradients grow sharply near
the particle, and, moreover, become increasingly more difficult to resolve
with larger $l$ (recall the $l$-mode gradient is $\propto l$ at large $l$).
The mode-sum formula, recall, converges rather slowly (like $\sim 1/l$),
and so requires one to compute a considerably large number of modes
(typically $\sim 20$ with even a moderate accuracy goal). This proves to
stretch the limit of what can be achieved today using finite differentiation on
a fixed mesh.

Several methods have been proposed to address this problem in the current
context. Sopuerta and collaborators \cite{Sopuerta:2005rd,Sopuerta:2005gz,
Canizares:2008dp,Canizares:2009ay} explored the use of finite-element
discretization. This
technique is particularly powerful in dealing with multi-scale problems, and,
being quasi-spectral, it benefits from an exponential numerical convergence.
So far it was applied successfully for generic orbits in Schwarzschild,
and higher-dimensional implementations (for Kerr studies) are currently
being considered. A related quasi-spectral scheme was recently suggested
by Field {\it et al.}~\cite{Field:2009kk}. Finally, Thornburg \cite{Thornburg:2009mw}
very recently developed an adaptive mesh-refinement algorithm for
Lousto-Price's finite differences scheme (with a global 4th-order convergence).
This was successfully implemented for a scalar charge in a circular
orbit in Schwarzschild, and generalizations are being considered.

\subparagraph{Puncture methods.}

In Kerr spacetime, one no longer benefits from a 1+1D separability. The
Lorenz-gauge perturbation equations are only separable into azimuthal
$m$-modes, each a function of $t,r,\theta$ in a 2+1D space. The $m$ modes
are not finite on the worldline, but rather they diverge there logarithmically
(see the discussion in Sec.\ II.C of Ref.\  \cite{Barack:2007jh}). Since the
2+1D numerical solutions are truly divergent, a direct finite-difference
treatment becomes problematic. However, since the singular behavior of
the perturbation can be approximated analytically, a simple remedy to this
problem suggests itself.

The idea, which has recently been studied independently by several group
\cite{Barack:2007jh,Vega:2007mc,Lousto:2008mb,Vega:2009qb}, is to utilize a new
perturbation variable for the numerical time-evolution, which we shall
call here the {\it residual} field. This field is constructed from the full
(retarded) perturbation by subtracting a suitable function---the ``puncture'' field,
given analytically---which approximates the singular part of the perturbation
well enough that the residual field is (at least) bounded and differentiable
at the particle.
The perturbation equations are then recast with the residual field as their
independent variable, and with a new source term (depending on the puncture
field and its derivatives) which now extends off the worldline but contains
no delta function. The equations are then solved for the residual field
in the time domain, using (e.g.) standard finite differentiation.

Several variants of this method have been studied and tested with
scalar-field codes in 1+1D \cite{Vega:2007mc} and 2+1D \cite{Barack:2007jh,
Lousto:2008mb}, and also proposed for use in full 3+1D \cite{:2007mn}.
The various schemes differ primarily in the
way they handle the puncture function far from the particle:
Barack and Golbourn \cite{Barack:2007jh} introduce a puncture with a
strictly compact support around the particle, Detweiler and Vega \cite{Vega:2007mc}
truncate it with a smooth attenuation function, and in Lousto and Nakano
\cite{Lousto:2008mb} the puncture is not truncated at all.
We will discuss the puncture method in more detail in Sec.\ \ref{sec:puncture}.

To obtain the necessary input for the SF mode-sum formula, the 2+1D (or 3+1D)
numerical solutions need to be decomposed into $l$ modes, in what then becomes
a somewhat awkward procedure (we decompose the field into separate $l$ modes
just to add these modes all up again after regularization). Fortunately,
there is a more direct alternative: Barack {\it et al.}~\cite{Barack:2007we}
showed how the SF can be constructed directly, in a simple way, from the $2+1$D
$m$-modes of the residual field (assuming only that these mode are differentiable
at the particle, which is achieved by designing a suitable puncture).
This direct ``$m$-mode regularization'' scheme, too, will be described in
Sec.\ \ref{sec:puncture}. It is hoped that this technique could provide a natural
framework for calculations in Kerr. It is yet to be applied in practice.

\newpage
\subsubsection{Particle representation in the frequency domain:
the high-frequency problem and its resolution}\label{sec:3.2.2}
\mbox{}\\

The $l$ modes required as input for the SF mode-sum formula can also be
obtained using a spectral treatment of the field equations. This has
the obvious advantage that one then only deals with ordinary differential
equations, although constructing the $l$ modes involves the additional step
of summing over sufficiently many frequency modes. (See \cite{Barton:2008eb}
for a recent analysis of the relative computational efficiencies of frequency
vs.\ time domain treatments.) As with the time-domain methods discussed above,
the representation of the particle in the frequency domain too brings about
technical complications, but these now take a different form.

To illustrate the problem, consider the toy model of a scalar charge
in Schwarzschild, allowing the particle to move on some bound (eccentric)
geodesic of the background, with radial location given as a function of time
by $r=r_{\rm p}(t)$. Decompose the scalar field in spherical harmonics,
and denote the multipolar mode functions by $\phi_{lm}(t,r)$.
The time-domain modes $\phi_{lm}$ are continuous along $r=r_{\rm p}(t)$
for each $l,m$. However, the derivatives $\phi_{lm,r}$ and $\phi_{lm,t}$ will
generally suffer a finite jump discontinuity across $r=r_{\rm p}(t)$
(recall our elementary example in Sec.\ \ref{sec:2.1}), which reflects
the presence of a source ``shell'' representing the $l,m$ mode of the scalar
charge. In particular, if the orbit is eccentric, the derivatives of
$\phi_{lm}$ will generally be discontinuous functions of $t$ at a fixed value of
$r$ along the orbit.

Now imagine trying to reconstruct $\phi_{lm}(t,r)$ (for some fixed $r$ along
the orbit) as a sum over its Fourier components:
\begin{equation} \label{eq370}
\phi_{lm}(t,r)=\sum_{\omega} R_{lm\omega}(r)e^{-i\omega t}.
\end{equation}
Since, for an eccentric orbit, $\phi_{lm}(t,r)$ is only a $C^0$ function of $t$
at the particle's worldline, it follows from standard Fourier theory \cite{James}
that the Fourier sum in Eq.\ (\ref{eq370}) will only converge there like
$\sim1/\omega$. The actual situation is even worse, because for
SF calculations we need not only $\phi_{lm}$ but also its derivatives.
Since (e.g.) $\phi_{lm,r}$ is a discontinuous function of $t$, we will
inevitably face here the well known ``Gibbs phenomenon'': the Fourier sum will
fail to converge to the correct value at $r\to r_{\rm p}(t)$. Of course,
the problematic behavior of the Fourier sum is simply a consequence of our
attempt to reconstruct a discontinuous function as a sum over smooth
harmonics.

From a practical point of view this would mean that
(i) at the coincidence limit, $r\to r_{\rm p}(t)$,
the sum over $\omega$ modes would fail to yield the correct one-sided values of
$\phi_{lm,r}$ however many $\omega$ modes are included in the sum; and
(ii) if we reconstruct $\phi_{lm,r}$ at a point $r=r_{0}$ off the worldline,
then the Fourier series should indeed formally converge---however, the number
of $\omega$ modes required for achieving a prescribed precision would grow
unboundedly as $r_{0}$ approaches $ r_{\rm p}(t)$, making it extremely
difficult to evaluate $\phi_{lm,r}$ at the coincidence limit.

This technical difficulty is rather generic, and will show also in calculations
of the local gravitational field. The situation is no different in
the Kerr case, because there too the mode-sum formula requires as
input the {\it spherical}-harmonic modes of the perturbation field, and for
each such mode the source is represented by a $\delta$-distribution on a thin
shell, which renders the field derivatives discontinuous across that shell.
The problem becomes even more severe when considering gravitational
perturbations via the Teukolsky formalism: Here, the $l,m$ modes of the
perturbation field (now the Newman--Penrose curvature scalar)
are not even continuous at the particle's orbit---a consequence of the fact that
the source term for Teukolsky's equation involves (second) derivatives of the
energy-momentum tensor. Again, a naive attempt to construct these multipoles as
a sum over their $\omega$ modes will be hampered by the Gibbs phenomenon.

A simple way around the problem was proposed recently in Ref.\
\cite{Barack:2008ms}. It was shown how the desired values of the field and
its derivatives at the particle can be constructed from a sum over properly
weighted {\it homogeneous} (source-free) radial functions $R_{lm\omega}(r)$,
instead of the actual inhomogeneous solutions of the frequency-domain equation.
The Fourier sum of such homogeneous radial functions, which are smooth everywhere,
converges exponentially fast. The Fourier sum of the derivatives, which are also smooth,
is likewise exponentially convergent. The validity of the method (and the exponential convergence) was
demonstrated in Ref.\ \cite{Barack:2008ms} with an explicit numerical calculation
in the scalar-field monopole case ($l=0$). It was later implemented in a frequency-domain
calculation of the monopole and dipole modes of the Lorenz gauge metric perturbation
for eccentric orbits in Schwarzschild \cite{BGSprep,BSprep,Barack:2009ey}.
The same method should be applicable for any of the other problems mentioned
above, including the calculation of EM and gravitational perturbations
using Teukolsky's equation.

The method of Ref.\ \cite{Barack:2008ms} (dubbed {\it method of
extended homogeneous solutions}) completely circumvents the problem of
slow convergence (or the lack thereof) in frequency domain calculations
involving point sources. It makes the frequency-domain approach an
attractive method-of-choice for some SF calculations.
The method is now being implemented in first calculations of the scalar-field
SF for Kerr orbits \cite{BWprep}.

\section{An Example: gravitational self-force in Schwarzschild via
1+1D evolution in Lorenz gauge}
\label{sec:Lorenz}

As an example of a fully worked out calculation of the SF, we review
here the work by Barack and Sago on eccentric geodesic in Schwarzschild
\cite{Barack:2007tm,BSprep}. This work represents a direct implementation
of the mode-sum formula in its original form (\ref{eq4:200}). The decomposed
Lorenz-gauge metric perturbation equations are integrated directly
using numerical evolution in 1+1D. The numerical algorithm employs a
straightforward 4th-order-convergent finite-difference scheme (a
variant of the Lousto--Price method) on a fixed staggered numerical mesh based
on characteristic coordinates. Below we briefly describe the perturbation
formalism, discuss the numerical implementation in some more detail, and
display some results.

\subsection{Lorenz-gauge perturbation formalism}

The linearized Einstein equations in the Lorenz gauge are given in Eq.\
(\ref{eq1:200}). On the right-hand side of these equations is a distributional
representation of the energy-momentum of a point particle of mass $\mu$,
which is moving on a timelike geodesic $z^{\mu}(\tau)$ of the background
spacetime. Mathematically, the linear set (\ref{eq1:200}) is a diagonal hyperbolic
system, which admits a well posed initial-value formulation on a spacelike Cauchy
hypersurface (see, e.g., Theorem 10.1.2 of \cite{Wald}). Furthermore, if the
Lorenz gauge conditions of Eq.\ (\ref{eq1:140}) are satisfied on the
initial Cauchy surface, then they are guaranteed to hold everywhere---assuming
that the field equations are satisfied everywhere and that the energy-momentum
satisfies $\nabla^{\beta}T_{\alpha\beta}=0$, as in the case of a geodesic particle.
We remind that the gauge conditions (\ref{eq1:140}) do not fully specify the
gauge: There is a residual gauge freedom within the family of Lorenz gauges,
$h_{\alpha\beta}\to h_{\alpha\beta}+\nabla_{\beta}\xi_{\alpha}+
\nabla_{\alpha}\xi_{\beta}$,
with any $\xi^{\mu}$ satisfying $\nabla^{\alpha}\nabla_{\alpha}\xi^{\mu}=0$.
It is easy to verify that both Eqs.\ (\ref{eq1:140}) and (\ref{eq1:200})
remain invariant under such gauge transformations.

Here we specialize to eccentric geodesics around a Schwarzschild black hole,
and employ a Schwarzschild coordinate system ($t,r,\theta,\varphi$) in which
the orbit is equatorial ($z^\theta=\pi/2$). Such orbits constitute a
two-parameter family; we may characterize each orbit by the radial ``turning
points'' $r_{\rm min}$ and $r_{\rm max}$, or alternatively by the
``semi-latus rectum''
$p\equiv 2(r_{\rm max}r_{\rm min})/(r_{\rm max}+r_{\rm min})$
and ``eccentricity''
$e\equiv (r_{\rm max}-r_{\rm min})/(r_{\rm max}+r_{\rm min})$.

Barack and Lousto \cite{Barack:2005nr} decomposed the metric
perturbation into tensor harmonics, in the form
\begin{equation}\label{eq400}
\bar h_{\alpha\beta}^{\rm ret}=\frac{\mu}{r}\sum_{l,m}\sum_{i=1}^{10}
\bar h^{(i)lm}(r,t)\,Y^{(i)lm}_{\alpha\beta}(\theta,\varphi),
\end{equation}
and similarly for the source $T_{\alpha\beta}$.
The harmonics $Y^{(i)lm}_{\alpha\beta}(\theta,\varphi)$ (whose components
are constructed from ordinary spherical harmonics and their first and second
derivatives) form a complete orthogonal basis for 2nd-rank covariant tensors
on a 2-sphere (see appendix A of \cite{Barack:2005nr}).
The time-radial functions $\bar h^{(i)lm}$ ($i=1,\ldots,10$) form our basic
set of perturbation fields, and serve as variables for the numerical
evolution.\footnote{To simplify the appearance of Eq.\ (\ref{eq400}) we
have used here a normalization of $\bar h^{(i)lm}$ which is slightly different
from that of \cite{Barack:2005nr}.} The tensor-harmonic
decomposition decouples Eq.\ (\ref{eq1:200}) with respect to $l,m$, although
not with respect to $i$: For each $l,m$, the variables $\bar h^{(i)lm}$
satisfy a coupled set of hyperbolic (in a 1+1D sense) scalar-like equations,
which may be written in the form
\begin{equation}\label{eq410}
\Box^{(2)} \bar h^{(i)lm}+
{\cal M}^{(i)l}_{\;(j)}\bar h^{(j)lm}=S^{(i)lm}\quad (i=1,\ldots,10).
\end{equation}
Here $\Box^{(2)}$ is the 1+1D scalar-field wave operator
$\partial^2_{uv}+ V(r)$, where $v$  and $u$ are the standard Eddington--Finkelstein
null coordinates, and $V(r)=\frac{1}{4}(1-2M/r)\left[2M/r^3+l(l+1)/r^2\right]$
is an effective potential.
The ``coupling'' terms ${\cal M}^{(i)l}_{\;(j)}\bar h^{(j)lm}$ involve
first derivatives of the $\bar h^{(j)lm}$'s at most (no second derivatives),
so that, conveniently, the set (\ref{eq410}) decouples at its principal part.
The decoupled source terms $S^{(i)lm}$ are each $\propto\delta[r-r_{\rm p}(\tau)]$
(no derivatives of $\delta$ function) and, as a result, the physical solutions
$\bar h^{(j)lm}$ are continuous even at the particle.
Explicit expressions for the coupling terms and the source terms in
Eq.\ (\ref{eq410}) can be found in Ref.\ \cite{Barack:2005nr}.

In addition to the evolution equations (\ref{eq410}), the functions
$\bar h^{(i)lm}$ also satisfy four 1st-order elliptic equations, which arise
from the separation of the gauge conditions (\ref{eq1:140}) into $l,m$ modes. In
the continuum initial-value problem, the solutions $\bar h^{(i)lm}$  satisfy
these ``constraints'' automatically if only they satisfy them on the initial
Cauchy surface. This is more difficult to guarantee in a numerical
finite-differences treatment, where (i) it is often impossible to prescribe
exact initial data that satisfy the constraints, and (ii) discretization
errors may amplify constraint violations during the numerical evolution.
Inspired by a remedy proposed for a similar problem in the context of
nonlinear Numerical Relativity \cite{Gundlach}, Ref.\ \cite{Barack:2005nr}
proposed the inclusion of ``divergence dissipation'' terms,
$\propto \nabla^{\beta}\bar h^{\rm ret}_{\alpha\beta}$,
in the original set (\ref{eq1:200}), so designed to guarantee that any violations
of the Lorenz gauge conditions are efficiently damped during the evolution.
These damping terms modify only the explicit form of the ${\cal M}$ terms
in Eq.\ (\ref{eq410}) as shown in \cite{Barack:2005nr}.

\subsection{Numerical implementation}

\begin{figure}[htb]
\input{epsf}
\centerline{\includegraphics[totalheight=6cm]{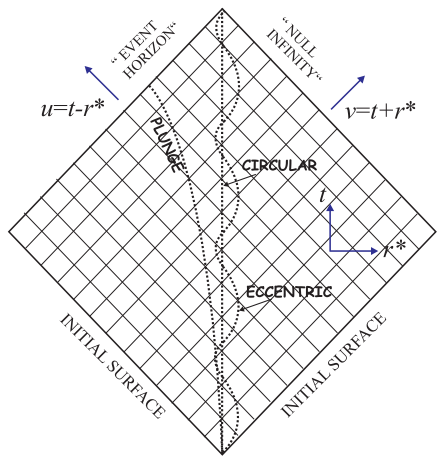}}
\caption{The numerical 1+1D domain in the Barack--Sago code
\cite{Barack:2007tm,BSprep}: A staggered mesh based on characteristic
(Eddington--Finkelstein) coordinates $v,u$. $t$ and $r^*$ are, respectively,
the Schwarzschild time and tortoise radial coordinates. The evolution proceeds
from characteristic initial data on two null surfaces. Illustrated are a few
sample geodesic orbits (radial plunge, circular, eccentric).
[{\it Graphics reproduced from Ref.\ }\cite{OrleansBook}.]
}
\label{fig3}       
\end{figure}

The code developed by Barack and Sago \cite{Barack:2007tm,BSprep} solves
the coupled set (\ref{eq410}) (with constraint dissipation terms incorporated
in the ${\cal M}$ terms) via time evolution. The numerical domain, covering a portion
of the external Schwarzschild geometry, is depicted in Fig.\ \ref{fig3}.
The numerical grid is based on Eddington--Finkelstein null coordinates $v,u$,
and initial data (the values of the 10 fields $\bar h^{(i)lm}$ for each $l,m$)
are specified on two characteristic initial surfaces $v=$const and $u$=const.
Equations (\ref{eq410}) are then discretized on this grid using a finite-difference
algorithm which is globally 4th-order convergent. The numerical integrator
solves for the various $\bar h^{(i)lm}$'s along consecutive $u=$const rays.
A particularly convenient feature of this setup is that no boundary conditions
need be specified (the characteristic grid has no causal boundaries).
Moreover, one need not be at all concerned with the determination of
faithful initial conditions: It is sufficient to set $\bar h^{(i)lm}=0$
on the initial surfaces and simply let the resulting spurious radiation
(which emanate from the intersection of the particle's worldline with the
initial surface) dissipate away over the evolution time. The early part
of the evolution, which is typically dominated by such spurious radiation,
is simply discarded.

The conservative modes $l=0$ and $l=1$ (respectively the monopole and dipole)
require a separate treatment, as they do not evolve stably using the above
numerical scheme. (A naive attempt to evolve these modes leads to numerical
instabilities which, so far, could not be cured.) Fortunately, the set
(\ref{eq410}) simplifies considerably for these modes, and solutions can be
obtained in a semi-analytic manner based on physical considerations.
Detweiler and Poisson \cite{Detweiler:2003ci} worked out the $l=0,1$ Lorenz-gauge
solutions for circular orbits, and their work is generalized to eccentric
orbits in Ref.\ \cite{BGSprep}, relying on the aforementioned method of extended
homogeneous solutions. The calculation of \cite{BGSprep} yields the
values of the fields $\bar h^{(i)lm}$ and their derivatives for $l=0,1$.

\begin{figure}[htb]
\input{epsf}
\centerline{\includegraphics[totalheight=6cm]{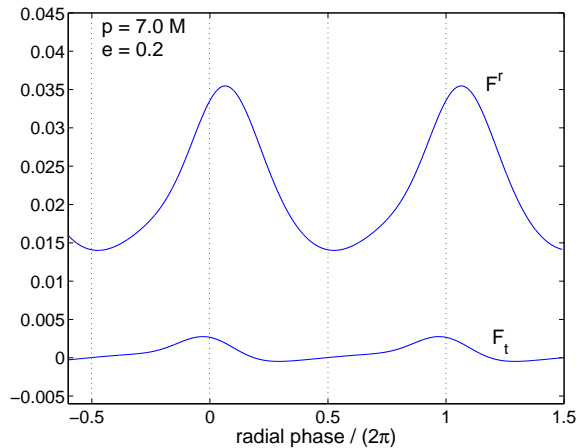}}
\caption{The gravitational SF [in units of $(\mu/M)^2$] along a
Schwarzschild geodesic with semi-latus rectum $p=7M$ and eccentricity $e=0.2$.
The upper and lower lines show $F_{\rm self}^r$ and $F^{\rm self}_t$,
respectively. Integer values on the horizontal axis correspond to
periapses ($r=r_{\rm min}$); note the slight retardation manifest in the
magnitude of the radial component. The data for these plots were obtained using
a direct integration of the metric perturbation equations in the Lorenz gauge,
in conjunction with the mode-sum method, as described in Sec.\ \ref{sec:Lorenz}.
[{\it Graphics reproduced from Ref.\ }\cite{OrleansBook}.]
}
\label{fig4}       
\end{figure}

To construct the $l$ modes $F^{\alpha l}_{{\rm ret}\pm}$---the necessary input
for the mode-sum formula (\ref{eq4:200})---one first substitutes the expansion
(\ref{eq400}) in the expression for the retarded force $F_{\rm ret}^{\alpha}$
[left-hand side expression in Eq.\ (\ref{eq4:20})], and then expands the result
in spherical harmonics. The outcome is a formula for $F^{\alpha l}_{{\rm ret}\pm}$
given in terms of the various fields $\bar h^{(i)l'm}$ and their derivatives
(evaluated at the particle and summed over $m$). The number of $l'$-modes 
$\bar h^{(i)l'm}$ contributing to each $l$-mode $F^{\alpha l}_{{\rm ret}\pm}$
depends on the off-worldline extension chosen for $\bar\nabla^{\alpha\beta\gamma}$.
In \cite{Barack:2007tm,BSprep} we used a convenient extension in which
[referring to Eq.\ (\ref{eq1:90}) and Fig.\ \ref{fig1}] the metric $g^{\alpha\beta}$ 
takes its value at the field point $x$ while the components $u^{\alpha}$
(in Schwarzschild coordinates) take the fixed value they have at $z$. 
In this extension we find that the only contributing $l'$-modes, for given
$l$, are $l-3\leq l'\leq l+3$. Explicit construction formulas for $F^{\alpha l}_{{\rm ret}\pm}$
can be found in \cite{Barack:2007tm,BSprep}. One then uses the numerical values of the
fields $\bar h^{(i)lm}$ and their derivatives, as calculated along the
orbit using the above code, to construct $F^{\alpha l}_{{\rm ret}\pm}$ for
sufficiently many $l$ modes. The mode-sum formula (\ref{eq4:200}) then gives
the SF. Figure \ref{fig4} shows the final result for an eccentric geodesic
with $p=7M$ and $e=0.2$.

The calculation we have just described represents a milestone in the SF
program: We now finally have a code able of tackling the generic
EMRI-relevant SF problem in Schwarzschild. (Indeed, to address the fully
generic physical problem would require a final generalization to the
Kerr case!) Using this code, and similar codes developed independently by
others \cite{Detweiler:2008ft}, we can now begin to explore the physical
consequences of the gravitational SF---particularly those effects associated
with its conservative piece. Work done so far includes (i) study of gauge
invariant SF effects on circular orbits \cite{Detweiler:2008ft};
(ii) comparison of SF calculations and results from PN theory in the weak
field regime \cite{Detweiler:2008ft,Sago:2008id};
(iii) comparison of SF results from different calculation schemes
and using different gauges \cite{Sago:2008id}; and
(iv) a calculation of the shift in the location and frequency of the Innermost
Stable Circular Orbit  (ISCO) due to the conservative piece of the SF \cite{Barack:2009ey}.
Work is in progress to calculate SF-related precession effects for eccentric
orbits. In Sec.\ \ref{sec:effects} we will describe some of the recent work
to explore the physical consequences of the gravitational SF.

\section{Towards self-force calculations in Kerr: the puncture method
and $m$-mode regularization}
\label{sec:puncture}

The time-domain Lorenz-gauge treatment of Sec.\ \ref{sec:Lorenz} relies
crucially on the separability of the field equations into (tensorial)
spherical harmonics, which is no longer possible in Kerr. In the Kerr
problem one can at best separate the metric perturbation equations into
azimuthal $m$-modes, using the substitution\footnote{A full separation of
variables in Kerr is possible within Teukolsky's formalism, which, alas,
brings about the metric reconstruction and gauge-related difficulties
discussed in previous chapters. A full separation of the metric perturbation
equations themselves, in Kerr, has not been formulated yet, to the best
of our knowledge.}
\begin{equation}\label{eq420}
\bar h_{\alpha\beta}=\sum_{m=-\infty}^{\infty}
\bar h^{m}_{\alpha\beta}(t,r,\theta)e^{im\varphi}.
\end{equation}
Then one faces solving the coupled set for the 2+1D variables $\bar h^{m}_
{\alpha\beta}$. Insofar as vacuum perturbations are considered, this computational
task is nowadays well within reach of even modest desktop computers. Indeed,
over the past decade, 2+1D numerical evolution has been a method of choice in
many studies of Kerr perturbations \cite{Krivan:1996da,Krivan:1997hc,
PazosAvalos:2004rp, LopezAleman:2003ik,Khanna:2003qv,Sundararajan:2007jg,
Barack:2007jh,Lousto:2008mb}. The main challenge, rather, has to do with
the inclusion of a $\delta$-function source in the 2+1D numerical domain.
This is problematic, because the 2+1D variables $\bar h^{m}_{\alpha\beta}$ suffer
a singularity at the location of the particle. The {\em puncture} method, which
we have mentioned briefly in Sec.\ \ref{sec:3.2.1}, overcomes this technical
difficulty. In this section we review this method (as implemented
by Barack and Golbourn \cite{Barack:2007jh}) in some more detail. We also
describe a new, ad hoc, regularization method---the {\em $m$-mode
regularization scheme}---which enables a straightforward construction
of the SF directly from the 2+1D numerical solutions, without resorting to
a multipole decomposition.

\subsection{Puncture method in 2+1D}

In Sec.\ \ref{sec:theory} we have split the (trace-reversed) metric perturbation
as $\bar h^{{\rm ret}}_{\alpha\beta}=\bar h^{\rm S}_{\alpha\beta}+\bar h^{\rm R}
_{\alpha\beta}$, with the gravitational SF then obtained from the smooth
field $\bar h^{\rm R}_{\alpha\beta}$ as prescribed in Eq.\ (\ref{eq1:600}).
Here we introduce a new splitting (defined below),
\begin{equation}\label{eq430}
\bar h^{{\rm ret}}_{\alpha\beta}=
\bar h^{\rm S}_{\alpha\beta}+\bar h^{\rm R}_{\alpha\beta}=
\bar h^{\rm punc}_{\alpha\beta}+\bar h^{\rm res}_{\alpha\beta},
\end{equation}
so that
\begin{equation}\label{eq440}
\bar h^{\rm R}_{\alpha\beta}=\bar h^{\rm res}_{\alpha\beta}
+(\bar h^{\rm punc}_{\alpha\beta}-\bar h^{\rm S}_{\alpha\beta}).
\end{equation}
We denote the $m$ modes of these various quantities, defined as in
Eq.\ (\ref{eq420}), by $\bar h^{\rm R,m}_{\alpha\beta}$, $\bar h^{\rm res,m}
_{\alpha\beta}$, etc. The new splitting is defined (in a non-unique way)
by introducing a {\it puncture} field $\bar h^{\rm punc}_{\alpha\beta}$,
given analytically, which approximates $\bar h^{\rm S}_{\alpha\beta}$ near
the particle well enough that the $m$ modes of the resulting {\it residual}
field $\bar h^{\rm res}_{\alpha\beta}$ are continuous and differentiable
on the particle's worldline (and elsewhere). A particular such
puncture is prescribed below [Eq.\ (\ref{eq500})]. The form of the puncture
function away from the particle can be chosen as convenient---e.g., in such
a way that it can be decomposed into $m$ modes explicitly, in analytic form.

The Lorenz-gauge perturbation equations (\ref{eq1:200}) are now written
in the form
\begin{equation}\label{eq450}
\nabla^{\gamma}\nabla_{\gamma} \bar h_{\alpha\beta}^{\rm res}
+2R^{\mu}{}_{\alpha}{}^{\nu}{}_{\beta}\bar h_{\mu\nu}^{\rm res}
=S^{\rm res}_{\alpha\beta},
\end{equation}
with
\begin{equation}\label{eq460}
S^{\rm res}_{\alpha\beta}\equiv
-16\pi T_{\alpha\beta}
-\nabla^{\gamma}\nabla_{\gamma} \bar h_{\alpha\beta}^{\rm punc}
-2R^{\mu}{}_{\alpha}{}^{\nu}{}_{\beta}\bar h_{\mu\nu}^{\rm punc},
\end{equation}
where $-16\pi T_{\alpha\beta}$ is the original (distributional) source term
appearing on the right-hand side of Eq.\ (\ref{eq1:200}).
The support of the source $S^{\rm res}_{\alpha\beta}$ now extends beyond
the particle's worldline, but it contains no $\delta$ function on the
worldline itself.\footnote{This can be shown by integrating $S^{\rm res}_{\alpha\beta}$
over a small 3-ball containing the particle (at a given time), and then
inspecting the limit as the radius of the ball tends to zero \cite{Barack:2007jh}.}
The equations are separated into $m$ modes, with the $m$-mode
source given by
\begin{equation}\label{eq470}
S^{\rm res,m}_{\alpha\beta}(t,r,\theta)\equiv
\frac{1}{2\pi}\int_{-\pi}^{\pi} S^{\rm res}_{\alpha\beta}(t,r,\theta,\varphi)\,
e^{-im\varphi}\, d\varphi.
\end{equation}
If the puncture is sufficiently simple, the source $S^{\rm res,m}_{\alpha\beta}$
can be evaluated in closed form (as in \cite{Barack:2007jh}).
The $m$-mode field equations for the variables $\bar h_{\alpha\beta}
^{\rm res,m}(t,r,\theta)$, which are everywhere continuous and differentiable,
can now be integrated numerically using straightforward finite differentiation on
a 2+1D grid.

\setcounter{footnote}{0}

To ease the imposition of boundary conditions for $\bar h_{\alpha\beta}
^{\rm res,m}$, it is convenient to suppress the support of the puncture
$\bar h_{\alpha\beta}^{\rm punc}$ away from the particle,
so that the physical boundary conditions for $\bar h_{\alpha\beta}^{\rm res}$
become practically identical to that of the full retarded field
$\bar h_{\alpha\beta}$. In \cite{Barack:2007jh} this is achieved in a simple
way by introducing an auxiliary ``worldtube'' around the particle's worldline
(in the 2+1D space): Inside this worldtube one solves the ``punctured''
$m$-mode equations for $\bar h_{\alpha\beta}^{\rm res,m}$, while outside
the worldtube one uses the original, retarded-field $m$-modes
$\bar h_{\alpha\beta}^{{\rm ret},m}$ as evolution variables; the value of the
evolution variables is adjusted on the boundary of the worldtube using
$\bar h^{{\rm ret},m}_{\alpha\beta}=\bar h^{\rm res,m}_{\alpha\beta}+
\bar h^{\rm punc,m}_{\alpha\beta}$.

Two very similar variants of the puncture scheme have been developed and
implemented independently by two groups \cite{Barack:2007jh,
Lousto:2008mb}---both for the toy model of a scalar charge on a circular
orbit around a Schwarzschild black hole (refraining from a spherical
harmonics decomposition and instead working in 2+1D). The method is
yet to be applied in Kerr and for gravitational perturbations.

\subsection{$m$-mode regularization}

The 2+1D numerical solutions obtained using the puncture method can be used
to construct the input modes $F^{\alpha l}_{\pm}$ for the mode-sum formula
(\ref{eq4:200}), but this would require the additional step of a
decomposition in spherical harmonics. It appears, however, that there is
a simple way to construct the SF directly from the residual modes
$\bar h_{\alpha\beta}^{\rm res,m}$, without resorting to a multipole decomposition.
Such ``$m$-mode regularization'' procedure was prescribed recently in
Ref.\ \cite{Barack:2007we} for the scalar, EM and gravitational
SFs. We describe it here as applied to the gravitational SF.

In analogy with Eq.\ (\ref{eq4:20}), let us define the ``force'' fields
\begin{equation} \label{eq480}
F^{\alpha}_{\rm res}(x)=\mu \bar\nabla^{\alpha\beta\gamma}_x
\bar h^{\rm res}_{\beta\gamma},
\quad\quad
F^{\alpha}_{\rm punc}(x)=\mu \bar\nabla^{\alpha\beta\gamma}_x
\bar h^{\rm punc}_{\beta\gamma}.
\end{equation}
Then, recalling Eqs.\ (\ref{eq1:600}) and (\ref{eq440}), we may write the SF
at a point $z$ along the orbit as
\begin{equation} \label{eq490}
F^{\alpha}_{\rm self}(z)=\lim_{x\to z}
\left[F^{\alpha}_{\rm res}(x)+\left(F^{\alpha}_{\rm punc}(x)
-F^{\alpha}_{\rm S}(x)\right)\right].
\end{equation}
Recall that the expression in square brackets
($=\mu \bar\nabla^{\alpha\beta\gamma}_x \bar h^{\rm R}_{\beta\gamma}$)
is a smooth (analytic) function of $x$, and so the limit
$x\to z$ is well defined. We proceed by prescribing a suitable puncture
function $\bar h^{\rm punc}_{\beta\gamma}$.

As in previous sections, let us parameterize the particle's (bound, timelike)
geodesic orbit by proper time $\tau$, and let $z^{\alpha}(\tau)$ describe the
orbit in Boyer--Lindquist coordinates. For an arbitrary spacetime point
$x^{\alpha}=(t,r,\theta,\varphi)$ outside the black hole, let $\Sigma_x$
be the spatial hypersurface $t$=const containing $x$, let $\bar\tau(t)$ be
the value of $\tau$ at which $\Sigma_x$ intersects the particle's worldline,
and denote $\bar z(t)\equiv z[\bar\tau(t)]$ and $\bar u^{\alpha}\equiv
u^{\alpha}(\bar z)$. This procedure assigns to any point $x$ outside the
black hole a unique point $\bar z$ on the worldline, with four-velocity
$\bar u^{\alpha}$. We denote the coordinate distance between a given point
$x$ and the point $\bar z$ associated with it by
$\delta x^{\alpha}\equiv x^{\alpha}-\bar z^{\alpha}(t)$,
with Boyer-Lindquist components $\delta x^{\alpha}=(0,\delta r,\delta\theta,
\delta\varphi)$.

Consider now the local form of the S field $\bar h^{\rm S}_{\alpha\beta}(x,z)$,
given in Eq.\ (\ref{eq1:170}), with $x$ being an arbitrary point near the
worldline, and $z$ being the worldline point $\bar z$ associated with $x$.
Recall in Eq.\ (\ref{eq1:170}) $\epsilon(x;z)$ is the normal geodesic distance
from $x$ to the worldline, and $\hat u^{\alpha}(x;z)$ is the four-velocity
parallelly propagated from $\bar z$ to $x$. The two quantities $\epsilon^2(x;\bar z)$
and $\hat u^{\alpha}(x;\bar z)$ are well defined and smooth functions of $x$
if $x$ is sufficiently close to the worldline [hence also close to $\bar z(t)$].
For $x\to \bar z$ (hence $\delta x\to 0$) we have the asymptotic expansions
\begin{equation}\label{eq492}
\epsilon^2(x,\bar z)=\bar S_0+\bar S_1+O(\delta x^4),
\quad\quad
\hat u_{\alpha}(x,\bar z)=\bar u_{\alpha}+\delta\bar u_{\alpha} +O(\delta x^2),
\end{equation}
where $\bar S_0$ and $\bar S_1$ are, respectively, quadratic and cubic in $\delta x$,
and $\delta\bar u_{\alpha}$ is linear in $\delta x$. Explicitly, these
expansion coefficients read
\begin{eqnarray} \label{eq494}
\bar S_0 &=& \bar P_{\alpha\beta}\delta x^{\alpha}\delta x^{\beta}  \\
\label{eq2260}
\bar S_1 &=& \bar P_{\alpha\lambda}\bar\Gamma^{\lambda}_{\beta\gamma}
\delta x^{\alpha}\delta x^{\beta}\delta x^{\gamma},\\
\label{eq228}
\delta\bar u_{\alpha} &=& \bar \Gamma^{\lambda}_{\alpha\beta}\bar u_{\lambda}\delta x^{\beta},
\end{eqnarray}
where
$\bar P_{\alpha\beta}\equiv \bar g_{\alpha\beta}+\bar u_{\alpha}\bar u_{\beta}$,
$\bar g_{\alpha\beta}\equiv g_{\alpha\beta}(\bar z)$ and
$\bar\Gamma^{\lambda}_{\beta\gamma}\equiv \Gamma^{\lambda}_{\beta\gamma}(\bar z)$
(for a derivation of $\bar S_1$, see, for example, Appendix A of \cite{Barack:2002mha}).
We now wish to regard Eqs.\ (\ref{eq494})--(\ref{eq228}) as {\em definitions} of the
quantities $\bar S_0$, $\bar S_1$ and $\delta\bar u_{\alpha}$, taken to be valid
globally, for any $x$ outside the black hole. [Of course, we keep in mind that
the asymptotic expansions (\ref{eq492}) are only valid sufficiently close to the
worldline.] We then define our puncture function, for any $x$, as follows:
\begin{equation}\label{eq500}
\bar h^{\rm punc}_{\alpha\beta}(x)=
\frac{4\mu\left(\bar u_{\alpha}\bar u_{\beta}+
\delta\bar u_{\alpha}\bar u_{\beta}+ \bar u_{\alpha}\delta\bar u_{\beta}\right)}
{(\bar S_0+\bar S_1)^{1/2}}.
\end{equation}
The support of this function can be attenuated far from the particle in order to control
its global properties, but such modifications will not affect our discussion here.

It is not difficult to show \cite{Barack:2007we} that, near the particle,
the difference between the puncture (\ref{eq500}) and the actual S field given
in Eq.\ (\ref{eq1:170}) (with $z$ taken to be $\bar z$) has the form
\begin{equation}\label{eq505}
\bar h^{\rm punc}_{\alpha\beta}(x)-\bar h^{\rm S}_{\alpha\beta}(x)=
\bar\epsilon_0^{-3}P_{\alpha\beta}^{(4)}(\delta x)+{\rm const}+O(\delta x^2),
\end{equation}
where $\bar\epsilon_0\equiv \bar S_0^{1/2}$, and $P_{\alpha\beta}^{(4)}$ is a
smooth function of the coordinate differences $\delta x^{\alpha}$,
of homogeneous order $(\delta x)^4$. [The function $P_{\alpha\beta}^{(4)}$
depends on the function $w_{\alpha\beta}^{\rm S}$ appearing Eq.\ (\ref{eq1:170}),
but here we will not need to know the explicit form of these function.]
It follows that the difference
between the corresponding ``force'' fields has the local form
\begin{equation}\label{eq510}
F^{\alpha}_{\rm punc}(x)-F^{\alpha}_{\rm S}(x)=
\bar\epsilon_0^{-5}P^{\alpha}_{(5)}(\delta x)+O(\delta x),
\end{equation}
where $P^{\alpha}_{(5)}$ is yet another smooth function, this time of homogeneous
order $(\delta x)^5$. Notice that $F^{\alpha}_{\rm punc}-F^{\alpha}_{\rm S}$
is bounded but generally discontinuous (direction-dependent) at $x\to\bar z$.
From Eq.\ (\ref{eq490}) it then follows that $F^{\alpha}_{\rm res}$ too
is bounded but discontinuous at $x\to\bar z$ [since the limit of the entire
expression in square brackets in (\ref{eq490}) is known to be finite and
direction-independent].

We now arrive at the crucial step of our discussion. In Eq.\ (\ref{eq490}),
for a given point $z$ along the particle's worldline,
we express the (analytic) function in square brackets as a formal sum over
$m$-modes, in the form
\begin{equation} \label{eq515}
F^{\alpha}_{\rm self}(z)=\lim_{x\to z}\sum_{m=-\infty}^{\infty}
\left[F^{\alpha,m}_{\rm res}(x)+\left(F^{\alpha,m}_{\rm punc}(x)
-F^{\alpha,m}_{\rm S}(x)\right)\right],
\end{equation}
where
\begin{equation} \label{eq520}
F^{\alpha,m}_{\rm res}(x)\equiv \frac{1}{2\pi}\int_{-\pi}^{\pi}
F^{\alpha}_{\rm res}(t,r,\theta\,\varphi';\bar z) e^{im(\varphi-\varphi')}
\, d\varphi',
\end{equation}
and similarly for $F^{\alpha,m}_{\rm punc}$ and $F^{\alpha,m}_{\rm S}$.
Since the $m$-mode sum is formally a Fourier expansion, and since the
Fourier expansion of an analytic function converges uniformly, we may
replace the order of limit and summation in Eq.\ (\ref{eq515}):
\begin{equation} \label{eq525}
F^{\alpha}_{\rm self}(z)=\sum_{m=-\infty}^{\infty}\lim_{x\to z}
\left[F^{\alpha,m}_{\rm res}(x)+\left(F^{\alpha,m}_{\rm punc}(x)
-F^{\alpha,m}_{\rm S}(x)\right)\right].
\end{equation}
From Eq.\ (\ref{eq510}), omitting the terms $O(\delta x)$ as they cannot
possibly affect the final value of the SF in Eq.\ (\ref{eq490}), we have
\begin{equation}\label{eq530}
\lim_{x\to z}\left(F^{\alpha,m}_{\rm punc}-F^{\alpha,m}_{\rm S}\right)=
\lim_{x\to \bar z}\frac{1}{2\pi}e^{im\varphi}\int_{-\pi}^{\pi}
\bar\epsilon_0^{-5}P^{\alpha}_{(5)}e^{-im\varphi'}\, d\varphi',
\end{equation}
where the integrand is evaluated at $\varphi=\varphi'$ and
where we have used the fact that $x\to z$ implies also $x\to \bar z$.
Crucially, one finds \cite{Barack:2007we} that the last integral {\em vanishes}
at the limit $x\to\bar z$, for any $m$, regardless of the explicit form of
$P^{\alpha}_{(5)}$. Hence, Eq.\ (\ref{eq525}) reduces to
\begin{equation} \label{eq535}
F^{\alpha}_{\rm self}(z)=\sum_{m=-\infty}^{\infty}F^{\alpha,m}_{\rm res}(z).
\end{equation}
Here the substitution $x=z$ is allowed since the limit $x\to z$ is known to be
well defined (and, in particular, direction-independent). Phrased differently,
the modes $F^{\alpha,m}_{\rm res}$ corresponding to our puncture (\ref{eq500})
are continuous at the particle, as desired.

It might perhaps seem suspicious that, in the above analysis, the sum over $m$-modes
$F^{\alpha,m}_{\rm punc}-F^{\alpha,m}_{\rm S}$ (which are all zero at $x=z$)
fails to recover the original function $F^{\alpha}_{\rm punc}-F^{\alpha}_{\rm S}$
(which is discontinuous at $x=z$ and hence indefinite there). This, however,
should not come as a surprise. Recall that the formal Fourier sum at a step
discontinuity (where the function itself is indefinite) is in fact convergent:
it yields the two-side average value of the function at the discontinuity.
Technically, this peculiarity of the formal Fourier expansion is due simply to
the non-interchangeability of the sum and limit at the point of discontinuity.

Relatedly, we should emphasize that the full (4D) residual field,
$F^{\alpha}_{\rm res}$, does {\em not} yield the correct SF upon taking
the limit $x\to z$: this field does not even have a well defined limit $x\to z$,
as we argued in the discussion below Eq.\ (\ref{eq510}). However, the
sum over the formal Fourier modes of $F^{\alpha}_{\rm res}$, which indeed
fails to reproduce $F^{\alpha}_{\rm res}$ at the particle, {\em does} turn
out to give the correct SF, as we have established in the discussion leading
to Eq.\ (\ref{eq535}).

It is possible to design an improved, higher-order-accurate puncture
$\bar h^{\rm punc}_{\alpha\beta}$, so that the SF is indeed simply given by
$F^{\alpha}_{\rm res}(z)$. Such an improvement is necessary for 3+1D
implementations of the puncture scheme, and is beneficial also in a 2+1D
framework as it enhances the differentiability of the residual variable
and improves the convergence of the $m$-mode sum in Eq.\ (\ref{eq535}).
However, such improvements entail an explicit calculation of higher-order
terms in the $S$ field [including the term $w_{\alpha\beta}^{\rm S}$ and
possibly higher-order terms in Eq.\ (\ref{eq1:170})]. A higher-order
puncture, suitable for 3+1D implementation, was devised recently by
Vega and collaborators \cite{Vega:2007mc,Vega:2009qb}.

Our result (\ref{eq535}) can be written explicitly in terms of the $m$ modes
of the residual field $\bar h^{\rm res}_{\beta\gamma}$, and easily so if we use
the fixed off-worldline extension of $\bar\nabla^{\alpha\beta\gamma}$ (the
choice of extension may affect the value of the individual $m$ modes,
but not the eventual value of the SF). We then have
\begin{equation} \label{eq540}
F^{\alpha}_{\rm self}=\mu \sum_{m=-\infty}^{\infty}
    \bar\nabla^{\alpha\beta\gamma}
    \left(\bar h^{\rm res,m}_{\beta\gamma}e^{im\varphi}\right),
\end{equation}
where, of course, the right-hand side is evaluated at the particle.
This formula prescribes a straightforward method for
constructing the SF in a 2+1D framework. In the puncture scheme we
effectively ``regularize'' the field equations themselves (not their
solutions, as in the standard $l$-mode regularization method),
by writing them in the form (\ref{eq450}) with a sufficiently accurate
puncture function---like the one we give in (\ref{eq500}). Once the
$m$-modes of the residual perturbation have been calculated, the SF is
constructed directly from these modes, via Eq.\ (\ref{eq540}), with no
need for further regularization. This ``$m$-mode regularization scheme'',
yet to be applied in actual calculations of the SF, offers a
simple and efficient framework for such calculations in Kerr spacetime.

\section{Physical effects of the self force}
\label{sec:effects}

With the development of the first gravitational SF codes in the past
few years, it became possible to start exploring quantitatively the physical
consequences of the SF. While the ultimate goal of the SF program remains
the description of the long-term orbital evolution, knowledge of the SF
along geodesic orbits already gives access to some interesting physics
associated with the $O(\mu)$ dynamics. Among the questions that one can
address: How does the finite mass of the particle affect the
rates of periapsis precession and orbital plane precession? How does it modify
the location and frequency of the ISCO? Are there other, less familiar
conservative SF effects that could manifest themselves in a characteristic
way in the emitted gravitational waveforms? Can transient resonances between
the radial and polar motion (for eccentric and inclined orbits in Kerr)
have important observational implications \cite{FHO}?

Identifying and quantifying concrete SF effects that are gauge invariant
and have a clear physical interpretation is also vital if one intends to
cross-validate SF calculations carried out in different gauges
\cite{Sago:2008id}, and in making a connection with results from PN
theory \cite{Detweiler:2008ft}. Quantitative information about concrete SF
effects [such as the $O(\mu)$ correction to the strong-field periapsis precession
rate as a function of the geodesic parameters] can be incorporated ``by hand''
into approximate (PN) models of EMRI orbital evolution \cite{Gair:2005ih,Huerta:2008gb}.
In this way, the study of SF effects for geodesic orbits can inform the
improvement of EMRI models even before a reliable and workable scheme
for the long-term evolution is at hand.

In this penultimate section we describe some of the recent initial
investigations into the physical consequences of the gravitational SF.

\subsection{Conservative and dissipative pieces of the SF}

In discussing the physical consequences of the SF it is very useful to distinguish
between ``conservative'' and ``dissipative'' effects. These are normally
defined as effects arising, correspondingly, from the ``time symmetric''
and ``time antisymmetric'' pieces of the SF \cite{Hinderer:2008dm}. To
formulate this more precisely, let us re-write Eq.\ (\ref{eq4:10}) as
\begin{equation} \label{eq8:10}
F^{\alpha}_{\rm self(ret)}(z)=\lim_{x\to z}
\left[F_{\rm ret}^{\alpha}(x)-F^{\alpha}_{\rm S}(x)\right],
\end{equation}
where the label `ret' is to remind us that the physical SF is derived
from the physical, {\it retarded } metric perturbation
[recall $F_{\rm ret}^{\alpha}=\mu \bar\nabla_x^{\alpha\beta\gamma}
\bar h^{\rm ret}_{\beta\gamma}$]. Then introduce the force
\begin{equation} \label{eq8:20}
F^{\alpha}_{\rm self(adv)}(z)\equiv \lim_{x\to z}
\left[F_{\rm adv}^{\alpha}(x)-F^{\alpha}_{\rm S}(x)\right],
\end{equation}
where $F_{\rm adv}^{\alpha}\equiv\mu \bar\nabla_x^{\alpha\beta\gamma}
\bar h^{\rm adv}_{\beta\gamma}$ (namely, $F^{\alpha}_{\rm self(adv)}$
is obtained from $F^{\alpha}_{\rm self(ret)}$ by replacing the retarded
perturbation with the advanced one). The conservative and dissipative
pieces of the SF are then defined through
\begin{equation} \label{eq8:30}
F^{\alpha}_{\rm cons}\equiv
\frac{1}{2}\left(F_{\rm self(ret)}^{\alpha}
+F_{\rm self(adv)}^{\alpha}\right),
\quad\quad
F^{\alpha}_{\rm diss}\equiv
\frac{1}{2}\left(F_{\rm self(ret)}^{\alpha}
-F_{\rm self(adv)}^{\alpha}\right).
\end{equation}
The physical SF is the sum of the two pieces:
$F^{\alpha}_{\rm self}=F^{\alpha}_{\rm self(ret)}=
F_{\rm cons}^{\alpha}+F_{\rm diss}^{\alpha}$.

The dissipative force, $F_{\rm diss}^{\alpha}$ is responsible (in particular)
for the long-term secular drift in the value of the intrinsic ``constants'' of
motion---the energy $E$, azimuthal angular momentum $L_z$ and Carter constant
$Q_c$ associated with the geodesic motion in Kerr. $F_{\rm cons}^{\alpha}$, on
the other hand, has no such long-term influence on the evolution of these orbital
parameters. Here ``long-term'' refers to a period of time much longer than the
longest orbital period.\footnote{Recall bound geodesics in Kerr are 3-periodic,
with three characteristic frequencies corresponding to the azimuthal, radial,
and longitudinal motion.}
It should be noted, however, that both $F_{\rm diss}^{\alpha}$ and
$F_{\rm cons}^{\alpha}$ affect the momentary values of the intrinsic
parameters ${E,L_z,Q_c}$ (in a gauge-dependent way);
in the case of $F_{\rm cons}^{\alpha}$ this effect ``averages out'' over
many orbital periods, whereas for $F_{\rm diss}^{\alpha}$ it accumulates.
It should also be noted (what is usually less well recognized) that
$F_{\rm cons}^{\alpha}$, too, gives rise to secular, long-term
effects---those associated with the evolution of the phase-type orbital
elements \cite{Pound:2005fs,Pound:2007ti,Pound:2007th}.

How does one go about separating the full SF into its conservative and
dissipative pieces in practice? In a frequency-domain framework this
poses no difficulty: In addition to calculating the physical SF $F_{\rm self(ret)}$,
one also calculates the advanced metric perturbation modes by suitably inverting
the boundary conditions in the relevant ordinary differential equations, and use
these modes to construct the quantity $F^{\alpha}_{\rm self(adv)}$
[e.g., using the mode-sum formula (\ref{eq4:200}), replacing
$F^{\alpha l}_{{\rm ret}\pm}\to F^{\alpha l}_{{\rm adv}\pm}$].
The conservative and dissipative pieces are then obtained using Eq.\
(\ref{eq8:30}). This procedure was implemented, for example, in the analysis
of Ref.\ \cite{DiazRivera:2004ik}. The situation is slightly different
in a time-domain framework, as one does not normally control the
boundary conditions during the time evolution (cf.\ Sec.\ \ref{sec:Lorenz}).
Haas \cite{Haas:Capra} proposed to obtain the necessary advanced SF data
by reversing the time-direction of the evolution: For a given orbit, one
solves the relevant time-domain field equations once evolving forward in
time as usual to obtain $F_{\rm self(ret)}$, and then again evolving backward
in time (starting with initial Cauchy data specified in the future) to
obtain $F_{\rm self(adv)}$.

There is a more computationally economic method for constructing
$F^{\alpha}_{\rm cons/diss}$ in the time domain, at least in the case
of orbits that are either equatorial or circular. This method, first
proposed and implemented in Ref.\ \cite{Barack:2009ey}, makes use of
the symmetries of Kerr geodesics, first noted in the current context by
Mino \cite{Mino:2003yg}. It can be shown (see, e.g., Sec.\ II.G of
\cite{Hinderer:2008dm}) that, at any given point in Kerr spacetime,
the following relation applies:
\begin{equation} \label{eq8:40}
F_{\rm self(adv)}^{\alpha}(u_t,u_r,u_{\theta},u_{\varphi})=
\epsilon_{(\alpha)}
F_{\rm self(ret)}^{\alpha}(u_t,-u_r,-u_{\theta},u_{\varphi}),
\end{equation}
where
\begin{equation} \label{eq8:45}
\epsilon_{(\alpha)}= (-1,1,1,-1)
\end{equation}
in Boyer-Lindquist coordinates (no summation over $\alpha$ on the right-hand
side). Here we are treating the SF as a function of the four-velocity at the
given spacetime point. Now consider a bound geodesic orbit in Kerr, which is
either circular (and possibly inclined) or equatorial (and possibly eccentric).
We parameterize this geodesic by the proper time $\tau$, and, if the orbit is
eccentric, take $\tau=0$ at one of the radial ``turning points'' (i.e., where
$dr/d\tau=0$) without loss of generality. Now examine the retarded and advanced
SFs $F_{\rm self(ret/adv)}^{\alpha}(\tau)$ along this geodesic.
It is clear from symmetry that
\begin{equation} \label{eq8:50}
F_{\rm self(ret/adv)}^{\alpha}(u_t,u_r,u_{\theta},u_{\varphi};-\tau)=
F_{\rm self(ret/adv)}^{\alpha}(u_t,-u_r,-u_{\theta},u_{\varphi};\tau).
\end{equation}
From Eq.\ (\ref{eq8:40}) it then follows that
\begin{equation} \label{eq8:60}
F_{\rm self(ret/adv)}^{\alpha}(\tau)=
\epsilon_{(\alpha)}
F_{\rm self(adv/ret)}^{\alpha}(-\tau),
\end{equation}
and Eq.\ (\ref{eq8:30}) gives
\begin{eqnarray} \label{eq8:70}
F^{\alpha}_{\rm cons}(\tau)&=&
\frac{1}{2}\left(F_{\rm self(ret)}^{\alpha}(\tau)+
\epsilon_{(\alpha)}F_{\rm self(ret)}^{\alpha}(-\tau)\right),
\\ \label{eq8:75}
F^{\alpha}_{\rm diss}(\tau)&=&
\frac{1}{2}\left(F_{\rm self(ret)}^{\alpha}(\tau)-
\epsilon_{(\alpha)}F_{\rm self(ret)}^{\alpha}(-\tau)\right).
\end{eqnarray}
Since in time-domain calculations the physical (retarded) SF is computed along
an entire radial period (at least), Eqs.\ (\ref{eq8:70}) and (\ref{eq8:75})
can be used to obtain
the conservative and dissipative components of this force anywhere along the orbit
without resorting to a calculation of the advanced perturbation.

The above trick can, of course, be implemented for any geodesic orbit in
Schwarzschild (as in this case the orbit can always be taken as ``equatorial'').
However, it cannot be applied for orbits in Kerr that are both eccentric
and inclined, because for such orbits the symmetry relation (\ref{eq8:50})
does not hold in general. Note that, for circular orbits, Eqs.\ (\ref{eq8:70})
and (\ref{eq8:75}) immediately yield
$F^{t}_{\rm cons}=F^{\varphi}_{\rm cons}=0$ as well as
$F^{r}_{\rm diss}=F^{\theta}_{\rm diss}=0$,
so in this case the $t$ and $\varphi$ components of the SF are purely
dissipative, while the $r$ and $\theta$ components are purely conservative.

Finally, it is useful to have at hand separate mode-sum formulas {\it \`{a} la}
(\ref{eq4:200}) for the conservative and dissipative pieces. To obtain
such formulas, first write a mode-sum expression for $F^{\alpha}_{\rm self(adv)}$
by replacing $F_{{\rm ret}\pm}^{\alpha l}\to F_{{\rm adv}\pm}^{\alpha l}$
in Eq.\ (\ref{eq4:200}). Here $F_{{\rm adv}\pm}^{\alpha l}$ are derived from
the modes of the advanced metric perturbation in just the same way as
$F_{{\rm ret}\pm}^{\alpha l}$ are derived from the modes of the retarded
perturbation. (The same regularization parameters apply to both retarded
and advanced forces, because the corresponding metric perturbations share
the same singular piece.) Then substitute the mode-sum expressions for
$F^{\alpha}_{\rm self(ret)}$ and $F^{\alpha}_{\rm self(adv)}$ in
Eq.\ (\ref{eq8:30}). This gives
\begin{eqnarray} \label{eq8:80}
F^{\alpha}_{\rm cons}&=&
\sum_{l=0}^{\infty}
\left(F_{{\rm cons}\pm}^{\alpha l}\mp LA^{\alpha}-B^{\alpha}\right),
\\ \label{eq8:85}
F^{\alpha}_{\rm diss}&=&
\sum_{l=0}^{\infty} F_{{\rm diss}}^{\alpha l\pm},
\end{eqnarray}
where we have used $C^{\alpha}=D^{\alpha}=0$ and introduced
$F_{{\rm cons}\pm}^{\alpha l}\equiv (F_{{\rm ret}\pm}^{\alpha l}
+F_{{\rm adv}\pm}^{\alpha l})/2$ and
$F_{{\rm diss}\pm}^{\alpha l}\equiv (F_{{\rm ret}\pm}^{\alpha l}
-F_{{\rm adv}\pm}^{\alpha l})/2$.
Notice that the dissipative component of the SF requires no regularization.
In fact, one can show that the mode sum in Eq.\ (\ref{eq8:85}) converges
exponentially fast. For that reason, the computation of the dissipative
piece of the SF via the mode-sum method is technically much less
challenging than that of the conservative piece.

\subsection{Dissipation of energy and angular momentum}\label{subsec:diss}

Perhaps the most familiar aspect of the self-interaction physics in the
binary context is the long-term radiative decay of the orbit. In the
language of the SF, we say that the work done on the particle by the
dissipative component of the SF converts orbital energy (and angular
momentum) into gravitational-wave energy (and angular momentum). The
relation between the SF and dissipation is immediately evident from the
equation of motion (\ref{eq1:190}), whose (Boyer-Lindquist) $t$ and
$\varphi$ components read, respectively,
\begin{equation} \label{eq8:90}
\mu \, \frac{du_t}{d\tau}=F^{\rm self}_t,
\quad\quad
\mu\, \frac{du_\varphi}{d\tau}=F^{\rm self}_\varphi.
\end{equation}
In the absence of SF (that is, in the test-particle limit), Eqs.\ (\ref{eq8:90})
tell us that $u_{t}$ and $u_{\varphi}$ are constants of the motion,
and we interpret these as the orbital energy, $E\equiv -u_t$, and
azimuthal angular momentum, $L_z\equiv u_{\varphi}$---both per unit
charge $\mu$.\footnote{This interpretation is motivated by the observation
that $-u_t$ and $u_{\varphi}$ are, in fact, projections of the four-velocity
onto the Killing vectors associated with the invariance of the Kerr background
under (respectively) time translations and spatial rotations.}
Equations (\ref{eq8:90}) also tell us how $E$ and $L_z$ change under the effect
of the SF. This effect is not entirely dissipative:
The SF includes periodic components which do not lead to a net long-term change
in the values of $E$ and $L_z$. The non-periodic component of $F_t$ and
$F_{\varphi}$ does lead to a secular drift in the values of $E$ and $L$,
interpreted as dissipation.

This non-periodic component is entirely contained
in the dissipative piece of $F^{\alpha}_{t}$ and $F^{\alpha}_{\varphi}$,
as expected. This is immediately clear from Eq.\ (\ref{eq8:70}) in the case of
equatorial or circular orbits in Kerr (and for all orbits in Schwarzschild):
The anti-symmetry of $F_t^{\rm cons}(\tau)$ and $F_\varphi^{\rm cons}(\tau)$
with respect to $\tau\to -\tau$ means that these pieces of the SF vanish upon
integration over one complete (radial or longitudinal) period, and thus produce
no net change in the values of $E$ and $L_z$ over that period (this, of course,
assumes that the orbit is very nearly geodesic over one period).
The same applies generically for all orbits in Kerr, if the orbital
integration is taken over sufficiently many periods
(formally, infinitely many). These statements are encapsulated in the relations
\begin{equation} \label{eq8:100}
\mu\langle\dot E\rangle=-\langle F_t/u^t\rangle
=-\langle F^{\rm diss}_t/u^t\rangle,
\quad\quad
\mu \langle\dot L_z\rangle=\langle F_\varphi/u^t\rangle
= \langle F^{\rm diss}_\varphi/u^t\rangle,
\end{equation}
where an overdot represents $d/dt$ (hence the factor $1/u^t=d\tau/dt$ on the
right-hand side), and $\langle\cdot\rangle$ indicates an average over
sufficiently long time. One always finds $\langle F^{\rm diss}_t\rangle>0$
and $\langle F^{\rm diss}_\varphi\rangle<0$ (see, e.g., \cite{Ori:1997be}), so that
$\langle\dot E\rangle<0$ and $\langle\dot L_z\rangle<0$, and orbital energy and
angular momentum are lost over time as one expects.
The time-average quantities in Eq.\ (\ref{eq8:100}) are expected to
be independent of the gauge even though the momentary values of
$F^{\rm diss}_t$ and $F^{\rm diss}_\varphi$ are themselves gauge dependent.

Implicit in the above discussion is the assumption that the orbit is
indeed ``available'' for the necessary time-averaging, i.e., that $E$ and
$L_z$ evolve very slowly over the required averaging time---otherwise the
averaging procedure would be meaningless. Such ``adiabaticity'' requirement
(formulated more accurately, e.g., in \cite{Hughes:1999bq}) is believed to
hold well in LISA-relevant EMRIs throughout the inspiral and until very
close to the innermost stable orbit---see, e.g., \cite{Hughes:1999bq}
for a more quantitative discussion of this point.

The loss of orbital energy and angular momentum is precisely balanced by
the flux of the corresponding quantities in the gravitational waves radiated
to infinity and down into the black hole. Using Eq.\ (\ref{eq8:100}) we can
express this balance in terms of the local SF:
\begin{eqnarray} \label{eq8:110}
\mu^{-1}\langle F^{\rm diss}_t/u^t\rangle&=&\langle \dot E^{\rm GW}\rangle_{\rm eh}
+ \langle \dot E^{\rm GW}\rangle_{\infty}
\nonumber\\
-\mu^{-1}\langle F^{\rm diss}_\varphi/u^t\rangle&=&\langle
\dot L_z^{\rm GW}\rangle_{\rm eh}
+ \langle \dot L_z^{\rm GW}\rangle_{\infty}.
\end{eqnarray}
The quantities on the right-hand side are the time-averaged asymptotic
fluxes down the event horizon (`eh') and out to infinity (`$\infty$'),
with the convention that the infinity fluxes are positive. This convention
also fixes the sign of the horizon fluxes for a given orbit, but it
should be noted that the latter can be either positive or---for certain
orbits in Kerr---negative. Negative horizon fluxes
reflect a superradiance behavior, whereby energy and angular momentum are, in
effect, transferred from the ergosphere of the Kerr hole {\em to} the orbit
\cite{TeukolskyPress,Hughes:1999bq,Glampedakis:2002ya}. The formal proof of
the balance equations (\ref{eq8:110}) involves the application of the Gauss
theorem in the 3-volume bounded between a 2-sphere at $r\to\infty$ and another
one just outside the horizon; how this is done is demonstrated, for example,
in appendix D of Ref.\ \cite{Detweiler:2008ft} (in the special case of circular
orbits in Schwarzschild).

The asymptotic fluxes $\langle\dot E^{\rm GW}\rangle_{\infty}$ and
$\langle \dot L_z^{\rm GW}\rangle_{\infty}$ are calculated across a large
2-sphere $r\to\infty$ residing in the ``wave zone'', where the background's
radius of curvature is much larger than the characteristic wavelength of the
emitted gravitational waves. In this case Isaacson's
effective energy-momentum tensor formulation \cite{Isaacson} applies,
and one can use it to calculate the fluxes at infinity. This is a standard
calculation in black hole perturbation theory, and since the early 1970's it
has been performed many times using frequency-domain methods (e.g, \cite{TeukolskyPress,
Poisson:1995vs,Hughes:1999bq,Glampedakis:2002ya}) and more recently within a time-domain
framework \cite{Martel:2003jj,Barack:2005nr,Barack:2007tm}. There is also a
standard prescription for calculating the horizon fluxes $\langle\dot E^{\rm GW}
\rangle_{\infty}$ and $\langle \dot L_z^{\rm GW}\rangle_{\infty}$; it is due
to Teukolsky and Press \cite{TeukolskyPress} (based on the horizon perturbation
formalism of Hawking and Hartle \cite{HH}) and is formulated in the frequency domain
in terms of curvature scalars. A time-domain formulation of the horizon fluxes was
more recently developed by Poisson \cite{Poisson:2004cw}.

The balance equations (\ref{eq8:110}), with (\ref{eq8:100}), allow us to
infer the leading-order long-term evolution of $E$ and $L_z$ without resorting
to an explicit calculation of the local SF (assuming adiabaticity). To get a
full description of the time-average orbital decay one must, in general, also
be able to calculate the evolution of the Carter constant $Q_c$. It is not
difficult to write down a relation analogous to (\ref{eq8:100}) which expresses
$\langle \dot Q_c\rangle$ directly in terms of the local SF \cite{Ori:1997be},
but there is no known analogue to (\ref{eq8:110}), relating $\langle \dot Q_c\rangle$
to the asymptotic flux of radiation. However, thanks to a breakthrough idea
by Mino \cite{Mino:2003yg,Mino:2005qj} and follow up work by Sago and collaborators
\cite{Sago:2005fn,Ganz:2007rf}, there is now known a practical formula for
$\langle \dot Q_c\rangle$ which does not require knowledge of the SF (it
does require knowledge of the local advanced and retarded modes of the
metric perturbation).

In conclusion, information on the time-averaged evolution of all three
intrinsic ``constants of motion'', $E$, $L_z$ and $Q_c$, is directly accessible
from the dissipative piece of the SF, but there are alternative methods (likely more
computationally efficient) for accessing this information without resorting
to the SF and to Eq.\ (\ref{eq8:110}). However, from the perspective of the SF
program development, the balance equations (\ref{eq8:110}) are still a very
useful tool for validating SF codes. Two independent tests are possible:
First, since a SF code always involves a calculation of the metric perturbation
itself, one can use this perturbation to construct the asymptotic fluxes of
energy and angular momentum, and then use Eqs.\ (\ref{eq8:110}) to test
the self-consistency of the code. Second, one can test the computed values
of the dissipative SF against asymptotic flux data available in the literature
(e.g., \cite{Martel:2003jj,Glampedakis:2002ya}). Such a two-fold test is
an important, reassuring check on the validity of a SF code. Explicit
calculations demonstrating the balance relations (\ref{eq8:110}) have been
carried out so far for circular \cite{Barack:2007tm} and eccentric \cite{BSprep}
orbits in Schwarzschild.

\subsection{Conservative effects on circular orbits in Schwarzschild}

Detweiler was first to explore quantitatively the conservative effects
of the SF---in the simple example of circular geodesics outside a Schwarzschild
black hole \cite{Detweiler:2008ft}. He pointed out two gauge-invariant quantities
that carry non-trivial information about the conservative SF dynamics
in this setting: the orbital frequency (with respect to time $t$),
\begin{equation} \label{eq8:200}
\Omega\equiv u^{\varphi}/u^{t},
\end{equation}
and the contravariant $t$ component of the four-velocity, denoted
\begin{equation} \label{eq8:210}
U\equiv u^t.
\end{equation}
Here $u^{\alpha}=dx^{\alpha}/d\tau$ denotes the value of the {\em perturbed}
four-velocity along the circular geodesic orbit, including SF corrections.
We have $u^r=0$ along the orbit, and we also take (without loss of generality)
$u^{\theta}=0$.
The said gauge invariance of $\Omega$ and $U$ is restricted to transformations
for which the gauge displacement vector $\xi^{\alpha}$ respects the helical
symmetry of the perturbed spacetime system, i.e., it satisfies
\begin{equation} \label{eq8:220}
(\partial_t + \Omega\partial_\phi)\xi^{\alpha}=0
\end{equation}
through $O(\mu)$, at least along the particle's worldline. It is easy to see
that all contravariant components $u^{\alpha}$ are invariant through $O(\mu)$ under such
transformations \cite{Sago:2008id}. The two nontrivial components $u^t$ and
$u^{\varphi}$ give two independent gauge invariants, which we can re-combine to form
Detweiler's gauge-invariant variables $\Omega$ and $U$.
\setcounter{footnote}{0}

While the physical significance of the frequency $\Omega$ is clear, that
of $U$ is slightly less obvious. Detweiler \cite{Detweiler:2008ft} discusses
two physical interpretations of $U$. First, it is a measure of the gravitational
red-shift experienced by photons emitted by the orbiting particle and observed
at a distance. Second, $U$ is intimately related to the helical Killing vector
of the perturbed spacetime [with its gauge invariance being simply the statement
that this Killing vector is invariant under gauge transformations satisfying
Eq.\ (\ref{eq8:200})]. Unfortunately, as we explain below, the
conservative SF corrections to $\Omega$ and $U$ are not expected to have
any short-term observable imprint on the gravitational waveforms
emitted from the circular orbits.

Explicit expressions for $\Omega$ and $U$, including SF terms, are easily
obtained from the $r$ component of the equation of motion (\ref{eq1:190}),
setting $u^r=0$ and $du^r/d\tau=0$. One finds \cite{Sago:2008id}
\begin{equation}\label{eq8:230}
\Omega =
\Omega_0 \left[
1 - \frac{r_0(r_0-3M)}{2\mu M} F_r,
\right],
\end{equation}
\begin{equation}\label{eq8:235}
U =U_0\left(1- \frac{r_0}{2\mu} F_r
\right),
\end{equation}
through $O(\mu)$,
where $r_0$ is the orbital radius (Schwarzschild $r$ coordinate),  and
$\Omega_0=(M/r_0^3)^{1/2}$ and $U_0\equiv (1-3M/r_0)^{-1/2}$
are the geodesic (unperturbed) values of $\Omega$ and $U$, respectively.
Recall that the radial component of the SF, $F_r$, is purely
conservative in the case of a circular orbit; hence the SF terms in
Eqs.\ (\ref{eq8:230}) and (\ref{eq8:235}) represent conservative corrections
to $\Omega$ and $U$.
Of course, the two quantities would also evolve dissipatively (this effect
is described by the $t$ and $\varphi$ components of the equation of motion),
but here we {\em ignore} the dissipative piece of the SF in order to study
the effect of its conservative piece in isolation. Equations (\ref{eq8:230})
and (\ref{eq8:235})
tell us that the effect of the conservative SF is to ``shift'' the values of
$\Omega$ and $U$ as compared with their non-perturbed geodesic values (for a
given orbital radius $r_0$). The conservative SF shift in $\Omega$,
$\Delta\Omega\equiv \Omega-\Omega_0$, was first calculated (numerically, as a
function of $r_0$) in Ref.\ \cite{Barack:2007tm}, along with the conservative
shift in the orbital energy $E$ (which is gauge dependent).

It should be understood that, despite the formal gauge invariance of $\Omega$,
the relation $\Delta\Omega(r_0)$ is, in fact, gauge dependent, because the
radius $r_0$ itself is gauge dependent (the calculation in \cite{Barack:2007tm}
was carried out in the Lorenz gauge). To illustrate this point, suppose that
we carry out two independent calculations of the SF in two different gauges,
and we wish to test our results by comparing the values of $\Omega$ using
Eq.\ (\ref{eq8:230}). The problem we would face is that a given value of $r_0$
would, in general, correspond to two physically distinct orbits, due to the
gauge ambiguity at $O(\mu)$; there is no way of relating the coordinate values
of the two physical radii in the two gauges without further information about the gauge
(in the form of the local metric perturbation, for example). The conservative SF
shift in $U$, $\Delta U(r_0)\equiv U-U_0$, is similarly gauge dependent, and thus
it too cannot be utilized usefully as a gauge-invariant measure of the SF effect.

One might hope to get around this problem by expressing one of our gauge
invariants in terms of the other. To this end, it is convenient to introduce
the {\em gauge invariant radius},
\begin{equation}\label{eq8:240}
R\equiv \left( M/\Omega^2 \right)^{1/3},
\end{equation}
and then express $U$ in terms of $R$. However, one then simply finds
\begin{equation}\label{eq8:250}
U(R) = (1-3M/R)^{-1/2}+O(\mu^2),
\end{equation}
which contains no explicit information about the SF. This is an obvious result:
An $O(\mu)$ term on the right-hand side of Eq.\ (\ref{eq8:250}) could only
involve $R$ and $F_r$, and since $R$ is gauge invariant while $F_r$ is not,
the occurrence of such a term would necessarily violate the gauge invariance
of $U$. We therefore have to maintain our conclusion that
$U$ and $\Omega$ (or $R$), each on its own or even both combined, do
not contain gauge-invariant information about the SF effect, unless, in
addition, we have access to local gauge information. For that reason, also, we
cannot expect to be able to identify the effect of the conservative SF
in a detected gravitational wave from a circular EMRI merely by measuring
the momentary values of $\Omega$ and/or $U$ (even if there was a way of
extracting $U$ from the waveform, which is unlikely).\footnote{
It is less clear how the conservative piece of the SF might affect the
long-term evolution of the circular orbit under the full SF (dissipation
included), and what influence it may have on the emitted gravitational
waveform---these questions await investigation.}

Fortunately, information about the local metric perturbation, not available
to the gravitational-wave astronomer, {\em is} available to the SF theorist
running a SF code. This allows the theorist to utilize $\Omega$ and $U$
usefully in quantifying the gauge-invariant content of the conservative SF,
as first shown by Detweiler \cite{Detweiler:2005kq,Detweiler:2008ft} and
further demonstrated in \cite{Sago:2008id}. In the following we outline the
method of Ref.\ \cite{Sago:2008id}.

Consider again a circular orbit of radius $r_0$ (in Schwarzschild), but
now interpret this, {\it \`{a} la} Detweiler and Whiting, as a geodesic
of the effective perturbed spacetime with metric $g_{\alpha\beta}+
h_{\alpha\beta}^{\rm R}$, where $h_{\alpha\beta}^{\rm R}$ is the
$R$-part of the physical perturbation associated with the particle. Let
$\tilde\tau$ be proper time along this geodesic.\footnote{Note our
notation here for $\tau$ and $\tilde\tau$ is reversed compared to that
of Ref.\ \cite{Sago:2008id}.}
To a given physical event along the orbit there thus correspond two proper
time values, $\tau$ and $\tilde\tau$, associated with the two interpretations.
(The two values will differ, in general, even if we calibrate $\tau$ and
$\tilde\tau$ to agree with one another at some initial moment.)
It is easily shown \cite{Sago:2008id} that, through $O(\mu)$,
\begin{equation}\label{eq8:260}
\frac{d\tau}{d\tilde\tau}=1+H,
\quad\quad {\rm with}\quad\quad
H\equiv \frac{1}{2}h_{\alpha\beta}^{\rm R}u^{\alpha}u^{\beta},
\end{equation}
where the perturbation is, of course, evaluated at the worldline point
in question.
[The four-velocity in $H$ can be defined interchangeably with
respect to either $\tau$ or $\tilde\tau$: as $h_{\alpha\beta}^{\rm R}$
is already $O(\mu)$, the difference would only affect terms of $O(\mu^2)$
which are anyway neglected in our discussion.]
Next, re-define the gauge invariants $U$ and $\Omega$ in terms of the
four-velocity $\tilde u^{\alpha}\equiv dx^{\alpha}/d\tilde\tau$:
\begin{equation}\label{eq8:270}
\tilde\Omega\equiv \tilde u^{\varphi}/\tilde u^{t}=u^{\varphi}/u^{t}=\Omega,
\quad\quad
\tilde U\equiv \tilde u^t=U(1+H),
\end{equation}
where we have used Eq.\ (\ref{eq8:260}). Finally, express $\tilde U$ in terms
of the gauge invariant radius $\tilde R=R$, and construct the $O(\mu)$
difference
$\Delta \tilde U(R) \equiv \tilde U(R) -(1-3M/R)^{-1/2}$.
The final result is \cite{Sago:2008id}
\begin{equation}\label{eq8:290}
\Delta \tilde U(R) = (1-3M/R)^{-1/2}H.
\end{equation}
Evidently, the metric perturbation combination $H$ is gauge-invariant, as
can be verified with an explicit calculation.

Equation (\ref{eq8:290}) provides a nontrivial gauge-invariant relation
which explicitly involves the $R$-part of the local metric perturbation
(although it does not involve directly the SF). Two theorists working in
two different gauges should be able to agree on the value of $\Delta \tilde U(R)$,
and such an agreement would constitute a nontrivial test of the calculation
of $h_{\alpha\beta}^{\rm R}$---and, to an extent, of the SF too.

Two such tests, based on Eq.\ (\ref{eq8:290}), were carried out so far.
In Ref.\ \cite{Sago:2008id} the results of Barack--Sago's Lorenz-gauge
SF code \cite{Barack:2007tm} were compared with those of Detweiler's
Regge-Wheeler SF code \cite{Detweiler:2008ft}. The two codes were shown to
agree on the values of $\Delta \tilde U(R)$ to within the computational
error (of merely $\sim 10^{-5}$ in fractional terms).
In Ref.\ \cite{Detweiler:2008ft}, Detweiler worked out a 2PN expression
for $\Delta \tilde U(R)$ and compared it with the numerical SF data---showing
an astonishing agreement down to radii as small as $R\sim 8M$. Later, a 3PN
expression derived by Le Tiec and collaborators \cite{LeTiec} showed an
even closer agreement. The 3PN expression approximates the ``exact'' SF
value of $\Delta\tilde U$ to within mere $\sim 1\%$ at $R=12M$ and $\sim 5\%$
at $R=7M$.

\subsection{ISCO shift}

To identify conservative SF effects with truly ``observable'' consequences
(by which we mean ones that can be measured in the gravitational waveform
at least in principle), we must move away from the simplicity of circular
orbits. We need not move very far, though. There is a simple effect with
a clear physical significance, which is manifest already in the dynamics of orbits
with infinitesimally small eccentricity: the conservative SF-induced shift in
the value of the orbital frequency $\Omega$ at the ISCO. This frequency shift,
which is gauge invariant [under transformations satisfying Eq.\ (\ref{eq8:220})],
was calculated recently in Ref.\ \cite{Barack:2009ey}, and we shall describe
this analysis briefly below.

Strictly speaking, the ISCO is defined in a precise way only at the test
particle limit, $\mu\to 0$. When $\mu$ is finite, dissipation ``smears''
the transition from inspiral to plunge and it is no longer precisely
localizable. Ori and Thorne showed \cite{Ori:2000zn} (for
circular orbits in Kerr, a result later generalized to other orbits by
O'Shaughnessy \cite{O'Shaughnessy:2002ez} and Sundararajan \cite{Sundararajan:2008bw})
that the ``width'' of the radiative transition regime (measured, e.g.,
in terms of the frequency bandwidth of the emitted gravitational wave) scales as
a low power of the mass ratio: $\sim(\mu/M)^{2/5}$. The same scaling was discovered 
independently by Buonanno and Damour for a binary of nonrotating black holes with
an arbitrary mass ratio \cite{Buonanno:2000ef}. Ori and Thorne's analysis ignores
the conservative effect of the SF, but the latter is expected to modify the
location of the ISCO only by an amount $\sim (\mu/M)$, which for very small
mass ratios we expect to be negligible compared with the width of the radiative
transition. Still, there is a strong motivation to study the conservative
effect of the SF on the ISCO. First, we would like to confirm the above
expectation and quantify it better. Second, the conservative $O(\mu)$ shift of
the ISCO frequency (dissipation ignored) is a {\em precisely specifiable}
gauge-invariant quantity, and as such it can serve as a convenient
``anchor'' point for comparison between different
calculation schemes. In particular, one can envisage it being used as a
strong-field benchmark for calibration of PN calculations. Indeed, the
value of the conservative ISCO frequency shift, for mass ratios not necessarily
small, has been utilized extensively in the past by PN theorists in assessing
the performance of various PN schemes (see, e.g., \cite{Damour:2000we,
Blanchet:2001id,Blanchet:2002mb}). The calculation in Ref.\ \cite{Barack:2009ey}
now provides this value {\em precisely}, at $O(\mu)$, in the Schwarzschild
case.

To outline the calculation, we first remind how the occurrence of an ISCO
is explained from the point of view of an effective radial potential.
Ignoring the SF for the moment, the radial component of
the equation of motion for a timelike geodesic in Schwarzschild takes
the familiar form
\begin{equation} \label{eq8:300}
\mu\, \ddot r=-\frac{\mu}{2}\,\frac{\partial
V_{\rm eff}(r,L_z)}{\partial r}\equiv{\cal F}_{\rm eff},
\end{equation}
where $V_{\rm eff}=(1-2M/r)(1+L_z^2/r^2)$ is an effective potential
for the radial motion, and throughout our current discussion an overdot
will denote $d/d\tau$ (not $d/dt$ as in Subsec.\ \ref{subsec:diss}).
The quantity ${\cal F}_{\rm eff}(r,L_z)$ can be interpreted as an effective
radial force acting on the geodesic test particle. A stable circular
orbit is associated with the (single) minimum of $V_{\rm eff}$
(for given $L_z$), which occurs only for $L_z>\sqrt{12M}$. The radius
of the circular orbit, $r=r_0(L_z)$, decreases monotonically with
decreasing $L_z$, and at the limiting value of $L_z=\sqrt{12M}$ we
have $r_0=6M$---the innermost stable circular orbit.

To better understand the dynamical significance of the ISCO it is
instructive to examine the behavior of a circular orbit under a
small-eccentricity perturbation. Writing $r(\tau)=r_0+e\,\delta r(\tau)$
and considering the linear variation of Eq.\ (\ref{eq8:300}) with respect
to the small eccentricity $e$, one readily obtains  \cite{Barack:2009ey}
\begin{equation} \label{eq8:310}
\ddot{\delta r}(\tau)=-\omega_0^2\,  \delta r(\tau),
\end{equation}
with
\begin{equation} \label{eq8:320}
\omega_0^2=\frac{M(r_0-6M)}{r_0^3(r_0-3M)}.
\end{equation}
Thus the radius of the perturbed orbit is a linear oscillator with
frequency $\omega_0$ [the subscript $0$ indicates geodesic (no-SF) value].
For $r_0>6M$ we have $\omega_0^2>0$, and the circular orbit is dynamically
stable under small-$e$ perturbations; the condition $\omega_0=0$ identifies
the ISCO at $r_0=6M$.\footnote{The fact that Eq.\ (\ref{eq8:320}) gives
$\omega_0^2>0$ also in the range $r_0<3M$ is irrelevant here, since there are no
timelike circular orbits in that range.} To avoid confusion, it should
be understood that the radial frequency $\omega_0$ is a characteristic
of the circular orbit (not the perturbed orbit), which, however, does
not manifest itself in the dynamics of the circular orbit itself.
Mathematically, $\omega_0$ is associated with the curvature of the
effective potential at its minimum; physically, it describes the radial
frequency of an eccentric orbit as the eccentricity tends to zero.
In our current discussion, we simply use the value of $\omega_0$ as
a convenient handle on the location of the ISCO.

\setcounter{footnote}{0}

Now consider the effect of the SF, ignoring its dissipative piece.
Equation (\ref{eq8:300}) becomes
\begin{equation} \label{eq8:330}
\mu\, \ddot r={\cal F}_{\rm eff}(r,L_z)+ F^r_{\rm cons}.
\end{equation}
The quantity $L_z$ ($=u_{\varphi}$ by its definition) is no longer
constant along the orbit; its (conservative) time-variation is
determined by the $\varphi$ component of the equation of motion, reading
\begin{equation} \label{eq8:340}
\mu\, \dot{L}_z=F^{\rm cons}_{\varphi}.
\end{equation}
To learn how the SF affects the radial frequency of slightly eccentric
orbits, we need to consider the linear variation of Eqs.\ (\ref{eq8:330})
and (\ref{eq8:340}) with respect to $e$. Through $O(e)$, the relevant
SF components along the orbit assume the general form
\begin{equation} \label{eq8:350}
F_{\rm cons}^r=F^r_0+e F^r_1\cos\omega_0\tau,
\quad\quad
F_\varphi^{\rm cons}=e\omega_0 F^1_{\varphi}\sin\omega_0\tau
\end{equation}
[this is shown in Ref.\ \cite{Barack:2009ey} based on Eq.\ (\ref{eq8:70})],
where the coefficients $F^r_0$, $F^r_1$ and $F^1_{\varphi}$ depend only
on $r_0$. In these expressions we can use the
unperturbed radial frequency $\omega_0$ instead of the perturbed
frequency since the $F$ coefficients are already $O(\mu^2)$, and it is
only the leading-order SF effect which concerns us here.
The linear variation procedure once again yields an equation of the form
(\ref{eq8:310}) where now the radial frequency, perturbed by the
conservative SF, is found to be
\begin{equation} \label{eq8:360}
\omega^2=\omega_0^2+
\mu^{-1}\left(\alpha F^r_0+\beta F^r_1+\gamma F^1_{\varphi}\right),
\end{equation}
with
$\alpha=-3r_0^{-1}(r_0-4M)/(r_0-3M)$, $\beta=r_0^{-1}$ and
$\gamma=-2r_0^{-4}[M(r_0-3M)]^{1/2}$.
This formula describes the $O(\mu)$ conservative shift in the radial
frequency off its geodesic value. Note it requires knowledge
of the SF through $O(e)$.

The perturbed ISCO radius, $r_0=6M+\Delta r_{\rm isco}$, is now obtained
from the condition $\omega(r_0)=0$. Recalling Eqs.\ (\ref{eq8:320}),
this gives
\begin{eqnarray} \label{eq8:370}
\Delta r_{\rm isco}
&=& \left. \mu^{-1}(r_0^3/M)(3M-r_0)(\alpha F^r_0+
\beta F^r_1+\gamma F^1_{\varphi})\right|_{r_0=6M}\nonumber\\
&=& \mu^{-1}\left.\left(216M^2F^r_{0} -108M^2 F^r_{1} +
\sqrt{3}\,F^1_{\varphi}\right)\right|_{r_0=6M},
\end{eqnarray}
where the substitution $r_0=6M$  is allowed since this only introduces
an error of $O(\mu^2)$ on the right-hand side. This equation describes
the $O(\mu)$ shift in the location of the ISCO due to the effect of the
conservative SF. This shift is well defined albeit gauge dependent.
However, the value of the azimuthal orbital frequency $\Omega$ at the
(shifted) location of the ISCO is gauge invariant. To obtain the shift
in $\Omega$ we simply substitute $r_0=6M+\Delta r_{\rm isco}$ in Eq.\
(\ref{eq8:230}), writing $\Omega=\Omega_{0,{\rm isco}}+\Delta\Omega_{\rm isco}$
where $\Omega_{0,{\rm isco}}\equiv\Omega_0(6M)=(6^{3/2}\,M)^{-1}$.
We find, through $O(\mu)$,
\begin{equation} \label{eq8:380}
\frac{\Delta\Omega_{\rm isco}}{\Omega_{0,{\rm isco}}}
= -\frac{1}{4}\Delta r_{\rm isco}/M - \frac{27}{2}M\mu^{-1} F_{0}^r(6M).
\end{equation}
In Ref.\ \cite{Barack:2009ey} the SF coefficient $F^r_0$, $F^r_1$ and
$F^1_{\varphi}$ are extracted numerically using the Lorenz-gauge SF
code reviewed here in Sec.\ \ref{sec:Lorenz}. With these numerical results,
Eqs.\ (\ref{eq8:370}) and (\ref{eq8:380}) give the final values
\begin{equation} \label{eq8:390}
\Delta r_{\rm isco}=-3.269\,\mu,
\quad\quad
\frac{\Delta\Omega_{\rm isco}}{\Omega_{0,{\rm isco}}}=0.4870\, \mu/M.
\end{equation}
We remind that the value of $\Delta r_{\rm isco}$ is specific to the Lorenz
gauge, but that for $\Delta\Omega_{\rm isco}$ is invariant under all gauge
transformations related to the Lorenz gauge through a transformation satisfying
Eq.\ (\ref{eq8:220}).

It is interesting to compare the conservative shift $\Delta\Omega_{\rm isco}$
with the frequency bandwidth of the dissipative transition across the ISCO,
calculated by Ori and Thorne in \cite{Ori:2000zn}. Denoting the latter (in the
Schwarzschild case) by $\Delta\Omega_{\rm diss}$, one finds
$\Delta\Omega_{\rm diss}/\Delta\Omega_{\rm isco}\sim 9\, (\mu/M)^{-3/5}$,
giving, for example, $\sim 35830$, $9000$ and $2261$ for mass ratios
$\mu/M=10^{-6}$, $10^{-5}$ and $10^{-4}$, respectively. This confirms our
expectation that the conservative shift in the ISCO is practically negligible
from the observational point of view. The main practical value of the results
in Eq.\ (\ref{eq8:390}) remains that they provide an accurate strong-field
benchmark to inform the development of approximate methods.

\section{Reflections and prospects}
\label{sec:conc}

We have attempted here to capture a snapshot of the activity surrounding
the development of reliable, efficient and accurate computational methods
for the gravitational SF in black hole spacetimes. The problem still attracts
considerable attention (more than half the items on our bibliographic list date $\geq 2005$),
with a multitude of different approaches being studied by different groups.
This multitude offers the opportunity for cross-validation of techniques and
results---a particularly important prospect given the intricate nature of the
numerics involved and the fact that SF calculations explore a new territory in
black hole physics, yet uncharted neither by PN theory nor by Numerical Relativity.

Indeed, the field has by now matured sufficiently that such cross-validation
exercises are becoming possible. As we described in Sec.\ \ref{sec:effects},
the last year has seen first quantitative comparisons between results from different
calculations carried out in different gauges and using different numerical methods.
We are now able to use SF codes to explore, for the first time, the conservative
post-geodesic dynamics of strong-field orbits around a Schwarzschild black hole
\cite{Detweiler:2005kq,Detweiler:2008ft,Sago:2008id,Barack:2009ey}.
We can compute physical gauge-invariant SF effects and test
them directly against results from PN theory in the weak-field regime
\cite{Detweiler:2005kq,Detweiler:2008ft,Sago:2008id,LeTiec}. Indeed, we can now
start to use strong-field SF results in order to calibrate PN methods and
assess their performance \cite{Barack:2009ey}. The exciting prospects
for synergistic interaction between SF and PN theories are beginning to
materialize, with much scope for important developments in the coming years.

At preset, the state of the art in SF calculations is a code that can
calculate the gravitational SF along any bound geodesic of the Schwarzschild
geometry (currently at substantial computational cost, which future
developments in the numerical technology may help reduce). This code,
as many others mentioned in our review, is an implementation of the
mode-sum regularization method (Sec.\ \ref{sec:modesum}), which has
proven a useful framework for calculations in Schwarzschild.

The Kerr problem, on the other hand, has hardly been tackled so far,
and it represents the next significant challenge. Although the standard
mode-sum method is in principle applicable to the gravitational SF in
Kerr spacetime, the details of its numerical implementation in this case
are yet to be worked out. It is possible that higher-dimensional techniques
(like the $m$-mode scheme discussed in Sec.\ \ref{sec:puncture}) could provide
an attractive alternative to standard mode-sum in the Kerr case.

In the short term, activity is likely to focus on the following tasks:
(i) continue to improve the computational efficiency of SF calculations
using advanced numerical techniques (mesh refinement, spectral methods, etc.);
(ii) tackle the Kerr problem;
(iii) use SF codes to study post-geodesic physical effects (such as the
finite-mass correction to the orbital precession rate), and in particular
assess the relative importance of conservative SF effects in the
EMRI problem;
(iv) explore what can be learned from a comparison between SF and PN
results.

Within the wider context of the LISA EMRI problem, the computation of the SF
on momentarily-geodesic particles is only one essential ingredient.
There is still much more to understand before a faithful
model of an astrophysical inspiral can be developed. Most crucially,
a reliable and workable method must be devised for calculating the long-term
evolution of the inspiral orbit. Work to address this problem has started
recently \cite{Pound:2007ti,Pound:2007th,Hinderer:2008dm,Gralla:2008fg}
but much further development is required.

\ack

Much of the material included in Sec.\ \ref{sec:modesum} and in the Appendix
is based on work with Amos Ori, and I am grateful to him for his mentorship
during the initial years of my work in the field, and for his continual
advice and input in the years thereafter. I have benefited greatly from
continuing collaborations with Carlos Lousto (Sec.\ \ref{sec:Lorenz}), Nori Sago
(Secs.\ \ref{sec:Lorenz},\ref{sec:puncture},\ref{sec:effects}), Lior Burko,
and Steve Detweiler.
Much of my understanding of the subject has formed through discussions with
participants of the annual Capra meetings, for which I am grateful.
I am particularly indebted to Eric Poisson for many useful discussions,
and for his encouragement over the years. Finally, I acknowledge the
generous support from STFC through grant number PP/E001025/1.

\appendix

\setcounter{section}{1}

\section*{Appendix: Derivation of the regularization parameters}
\label{App}

We describe here the derivation of the regularization parameters for the
gravitational SF in Kerr. The values of these parameters [stated in Sec.\
\ref{sec:modesum}, Eqs.\ (\ref{C}), (\ref{A}) and (\ref{B})] were first
published in Ref.\ \cite{Barack:2002mh}, but the detailed derivation has
not appeared in print so far. We reproduce it here using the original
method of Ref.\ \cite{Barack:2002mh}.

Let the arbitrary timelike geodesic $\Gamma$ be the worldline of a particle
with mass $\mu$ in a Kerr geometry with mass $M\gg\mu$ and arbitrary spin $aM$.
We wish to calculate the regularization parameters for the gravitational SF
acting on the particle at an arbitrary point $z$ along $\Gamma$ with
Boyer--Lindquist coordinates $(t_0,r_0,\theta_0,\varphi_0)$.
Without loss of generality we shall take $\varphi_0=0$.

We remind that the parameters $\pm A^{\alpha}$, $B^{\alpha}$ and $C^{\alpha}$
are the leading-order coefficients in the formal expansion of the $l$-modes
$F^{\alpha l}_{{\rm ret}\pm}(z)$ at large $l$ [recall Eq.\ (\ref{eq4:160})].
We also remind that the difference
$F^{\alpha l}_{{\rm ret}\pm}(z)-F^{\alpha l}_{{\rm S}\pm}(z)$ is expected to
die off at large $l$ faster than any power of $1/l$ [recall the discussion
surrounding Eq.\ (\ref{eq4:180})]. Therefore, the values of
$A^{\alpha}$, $B^{\alpha}$ and $C^{\alpha}$ can also---more conveniently---be
deduced from the large-$l$ asymptotics of the S field modes:
\begin{equation}\label{eqA:10}
F^{\alpha l}_{{\rm S}\pm}(z)=\pm A^{\alpha}L+B^{\alpha}+C^{\alpha}/L
+O(L^{-2}).
\end{equation}
Once $A^{\alpha}$, $B^{\alpha}$ and $C^{\alpha}$ are known, the parameter
$D^{\alpha}$ is given as the residual quantity [Eq.\ (\ref{eq4:210})]
\begin{equation} \label{eqA:20}
D^{\alpha}\equiv \sum_{l=0}^{\infty}
\left[F_{{\rm S}\pm}^{\alpha l}(z)\mp LA^{\alpha}-B^{\alpha}-C^{\alpha}\right].
\end{equation}

Our starting point will be the local expression for the S-field
$\bar h_{\alpha\beta}^{\rm S}$, given in Eq.\ (\ref{eq1:170}).
Referring back to Fig.\ \ref{fig1}, we denote the {\em Boyer--Lindquist}
coordinate difference between point $x$
(an arbitrary point in the immediate vicinity of $z$) and point $z$ by
$\delta x^{\alpha}\equiv x^{\alpha}-z^{\alpha}$. In Eq.\ (\ref{eq1:170})
the quantities $\epsilon^2(x;z)$ and $\hat u_{\alpha}(x;z)$ are smooth
functions of $\delta x$, and we may expand them in the form
\begin{equation}\label{eqA:30}
\epsilon^2=S_0+S_1+O(\delta x^4),
\quad\quad
\hat u_{\alpha}=u_{\alpha}+\delta u_{\alpha} +O(\delta x^2),
\end{equation}
where $S_0$ and $S_1$ are, respectively, quadratic and cubic in $\delta x$,
and $\delta u_{\alpha}$ is linear in $\delta x$. Explicitly, these
expansion terms read
\begin{eqnarray} \label{eqA:40}
S_0 &=& (g_{\alpha\beta}+u_{\alpha}u_{\beta})\delta x^{\alpha}\delta x^{\beta}
= P_{\alpha\beta}\delta x^{\alpha}\delta x^{\beta},  \\
\label{eq226}
S_1 &=& \Gamma^{\lambda}_{\alpha\beta}P_{\lambda\gamma}
\delta x^{\alpha}\delta x^{\beta}\delta x^{\gamma}
= P_{\alpha\beta\gamma}\delta x^{\alpha}\delta x^{\beta}\delta x^{\gamma},  \\
\label{eq227}
\delta u_{\alpha} &=& \Gamma^{\lambda}_{\alpha\beta}u_{\lambda}\delta x^{\beta},
\end{eqnarray}
where the background metric and connections are evaluated at $z$, and the coefficients
$P_{\alpha\beta}$ and $P_{\alpha\beta\gamma}$ are those defined in Eq.\ (\ref{eqA70})
(for a derivation of $S_1$ see, for example, Appendix A of \cite{Barack:2002mha}).

We now substitute the above expansions (\ref{eqA:30}) in Eq.\ (\ref{eq1:170}),
and subsequently construct the field $F^{\alpha}_{\rm S}(x)=\mu
\bar\nabla_x^{\alpha\beta\gamma}\bar h^{\rm S}_{\beta\gamma}(x)$, where, recall,
the operator $\bar\nabla_x^{\alpha\beta\gamma}$ is the one defined immediately
below Eq.\ (\ref{eq3:10}).
The resulting expression for $F^{\alpha}_{\rm S}(x)$ can be written down as
a sum of terms sorted according to how they scale with $\delta x$:
\begin{equation}\label{eqA:50}
\mu^{-2}F_{\rm S}^{\alpha}(x)
=\frac{P^{\alpha}_{(1)}(\delta x)}{\epsilon_0^3}+
\frac{P^{\alpha}_{(4)}(\delta x)}{\epsilon_0^5}+
\frac{P^{\alpha}_{(7)}(\delta x)}{\epsilon_0^7}+O(\delta x).
\end{equation}
Here $\epsilon_0\equiv S_0^{1/2}$, and $P^{\alpha}_{(n)}$ denotes
a certain multilinear function of the coordinate differences
$\delta x^{\alpha}$, of homogeneous order $n$ in $\delta x$.
Note that the first term on the right-hand side scales as $\delta x^{-2}$,
the second as $\delta x^{-1}$, and the third as $\delta x^{0}$.
The $O(\delta x)$ remainder disappears at the limit $x\to z$ and cannot
affect the value of the final SF; it is therefore safe to ignore it.
The explicit form of $P^{\alpha}_{(7)}$ will not be needed in our analysis.
The two other coefficients read
\begin{equation} \label{eqA:60}
P^{\alpha}_{(1)}=-2u_{\beta}u_{\gamma}\bar\nabla_x^{\alpha\beta\gamma}S_0
=-P^{\alpha}_{\ \beta}\delta x^\beta
\end{equation}
and
\begin{eqnarray} \label{eqA:70}
\fl \quad P^{\alpha}_{(4)} &=&
u_{\beta}u_{\gamma}\left(3S_1\bar\nabla_x^{\alpha\beta\gamma}S_0
    -2S_0\bar\nabla_x^{\alpha\beta\gamma}S_1\right)
-2(\delta u_{\beta}u_{\gamma}+u_{\beta}\delta u_{\gamma})
    S_0\bar\nabla_x^{\alpha\beta\gamma}S_0 \nonumber\\
\fl \quad  &=& \frac{1}{2}\left[
        P^{\alpha}_{\ \delta}(3P_{\rho\beta\gamma}+2P_{\rho\beta}P_{\gamma})
        -P^{\alpha\lambda}(2P_{\lambda\rho\beta}+P_{\rho\beta\lambda})P_{\gamma\delta}
    \right]\delta x^{\rho}\delta x^{\beta}\delta x^{\gamma}
    \delta x^{\delta}.
\end{eqnarray}
The second equality in each of (\ref{eqA:60}) and (\ref{eqA:70}) was evaluated
with the help of the following identities, which are easily confirmed:
\begin{equation} \label{eqA:80}
u_{\beta}u_{\gamma}\bar\nabla_x^{\alpha\beta\gamma}
=\frac{1}{4}P^{\alpha\delta}\nabla_{\delta},
\quad\quad\quad
P^{\alpha\beta}P_{\beta\gamma}=P^{\alpha}_{\ \gamma},
\end{equation}
\begin{equation} \label{eqA:90}
\nabla_{\delta}S_0=2P_{\delta\gamma}\delta x^{\gamma},
\quad\quad\quad
\nabla_{\delta} S_1=(2P_{\delta\alpha\beta}+P_{\alpha\beta\delta})
\delta x^{\alpha}\delta x^{\beta},
\end{equation}
\begin{equation} \label{eqA:100}
(\delta u_{\beta}u_{\gamma}+u_{\beta}\delta u_{\gamma})
\bar\nabla_x^{\alpha\beta\gamma}S_0
=-\delta u_{\gamma}u^{\gamma}P^{\alpha}_{\ \beta}\delta x^{\beta}
=-\frac{1}{2}P^{\alpha}_{\ \beta}P_{\gamma}\delta x^{\beta} \delta x^{\gamma} ,
\end{equation}
where $P_{\beta}$ is given in Eq.\ (\ref{eqA70}).

To obtain the regularization parameters, we need to consider the decomposition
of $F_{\rm S}^{\alpha}(x)$ in spherical harmonics (and then take $x\to z$).
Notwithstanding the spheroidicity of the Kerr background, the spherical
harmonics $Y_{lm}(\theta,\varphi)$ are defined as usual on surfaces of constant
$r$ and $t$, where $(t,r,\theta,\varphi)$ are Boyer--Lindquist coordinates.
To make this $l$-mode decomposition easier, we introduce new coordinates:
In general (i.e., for general $\theta_0$), the multipole decomposition will
involve all azimuthal numbers $|m|\leq l$ for each $l$. However, if we
choose a new set of angular coordinates, $(\theta,\varphi)\to(\theta',\varphi')$,
such that in the new coordinates $z$ is positioned {\em at the pole} (i.e., $\theta'_0=0$),
then the only contribution to each $l$ mode in the new system would come from
the axially-symmetric $m=0$ mode alone. Yet, due to the invariance of the Legendre
functions under rotations on the $2$-sphere, the overall $l$ mode contribution
(summed over $m$) would be the same in both systems. A transformation
$(\theta,\varphi)\to(\theta',\varphi')$ is given explicitly by
\begin{eqnarray}\label{eqA:110}
\cos\theta'  =\cos\varphi \sin\theta \sin\theta_0+ \cos\theta \cos\theta_0,
\nonumber\\
\tan(\varphi'-\varphi'_u) = \frac{\sin\varphi \sin\theta}
{\cos\varphi \sin\theta \cos\theta_0 - \cos\theta\sin\theta_0},
\end{eqnarray}
where $\varphi'_u$ indicates the azimuthal direction of the particle's velocity
in the new polar system (i.e., in the new system the velocity is momentarily
tangent to the longitude line $\varphi'={\rm const}=\varphi'_u$ on the $2$-sphere).
The angle $\varphi'_u$ depends on $u^{\theta}$ and $u^{\varphi}$, but the
explicit relation will not be needed here.

Since the new angular coordinates are not well behaved at $z$, we also
introduce Cartesian-like local coordinates based around $z$:
\begin{eqnarray}\label{eqA:120}
x\equiv \rho(\theta')\cos(\varphi'-\varphi'_u), \quad\quad
y\equiv \rho(\theta')\sin(\varphi'-\varphi'_u),
\end{eqnarray}
where we require $\rho=\theta'+O(\theta'^3)$ near $z$. Note that at $z$ we
have $x=y=0$ as well as $u^y=0$.
For later convenience, we make here the concrete choice
\begin{equation}\label{eqA:125}
\rho= 2\sin(\theta'/2),
\end{equation}
giving
\begin{equation}\label{eqA:127}
\cos\theta'=1-\frac{1}{2}(x^2+y^2).
\end{equation}

In the following we will need the
transformation $(\theta,\varphi)\leftrightarrow (x,y)$ only in the
immediate neighborhood of $z$. To sufficient order, we find
\begin{eqnarray}\label{eqA:130}
\delta\theta&=&x+(1/2)\cot\theta_0 y^2 +O(\delta x^3), \nonumber\\
\delta\varphi&=&(\sin\theta_0)^{-1}(y-\cot\theta_0 xy)+O(\delta x^3),
\end{eqnarray}
where, recall, $\delta\theta=\theta-\theta_0$ and $\delta\varphi=
\varphi-\varphi_0=\varphi$ are the Boyer--Lindquist coordinate differences,
and $O(\delta x^3)$ represents terms at least cubic in $x$ and $y$.
We denote
\begin{equation}\label{eqA:140}
X^{\alpha}=(X^0,X^1,X^2,X^3)\equiv(\delta t,\delta r,x,y/\sin\theta_0),
\end{equation}
noting $X^{\alpha}=0$ at the particle.
The direct transformation from $X^{\alpha}$ to the Boyer--Lindquist coordinate
differences $\delta x^{\alpha}$ can be written to sufficient order in the compact form
\begin{eqnarray}\label{eqA:150}
\delta x^{\alpha}=
X^{\alpha}+C_{\mu\nu}^{\alpha}X^{\mu}X^{\nu},
\end{eqnarray}
where the constant coefficients $C_{\mu\nu}^{\alpha}$
are those defined in Eq.\ (\ref{CC}).

Our next step is to express the S-field force $F_{\rm S}^{\alpha}(x)$ in terms
of the $X^{\alpha}$ coordinates (note, however, that we do {\em not} consider
here the $X^{\alpha}$ components of $F_{\rm S}^{\alpha}(x)$, but rather the
Boyer--Lindquist components; the $X^{\alpha}$ coordinates are merely introduced to
assist with the Legendre integration below). We substitute Eq.\ (\ref{eqA:150}) in
Eq.\ (\ref{eqA:50}) and then re-expand in the coordinate differences $X^{\alpha}$.
The result takes the form
\begin{equation}\label{eqA:160}
\mu^{-2}F_{\rm S}^{\alpha}(x)=F_{A}^{\alpha}+F_{B}^{\alpha}+F_{C}^{\alpha}
+O(\delta x),
\end{equation}
where
\begin{equation}\label{eqA:170}
F^{\alpha}_{A}=-\frac{P^{\alpha}_{\ \mu}X^{\mu}}{\epsilon_{0X}^{3}},
\end{equation}
\begin{equation}\label{eqA:180}
F^{\alpha}_{B}=
\frac{P^{\alpha}_{\ \rho\beta\gamma\delta}
X^{\rho}X^{\beta}X^{\gamma}X^{\delta}}{\epsilon_{0X}^{5}},
\end{equation}
\begin{equation}\label{eqA:190}
F^{\alpha}_{C}=
\frac{P^{\alpha}_{\ \mu\nu\lambda\sigma\beta\gamma\delta}
X^{\mu}X^{\nu}X^{\lambda}X^{\sigma}X^{\beta}X^{\gamma}X^{\delta}}
{\epsilon_{0X}^{7}}.
\end{equation}
Here
\begin{equation}\label{eqA:195}
\epsilon_{0X}^2\equiv P_{\alpha\beta}X^{\alpha}X^{\beta},
\end{equation}
$P^{\alpha}_{\ \mu\nu\lambda\sigma\beta\gamma\delta}$ are constant
coefficients whose explicit values will not be needed, and
\begin{eqnarray}\label{eqA:200}
\fl \quad\quad\quad\quad P^{\alpha}_{\ \rho\beta\gamma\delta} &=&
 \frac{1}{2}\left[
        P^{\alpha}_{\ \delta}(3P_{\rho\beta\gamma}+2P_{\rho\beta}P_{\gamma})
        -P^{\alpha\lambda}(2P_{\lambda\rho\beta}+P_{\rho\beta\lambda})P_{\gamma\delta}
    \right] \nonumber\\
\fl \quad\quad\quad\quad &&+ (3P^{\alpha}_{\ \rho}P_{\beta\lambda}
    -P^{\alpha}_{\ \lambda}P_{\rho\beta})
    C_{\gamma\delta}^{\lambda}.
\end{eqnarray}
[In the last expression, the first line is simply the coefficient of
$P^{\alpha}_{(4)}$ from Eq.\ (\ref{eqA:70}), and the second line arises from
expanding the term $\epsilon_0^{-3}P^{\alpha}_{(1)}$ in Eq.\ (\ref{eqA:50})
in powers of $X^{\alpha}$.] Note in Eq.\ (\ref{eqA:160}) that the terms
$F^{\alpha}_{A}$, $F^{\alpha}_{B}$ and $F^{\alpha}_{C}$ scale as $\sim X^{-2}$,
$X^{-1}$ and $X^{-0}$, respectively. The terms included in $O(\delta x)$
vanish at $x\to z$, and are therefore irrelevant for calculating the
regularization parameters---we can hence safely ignore them in what follows.

To obtain the regularization parameters, via Eqs.\ (\ref{eqA:10}) and (\ref{eqA:20}),
we now need to consider the spherical-harmonic $l$ modes $F_{\rm S\pm}^{\alpha l}(z)$.
These are constructed, by definition, through
\begin{equation}\label{eqA:220}
F^{\alpha l}_{\rm S\pm}=
\lim_{\delta r\to 0^{\pm}}
\frac{L}{2\pi}\int_{1}^{-1}d\cos\theta'\int_0^{2\pi}d\varphi' P_l(\cos\theta')
F^{\alpha}_{\rm S}(r,\theta',\varphi';z),
\end{equation}
where $P_l(\cos\theta')$ is the Legendre polynomial and, recall, $L=l+1/2$.
Here we have already taken the limits $t\to t_0$, $\theta\to\theta_0$ and
$\varphi\to\varphi_0$, thereby choosing to approach the particle along the
radial direction, recalling that we expect two different one-side values
(hence the label `$\pm$'). Let us write
\begin{equation}\label{eqA:230}
\mu^{-2}F^{\alpha l}_{\rm S\pm}=F^{\alpha l}_{A\pm}+F^{\alpha l}_{B}+F^{\alpha l}_{C},
\end{equation}
where the three terms on the right-hand side represent the respective
contributions to $F^{\alpha l}_{\rm S\pm}$ from the three terms
$F^{\alpha}_{A,B,C}$ in Eq.\ (\ref{eqA:160}).
As explained in Ref.\ \cite{Barack:2002mha}, the second and third terms
(scaling near the particle as $\sim\delta x^{-1}$
and $\sim\delta x^{0}$, respectively) are sufficiently regular to allow
interchanging the order of the limit and integration in Eq.\ (\ref{eqA:220}).
For the same reason, the radial limit is well defined and the two-side ambiguity
does not occur. As we discuss below, such ambiguity does appear when considering
$F^{\alpha l}_{A\pm}$---hence the label $\pm$.
In what follows we consider each of the three contributions to
$F^{\alpha l}_{\rm S\pm}$ in turn.

We start with $F^{\alpha l}_{C}$. This term is obtained by replacing
$F^{\alpha}_{\rm S}$ in Eq.\ (\ref{eqA:220}) with $F_C^{\alpha}$ from
Eq.\ (\ref{eqA:190}), and we are allowed to pull the limit
$\delta r\to 0^\pm$ over the integrals. The outcome has the form
\begin{equation}\label{eqA:235}
F^{\alpha l}_{C}=\frac{L}{2\pi}\int_{1}^{-1}d\cos\theta'\int_0^{2\pi}d\varphi'
P_l(\cos\theta')\hat\epsilon_{0X}^{-7}(x,y)\tilde P^{\alpha}_{(7)}(x,y),
\end{equation}
where $\tilde P^{\alpha}_{(7)}(x,y)$ is a certain polynomial of homogeneous
order 7 in the two ``angular'' coordinates $(x,y)=(X^2,X^3\sin\theta_0)$,
and
\begin{equation}\label{eqA:237}
\hat\epsilon_{0X}\equiv\epsilon_{0X}(\delta r\to 0,\delta t\to 0)
=\left[[g_{xx}(z)+u_x^2]x^2+g_{yy}(z)y^2\right]^{1/2}.
\end{equation}
In the last equality we used Eq.\ (\ref{eqA:195}) with $P_{\alpha\beta}=
g_{\alpha\beta}+u_{\alpha}u_{\beta}$, recalling $u_y=0$ along with
$g_{xy}(z)=0$.
We observe that $\hat\epsilon_{0X}$ is an {\em even} function of both $x$ and $y$,
and, recalling Eq.\ (\ref{eqA:127}), so is $\cos\theta'$. However, each
of the terms in $\tilde P^{\alpha}_{(7)}(x,y)$ (such as $\propto x^2y^5$,
$\propto x^4y^3$, etc.) is {\em odd} in either $x$ or $y$.
It then immediately follows from symmetry that the integral in
Eq.\ (\ref{eqA:235}) vanishes:
\begin{equation}\label{eqA:240}
F^{\alpha l}_{C}=0.
\end{equation}

Next consider $F_{\alpha l}^{B}$. First, taking the limits $\delta r\to 0$
and $\delta t\to 0$, we have from Eq.\ (\ref{eqA:180})
\begin{equation}\label{eqA:250}
F^{\alpha}_{B}(\delta r,\delta t\to 0)=\hat\epsilon_{0X}^{-5}P^{\alpha}_{\; abcd}
X^{a}X^{b}X^{c}X^{d},
\end{equation}
where roman indices run over the angular coordinates $(X^2,X^3)$ only.
Then, using $(X^2,X^3)=(x,y/\sin\theta_0)=\rho(\cos\varphi',
\sin\varphi'/\sin\theta_0)\equiv \rho w^a$,
we write $X^a=\rho w^a$ and $\hat\epsilon_{0X}^2=\rho^2 P_{ab}w^a w^b$.
This allows us to write the Legendre integral for $F^{\alpha l}_{B}$ in
a factorized form:
\begin{equation}\label{eqA:260}
F^{\alpha l}_{B}=
\left[\frac{L}{2\pi}\int_{1}^{-1} d\cos\theta'\, \frac{P_l(\cos\theta')}
{\rho(\theta')}\right]
\left[\int_{0}^{2\pi}d\varphi'\,
\frac{P^{\alpha}_{\; abcd}w^{a}w^{b}w^{c}w^{d}}
{(P_{ab}w^a w^b)^{5/2}}\right].
\end{equation}
Here the $\theta'$ integral is elementary: The entire term in the
first set of square brackets reads simply $(2\pi)^{-1}$ (for any $l$).
The term in the second set of square brackets is $P^{\alpha}_{\; abcd}I^{abcd}$,
where $I^{abcd}$ are the ($l$-independent) integrals given in Eq.\ (\ref{Iabcd}).
These integrals, recall, are not elementary, but they can be expressed
explicitly in terms of complete elliptic integrals, as in Eq.\ (\ref{eqIabcd}).
In conclusion, we find
\begin{equation}\label{eqA:270}
F^{\alpha l}_{B}=(2\pi)^{-1}P^{\alpha}_{\; abcd}I^{abcd}.
\end{equation}
Importantly, the term $F^{\alpha l}_{B}$ contributes to each $l$-mode
$F^{\alpha l}_{\rm S}$ a {\em constant amount}, independent of $l$.

We have one more contribution to $F^{\alpha l}_{\rm S}$ to consider:
that of $F^{\alpha l}_{\rm A\pm}$. Recalling Eq.\ (\ref{eqA:170}), we have
\begin{equation}\label{eqA:280}
F^{\alpha l}_{\rm A\pm}=\lim_{\delta r\to 0^{\pm}}
\frac{L}{2\pi}\int dx\,dy\, P_l(\cos\theta')
\epsilon_{0X}^{-3}P^{\alpha}_{\ \mu}X^{\mu},
\end{equation}
where the integral is taken over the 2-sphere, we have used the
Jacobian $\partial(\cos\theta',\varphi')/\partial(x,y)=-1$,
and the integrand is understood to be already evaluated at $t=t_0$.
The Legendre polynomial can be written in the form
$P_l(\cos\theta')=1+\rho^2 H(\rho)$, where $H(\rho)$ admits a
regular Taylor expansion in $\rho^2$. Consider first the contribution
of the term $\propto \rho^2 H(\rho)$ to $F^{\alpha l}_{\rm A\pm}$: The
corresponding integrand in Eq.\ (\ref{eqA:280}) has the form
$\epsilon_{0X}^{-3}H(\rho)\tilde P^{\alpha}_{(3)}(\delta r,x,y)$
(recalling $\rho^2=x^2+y^2$), where
$\tilde P^{\alpha}_{(3)}$ is a polynomial of homogeneous order 3 in
$\delta r,x,y$. This contribution to the integrand is therefore bounded, and we are
allowed to swap the limit and integral just as we did with $F^{\alpha l}_{\rm C}$.
The resulting integral then vanishes by virtue of the same
symmetry argument applied in the case of $F^{\alpha l}_{\rm C}$:
Both functions $\hat\epsilon_{0X}^{-3}$ and $H(\rho)$ are even in each of $x$ and
$y$, while all the possible terms in $P^{\alpha}_{(3)}(0,x,y)$ are odd in either
$x$ or $y$.

We are thus left with the contribution
\begin{equation}\label{eqA:290}
F^{\alpha l}_{\rm A\pm}=\lim_{\delta r\to 0^{\pm}}
\frac{L}{2\pi}\int dx\,dy\, \epsilon_{0X}^{-3}P^{\alpha}_{\ \mu}X^{\mu}.
\end{equation}
To evaluate this integral, we divide the integration domain (the 2-sphere)
into two regions: Let $D_{\rm in}$ denote the square $-h<x,y<h$ for some particular
$0<h<1$ (say, $h=1/10$); and let $D_{\rm out}$ denote the remaining integration
area. Consider first the contribution to $F^{\alpha l}_{\rm A\pm}$ from
$D_{\rm out}$: Since in this domain the integrand is regular (the only
singularity is at $x=y=0$, which is in $D_{\rm in}$), we are allowed to
interchange the limit and integration. As a result, the integrand takes
the form $\hat\epsilon_{0X}^{-3}P^{\alpha}_{\ b}X^{b}$, and the integral
over $D_{\rm out}$ vanishes by virtue of the odd symmetry. The remaining
piece of the integral, over $D_{\rm in}$, is
\begin{equation}\label{eqA:295}
F^{\alpha l}_{\rm A\pm}=\lim_{\delta r\to 0^{\pm}}
\frac{L}{2\pi}\int_{-h}^{h}\int_{-h}^{h} dx\,dy\,
\epsilon_{0X}^{-3}P^{\alpha}_{\ \mu}X^{\mu}.
\end{equation}
This integral takes a simpler form if we express it in terms of the
rescaled coordinates $\tilde x\equiv x/\delta r$, $\tilde y\equiv y/\delta r$,
and $\tilde X^{\mu}\equiv X^{\mu}/\delta r=(0,1,\tilde x,\tilde y/\sin\theta_0)$
(where we have already taken $\delta t\to 0$). For given $\delta r\ne 0$ we have
$\epsilon_{0X}=(P_{\alpha\beta}X^{\alpha}X^{\beta})^{1/2}=
\pm\delta r(P_{\alpha\beta}\tilde X^{\alpha}\tilde X^{\beta})^{1/2}$,
where the sign corresponds to the sign of $\delta r$. Hence we obtain
\begin{eqnarray}\label{eqA:297}
F^{\alpha l}_{\rm A\pm}&=&\pm\lim_{\delta r\to 0^{\pm}}
\frac{L}{2\pi}\int_{-h/\delta r}^{h/\delta r}\int_{-h/\delta r}^{h/\delta r}
d\tilde x\,d\tilde y\, \frac{P^{\alpha}_{\ \mu}\tilde X^{\mu}}
{(P_{\alpha\beta}\tilde X^{\alpha}\tilde X^{\beta})^{3/2}}=
\nonumber\\
&&\pm\frac{L}{2\pi}\int_{-\infty}^{\infty}\int_{-\infty}^{\infty}
d\tilde x\,d\tilde y\, \frac{P^{\alpha}_{\ \mu}\tilde X^{\mu}}
{(P_{\alpha\beta}\tilde X^{\alpha}\tilde X^{\beta})^{3/2}},
\end{eqnarray}
where the last equality holds because the integrand, expressed in terms of
the tilde variables, no longer depends on $\delta r$. This integral is now
elementary, and evaluating it gives
\begin{equation}\label{eqA:300}
F^{\alpha l}_{\rm A\pm}=\pm L A^{\alpha},
\end{equation}
where the signs correspond to $\delta r\to 0^{\pm}$, and where the various
Boyer-Lindquist components of $A^{\alpha}$ are given in Eq.\ (\ref{A}).
[Section V.D of Ref.\ \cite{Barack:2002mha} explains in detail how the
integral in Eq.\ (\ref{eqA:297}) is evaluated in the special case of
Schwarzschild, and the method is directly applicable to Kerr.]
Notice that $F^{\alpha l}_{\rm A\pm}$ is found to depend on $l$ only through
the prefactor $L$.

In summary, collecting the results (\ref{eqA:160}), (\ref{eqA:240}),
(\ref{eqA:270}) and (\ref{eqA:300}), we find that the $l$ modes
$F^{\alpha l}_{\rm S}$ are given {\em precisely} by
\begin{eqnarray} \label{eqA:310}
F^{\alpha l}_{\rm S\pm}=\pm LA^{\alpha}+B^{\alpha},
\end{eqnarray}
where $A^{\alpha}$ and $B^{\alpha}\equiv (2\pi)^{-1}P^{\alpha}_{\; abcd}I^{abcd}$
are $l$-independent. Comparing with Eq.\ (\ref{eqA:10}), we identify
$A^{\alpha}$ and $B^{\alpha}$ as the first two of the regularization parameters.
This comparison also implies $C^{\alpha}=0$. Furthermore, we find that each
of the individual terms in the sum over $l$ in Eq.\ (\ref{eqA:20}) vanishes,
giving also $D^{\alpha}=0$. This completes the derivation of all regularization
parameters.

We comment on a potentially confusing aspect of the above analysis:
We have discarded in Eq.\ (\ref{eqA:160}) the $O(\delta x)$ terms of
$F_{\rm S\pm}^{\alpha}(x)$, which vanish at $x\to z$.
Clearly, the multipole expansion of these neglected terms could contribute
to $F^{\alpha l}_{\rm S\pm}$  [e.g., they may well add a term $\propto L^{-2}$
in Eq.\ (\ref{eqA:310})]. Such terms, however, must add up to zero upon
summation over $l$ (when evaluated at the particle). They hence affect
neither the value of $D^{\alpha}$ in Eq.\ (\ref{eqA:20}), nor the value of
the final SF in Eq.\ (\ref{eq4:200}).

\section*{References}

%
%

\begin{thebibliography}{99.}%


\bibitem{Capra12}
Capra 12 website:
http://www.astro.indiana.edu/~jthorn/capra12/

\bibitem{LISA}
LISA mission website: http://lisa.nasa.gov/

\bibitem{AmaroSeoane:2007aw}
  P.~Amaro-Seoane, J.~R.~Gair, M.~Freitag, M.~Coleman Miller, I.~Mandel, C.~J.~Cutler and S.~Babak,
  Class.\ Quant.\ Grav.\  {\bf 24}, R113 (2007)
  [arXiv:astro-ph/0703495].

\bibitem{Anderson:2003qa}
  P.~R.~Anderson and B.~L.~Hu,
  Phys.\ Rev.\  D {\bf 69}, 064039 (2004)
  [Erratum-ibid.\  D {\bf 75}, 129901 (2007)]
  [arXiv:gr-qc/0308034].

\bibitem{Anderson:2005ds}
  P.~R.~Anderson, A.~Eftekharzadeh and B.~L.~Hu,
  Phys.\ Rev.\  D {\bf 73}, 064023 (2006)
  [arXiv:gr-qc/0507067].

\bibitem{Anderson:2005gb}
  W.~G.~Anderson and A.~G.~Wiseman,
  Class.\ Quant.\ Grav.\  {\bf 22}, S783 (2005)
  [arXiv:gr-qc/0506136].

\bibitem{Anderson:2004eg}
  W.~G.~Anderson, E.~E.~Flanagan and A.~C.~Ottewill,
  Phys.\ Rev.\  D {\bf 71}, 024036 (2005)
  [arXiv:gr-qc/0412009].

\bibitem{Babak:2009ua}
  S.~Babak, J.~R.~Gair and E.~K.~Porter,
  Class.\ Quant.\ Grav.\  {\bf 26}, 135004 (2009)
  [arXiv:0902.4133 [gr-qc]].

\bibitem{Baker:2005vv}
  J.~G.~Baker {\it et al.},
  Phys.\ Rev.\ Lett.\ {\bf 96}, 111102 (2006).

\bibitem{Barack:2000eh}
  L.~Barack,
  Phys.\ Rev.\  D {\bf 62}, 084027 (2000)
  [arXiv:gr-qc/0005042].

\bibitem{Barack:2001bw}
  L.~Barack,
  Phys.\ Rev.\  D {\bf 64}, 084021 (2001)
  [arXiv:gr-qc/0105040].

\bibitem{OrleansBook}
  L.~Barack
  in {\it Mass and motion in general relativity}, ed.\ L.~Blanchet, A.~Spallicci and
  B.~Whiting, Springer (2009).

\bibitem{Barack:2000zq}
  L.~Barack and L.~M.~Burko,
  Phys.\ Rev.\  D {\bf 62}, 084040 (2000)
  [arXiv:gr-qc/0007033].

\bibitem{Barack:2003fp}
  L.~Barack and C.~Cutler,
  Phys.\ Rev.\  D {\bf 69}, 082005 (2004)
  [arXiv:gr-qc/0310125].

\bibitem{Barack:2004wc}
  L.~Barack and C.~Cutler,
  Phys.\ Rev.\  D {\bf 70}, 122002 (2004)
  [arXiv:gr-qc/0409010].

\bibitem{Barack:2006pq}
  L.~Barack and C.~Cutler,
  Phys.\ Rev.\  D {\bf 75}, 042003 (2007)
  [arXiv:gr-qc/0612029].

\bibitem{Barack:2007jh}
  L.~Barack and D.~A.~Golbourn,
  Phys.\ Rev.\  D {\bf 76}, 044020 (2007)
  [arXiv:0705.3620 [gr-qc]].

\bibitem{Barack:2002ku}
  L.~Barack and C.~O.~Lousto,
  Phys.\ Rev.\  D {\bf 66}, 061502 (2002)
  [arXiv:gr-qc/0205043].

\bibitem{Barack:2005nr}
  L.~Barack and C.~O.~Lousto,
  Phys.\ Rev.\  D {\bf 72}, 104026 (2005)
  [arXiv:gr-qc/0510019].

\bibitem{Barack:1999wf}
  L.~Barack and A.~Ori,
  Phys.\ Rev.\  D {\bf 61}, 061502 (2000)
  [arXiv:gr-qc/9912010].

\bibitem{Barack:2001ph}
  L.~Barack and A.~Ori,
  Phys.\ Rev.\  D {\bf 64}, 124003 (2001)
  [arXiv:gr-qc/0107056].

\bibitem{Barack:2002mha}
  L.~Barack and A.~Ori,
  Phys.\ Rev.\  D {\bf 66}, 084022 (2002)
  [arXiv:gr-qc/0204093].

\bibitem{Barack:2002bt}
  L.~Barack and A.~Ori,
  Phys.\ Rev.\  D {\bf 67}, 024029 (2003)
  [arXiv:gr-qc/0209072].

\bibitem{Barack:2002mh}
  L.~Barack and A.~Ori,
  Phys.\ Rev.\ Lett.\  {\bf 90}, 111101 (2003)
  [arXiv:gr-qc/0212103v2].

\bibitem{Barack:2007tm}
  L.~Barack and N.~Sago,
  Phys.\ Rev.\  D {\bf 75}, 064021 (2007)
  [arXiv:gr-qc/0701069].

\bibitem{Barack:2009ey}
  L.~Barack and N.~Sago,
  Phys.\ Rev.\ Lett.\  {\bf 102}, 191101 (2009)
  [arXiv:0902.0573 [gr-qc]].

\bibitem{Barack:2001gx}
  L.~Barack, Y.~Mino, H.~Nakano, A.~Ori and M.~Sasaki,
  Phys.\ Rev.\ Lett.\  {\bf 88}, 091101 (2002)
  [arXiv:gr-qc/0111001].

\bibitem{Barack:2007we}
  L.~Barack, D.~A.~Golbourn and N.~Sago,
  Phys.\ Rev.\  D {\bf 76}, 124036 (2007)
  [arXiv:0709.4588 [gr-qc]].

\bibitem{Barack:2008ms}
  L.~Barack, A.~Ori and N.~Sago,
  Phys.\ Rev.\  D {\bf 78}, 084021 (2008)
  [arXiv:0808.2315 [gr-qc]].

\bibitem{BGSprep} L.~Barack, D. A. Golbourn and N.~Sago, in preparation.

\bibitem{Barausse:2007dy}
  E.~Barausse and L.~Rezzolla,
  Phys.\ Rev.\  D {\bf 77}, 104027 (2008)
  [arXiv:0711.4558 [gr-qc]].

\bibitem{Barton:2008eb}
  J.~L.~Barton, D.~J.~Lazar, D.~J.~Kennefick, G.~Khanna and L.~M.~Burko,
  Phys.\ Rev.\  D {\bf 78}, 064042 (2008)
  [arXiv:0804.1075 [gr-qc]].

\bibitem{Berndtson:2009hp}
  M.~V.~Berndtson,
  arXiv:0904.0033 [gr-qc].

\bibitem{Bishop:2003bs}
  N.~T.~Bishop, R.~Gomez, S.~Husa, L.~Lehner and J.~Winicour,
  Phys.\ Rev.\  D {\bf 68}, 084015 (2003)
  [arXiv:gr-qc/0301060].

\bibitem{Blanchet:2001id}
  L.~Blanchet,
  Phys.\ Rev.\  D {\bf 65}, 124009 (2002)
  [arXiv:gr-qc/0112056].

\bibitem{Blanchet:2002av}
  L.~Blanchet,
  Living Rev.\ Rel.\  {\bf 9}, 4 (2006)
  [arXiv:gr-qc/0202016].

\bibitem{Blanchet:2002mb}
  L.~Blanchet and B.~R.~Iyer,
  Class.\ Quant.\ Grav.\  {\bf 20} (2003) 755
  [arXiv:gr-qc/0209089].

\bibitem{Buonanno:2000ef}
  A.~Buonanno and T.~Damour,
  Phys.\ Rev.\  D {\bf 62}, 064015 (2000)
  [arXiv:gr-qc/0001013].

\bibitem{Burko:1999zy}
  L.~M.~Burko,
  Class.\ Quant.\ Grav.\  {\bf 17}, 227 (2000)
  [arXiv:gr-qc/9911042].

\bibitem{Burko:2000xx}
  L.~M.~Burko,
  Phys.\ Rev.\ Lett.\  {\bf 84}, 4529 (2000)
  [arXiv:gr-qc/0003074].

\bibitem{Burko:2006ua}
  L.~M.~Burko and G.~Khanna,
  Europhys.\ Lett.\  {\bf 78}, 60005 (2007)
  [arXiv:gr-qc/0609002].

\bibitem{Burko:2001kr}
  L.~M.~Burko and Y.~T.~Liu,
  Phys.\ Rev.\  D {\bf 64}, 024006 (2001)
  [arXiv:gr-qc/0103008].

\bibitem{Burko:2000yx}
  L.~M.~Burko, Y.~T.~Liu and Y.~Soen,
  Phys.\ Rev.\  D {\bf 63}, 024015 (2001)
  [arXiv:gr-qc/0008065].

\bibitem{Burko:2002ge}
  L.~M.~Burko, A.~I.~Harte and E.~Poisson,
  Phys.\ Rev.\  D {\bf 65}, 124006 (2002)
  [arXiv:gr-qc/0201020].

\bibitem{Campanelli:2005dd}
  M.~Campanelli {\it et al.},
  Phys.\ Rev.\ Lett.\ {\bf 96}, 111101 (2006).

\bibitem{Canizares:2008dp}
  P.~Canizares and C.~F.~Sopuerta,
  J.\ Phys.\ Conf.\ Ser.\  {\bf 154}, 012053 (2009)
  [arXiv:0811.0294 [gr-qc]].

\bibitem{Canizares:2009ay}
  P.~Canizares and C.~F.~Sopuerta,
  Phys.\ Rev.\  D {\bf 79}, 084020 (2009)
  arXiv:0903.0505 [gr-qc].


\bibitem{Casals:2009zh}
  M.~Casals, S.~R.~Dolan, A.~C.~Ottewill and B.~Wardell,
  Phys.\ Rev.\  D {\bf 79}, 124043 (2009)
  [arXiv:0903.0395 [gr-qc]].

\bibitem{Chrz}
P. L. Chrzanowski, Phys.\ Rev.\ D {\bf 11}, 2042 (1975).

\bibitem{Collins:2004ex}
  N.~A.~Collins and S.~A.~Hughes,
  Phys.\ Rev.\  D {\bf 69}, 124022 (2004)
  [arXiv:gr-qc/0402063].

\bibitem{Damour:2000we}
  T.~Damour, P.~Jaranowski and G.~Schaefer,
  Phys.\ Rev.\  D {\bf 62}, 084011 (2000)
  [arXiv:gr-qc/0005034].

\bibitem{Detweiler:2000gt}
  S.~Detweiler,
  Phys.\ Rev.\ Lett.\  {\bf 86}, 1931 (2001)
  [arXiv:gr-qc/0011039].

\bibitem{Detweiler:2005kq}
  S.~Detweiler,
  Class.\ Quant.\ Grav.\  {\bf 22}, S681 (2005)
  [arXiv:gr-qc/0501004].

\bibitem{Detweiler:2008ft}
  S.~Detweiler,
  Phys.\ Rev.\  D {\bf 77}, 124026 (2008)
  [arXiv:0804.3529 [gr-qc]].

\bibitem{Detweiler:2003ci}
  S.~Detweiler and E.~Poisson,
  Phys.\ Rev.\  D {\bf 69}, 084019 (2004)
  [arXiv:gr-qc/0312010].

\bibitem{Detweiler:2002mi}
  S.~Detweiler and B.~F.~Whiting,
  Phys.\ Rev.\  D {\bf 67}, 024025 (2003)
  [arXiv:gr-qc/0202086].

\bibitem{Detweiler:2002gi}
  S.~Detweiler, E.~Messaritaki and B.~F.~Whiting,
  Phys.\ Rev.\  D {\bf 67}, 104016 (2003)
  [arXiv:gr-qc/0205079].

\bibitem{DeWitt:1960fc}
  B.~S.~DeWitt and R.~W.~Brehme,
  Annals Phys.\  {\bf 9}, 220 (1960).

\bibitem{DeWitt*2}
  B.~S.~DeWitt and C.~M.~DeWitt, Physics (Long Island City, NY) 1, 3 (1964).

\bibitem{DiazRivera:2004ik}
  L.~M.~Diaz-Rivera, E.~Messaritaki, B.~F.~Whiting and S.~Detweiler,
  Phys.\ Rev.\  D {\bf 70}, 124018 (2004)
  [arXiv:gr-qc/0410011].

\bibitem{Dirac}
  P.~A.~M.~Dirac, Proc.\ R. Soc.\ London, Ser.\ A{\bf 167},148 (1938).

\bibitem{Drasco:2006ws}
  S.~Drasco,
  Class.\ Quant.\ Grav.\  {\bf 23}, S769 (2006)
  [arXiv:gr-qc/0604115].

\bibitem{Drasco:2005is}
  S.~Drasco, E.~E.~Flanagan and S.~A.~Hughes,
  Class.\ Quant.\ Grav.\  {\bf 22}, S801 (2005)
  [arXiv:gr-qc/0505075].

\bibitem{Drasco:2005kz}
  S.~Drasco and S.~A.~Hughes,
  Phys.\ Rev.\  D {\bf 73}, 024027 (2006)
  [arXiv:gr-qc/0509101].

\bibitem{Field:2009kk}
  S.~E.~Field, J.~S.~Hesthaven and S.~R.~Lau,
  arXiv:0902.1287 [gr-qc].

\bibitem{FHO}
 E.~E.~Flanagan, T.~Hinderer and A.~Ori, private communication.

\bibitem{Friedman}
   J.~Friedman and A.~Shah, presentation at the
  12th Capra meeting on radiation reaction; posted on the meeting's website
  \cite{Capra12}.

\bibitem{Gair:2008bx}
  J.~R.~Gair,
  Class.\ Quant.\ Grav.\  {\bf 26}, 094034 (2009)
  [arXiv:0811.0188 [gr-qc]].

\bibitem{Gair:2005ih}
  J.~R.~Gair and K.~Glampedakis,
  Phys.\ Rev.\  D {\bf 73}, 064037 (2006)
  [arXiv:gr-qc/0510129].

\bibitem{Gair:2004iv}
  J.~R.~Gair, L.~Barack, T.~Creighton, C.~Cutler, S.~L.~Larson, E.~S.~Phinney and M.~Vallisneri,
  Class.\ Quant.\ Grav.\  {\bf 21}, S1595 (2004)
  [arXiv:gr-qc/0405137].

\bibitem{Gair:2007kr}
  J.~R.~Gair, C.~Li and I.~Mandel,
  Phys.\ Rev.\  D {\bf 77}, 024035 (2008)
  [arXiv:0708.0628 [gr-qc]].

\bibitem{Gair:2008ec}
  J.~R.~Gair, I.~Mandel and L.~Wen,
  Class.\ Quant.\ Grav.\  {\bf 25}, 184031 (2008)
  [arXiv:0804.1084 [gr-qc]].

\bibitem{Ganz:2007rf}
  K.~Ganz, W.~Hikida, H.~Nakano, N.~Sago and T.~Tanaka,
  arXiv:gr-qc/0702054.

\bibitem{Geroch}
   R.~Geroch and J. Traschen,
   Phys.\ Rev.\  D {\bf 36}, 1017 (1987).

\bibitem{Glampedakis:2002ya}
  K.~Glampedakis and D.~Kennefick,
  Phys.\ Rev.\  D {\bf 66}, 044002 (2002)
  [arXiv:gr-qc/0203086].

\bibitem{Gonzalez:2008bi}
  J.~A.~Gonzalez, U.~Sperhake and B.~Brugmann,
  Phys.\ Rev.\  D {\bf 79}, 124006 (2009)
  [arXiv:0811.3952 [gr-qc]].

\bibitem{Gralla:2008fg}
  S.~E.~Gralla and R.~M.~Wald,
  Class.\ Quant.\ Grav.\  {\bf 25}, 205009 (2008)
  [arXiv:0806.3293 [gr-qc]].

\bibitem{Gralla:2009uf}
  S.~E.~Gralla and R.~M.~Wald,
  arXiv:0907.0414 [gr-qc].


\bibitem{Gundlach}
C. Gundlach, G. Calabrese, I. Hinder, and J. M. Martin-Garc{\'{\i}}a,
Class.\ Quant.\ Grav.\ {\bf 22}, 3767 (2005).

\bibitem{Haas:2007kz}
  R.~Haas,
  Phys.\ Rev.\  D {\bf 75}, 124011 (2007)
  [arXiv:0704.0797 [gr-qc]].

\bibitem{Haas:Capra}
   R.~Haas, presentation at the 11th Capra meeting on radiation reaction,
   Orleans, June 2008; posted on http://web.cnrs-orleans.fr/osuc/conf/

\bibitem{Haas:2004kw}
  R.~Haas and E.~Poisson,
  Class.\ Quant.\ Grav.\  {\bf 22}, S739 (2005)
  [arXiv:gr-qc/0411108].

\bibitem{Haas:2006ne}
  R.~Haas and E.~Poisson,
  Phys.\ Rev.\  D {\bf 74}, 044009 (2006)
  [arXiv:gr-qc/0605077].

\bibitem{HH}
  S. W. Hawking and J. B. Hartle, Commun. Math. Phys.
  27, 283 (1972).

\bibitem{Hikida:2003pi}
  W.~Hikida, S.~Jhingan, H.~Nakano, N.~Sago, M.~Sasaki and T.~Tanaka,
  Prog.\ Theor.\ Phys.\  {\bf 111}, 821 (2004)
  [arXiv:gr-qc/0308068].

\bibitem{Hikida:2004jw}
  W.~Hikida, S.~Jhingan, H.~Nakano, N.~Sago, M.~Sasaki and T.~Tanaka,
  Prog.\ Theor.\ Phys.\  {\bf 113}, 283 (2005)
  [arXiv:gr-qc/0410115].

\bibitem{Hinderer:2008dm}
  T.~Hinderer and E.~E.~Flanagan,
  Phys.\ Rev.\  D {\bf 78}, 064028 (2008)
  [arXiv:0805.3337 [gr-qc]].

\bibitem{Hobbs:1968}
  J.~M.~Hobbs,
  Annals Phys.\  {\bf 47}, 141 (1968).

\bibitem{Hopman:2006pv}
  C.~Hopman,
  AIP Conf.\ Proc.\  {\bf 873}, 241 (2006)
  [arXiv:astro-ph/0608460].

\bibitem{Huerta:2008gb}
  E.~A.~Huerta and J.~R.~Gair,
  arXiv:0812.4208 [gr-qc].

\bibitem{Hughes:1999bq}
  S.~A.~Hughes,
  Phys.\ Rev.\  D {\bf 61}, 084004 (2000)
  [Erratum-ibid.\  D {\bf 63}, 049902 (2001\ ERRAT,D65,069902.2002\ ERRAT,D67,089901.2003)]
  [arXiv:gr-qc/9910091].

\bibitem{Hughes:2005qb}
  S.~A.~Hughes, S.~Drasco, E.~E.~Flanagan and J.~Franklin,
  Phys.\ Rev.\ Lett.\  {\bf 94}, 221101 (2005)
  [arXiv:gr-qc/0504015].

\bibitem{Isaacson} R. A. Isaacson, Phys. Rev. 166, 1272 (1968).

\bibitem{James}
  G. James, {\it Advanced Modern Engineering Mathematics}
  (Pearson, Harlow, 2004), 3rd ed., Sec.\ 4.2.8.

\bibitem{:2007mn}
  P.~Jaranowski {\it et al.},
  Class.\ Quant.\ Grav.\  {\bf 25}, 114020 (2008)
  [arXiv:0710.5658 [gr-qc]].

\bibitem{Keidl:2006wk}
  T.~S.~Keidl, J.~L.~Friedman and A.~G.~Wiseman,
  Phys.\ Rev.\  D {\bf 75}, 124009 (2007)
  [arXiv:gr-qc/0611072].

\bibitem{Khanna:2003qv}
  G.~Khanna,
  Phys.\ Rev.\  D {\bf 69}, 024016 (2004)
  [arXiv:gr-qc/0309107].

\bibitem{Krivan:1996da}
  W.~Krivan, P.~Laguna and P.~Papadopoulos,
  Phys.\ Rev.\  D {\bf 54}, 4728 (1996)
  [arXiv:gr-qc/9606003].

\bibitem{Krivan:1997hc}
  W.~Krivan, P.~Laguna, P.~Papadopoulos and N.~Andersson,
  Phys.\ Rev.\  D {\bf 56}, 3395 (1997)
  [arXiv:gr-qc/9702048].

\bibitem{LeTiec}
  A.~ Le Tiec, presentation at the
  12th Capra meeting on radiation reaction; posted on the meeting's website
  \cite{Capra12}.

\bibitem{LopezAleman:2003ik}
  R.~Lopez-Aleman, G.~Khanna and J.~Pullin,
  Class.\ Quant.\ Grav.\  {\bf 20}, 3259 (2003)
  [arXiv:gr-qc/0303054].

\bibitem{Lousto:1999za}
  C.~O.~Lousto,
  Phys.\ Rev.\ Lett.\  {\bf 84}, 5251 (2000)
  [arXiv:gr-qc/9912017].

\bibitem{Lousto:2005ip}
  C.~O.~Lousto,
  Class.\ Quant.\ Grav.\  {\bf 22}, S543 (2005)
  [arXiv:gr-qc/0503001].

\bibitem{CQGspecial}
  {\it Gravitational radiation from binary black holes:
  advances in the perturbative approach}, Class.\ Quant.\ Grav. {\bf 22} (2005),
  ed.\ C. O Lousto.

\bibitem{Lousto:2008mb}
  C.~O.~Lousto and H.~Nakano,
  Class.\ Quant.\ Grav.\  {\bf 25}, 145018 (2008)
  [arXiv:0802.4277 [gr-qc]].

\bibitem{Lousto:1997wf}
  C.~O.~Lousto and R.~H.~Price,
  Phys.\ Rev.\  D {\bf 56}, 6439 (1997)
  [arXiv:gr-qc/9705071].

\bibitem{Martel:2003jj}
  K.~Martel,
  Phys.\ Rev.\  D {\bf 69}, 044025 (2004)
  [arXiv:gr-qc/0311017].

\bibitem{Martel:2001yf}
  K.~Martel and E.~Poisson,
  Phys.\ Rev.\  D {\bf 66}, 084001 (2002)
  [arXiv:gr-qc/0107104].

\bibitem{Miller:2009wv}
  M.~C.~Miller {\it et al.},
  arXiv:0903.0285 [astro-ph.GA].

\bibitem{Mino:2003yg}
  Y.~Mino,
  Phys.\ Rev.\  D {\bf 67}, 084027 (2003)
  [arXiv:gr-qc/0302075].

\bibitem{Mino:2005qj}
  Y.~Mino,
  Class.\ Quant.\ Grav.\  {\bf 22}, S717 (2005).

\bibitem{Mino:2006em}
  Y.~Mino,
  Prog.\ Theor.\ Phys.\  {\bf 115}, 43 (2006)
  [arXiv:gr-qc/0601019].

\bibitem{Mino:2007ft}
  Y.~Mino,
  Phys.\ Rev.\  D {\bf 77}, 044008 (2008)
  [arXiv:0711.3007 [gr-qc]].

\bibitem{MN1998}
Y. Mino and H. Nakano, Prog. Theore. Phys. 100, 507, (1998);

\bibitem{Mino:1996nk}
  Y.~Mino, M.~Sasaki and T.~Tanaka,
  Phys.\ Rev.\  D {\bf 55}, 3457 (1997)
  [arXiv:gr-qc/9606018].

\bibitem{Mino:2001mq}
  Y.~Mino, H.~Nakano and M.~Sasaki,
  Prog.\ Theor.\ Phys.\  {\bf 108}, 1039 (2003)
  [arXiv:gr-qc/0111074].

\bibitem{MTW} C. W. Misner, K. S. Thorne, and J. A. Wheeler,
  {\it Gravitation} (Freeman, San Francisco, 1973).

\bibitem{Moncrief} V. Moncrief, Ann.\ Phys.\ {\bf 88}, 323 (1974).

\bibitem{Nakano:2000ne}
  H.~Nakano and M.~Sasaki,
  Prog.\ Theor.\ Phys.\  {\bf 105}, 197 (2001)
  [arXiv:gr-qc/0010036].

\bibitem{Nakano:2001kw}
  H.~Nakano, Y.~Mino and M.~Sasaki,
  Prog.\ Theor.\ Phys.\  {\bf 106}, 339 (2001)
  [arXiv:gr-qc/0104012].

\bibitem{Nakano:2003he}
  H.~Nakano, N.~Sago and M.~Sasaki,
  Phys.\ Rev.\  D {\bf 68}, 124003 (2003)
  [arXiv:gr-qc/0308027].

\bibitem{Ori:1997be}
  A.~Ori,
  Phys.\ Rev.\  D {\bf 55}, 3444 (1997).

\bibitem{OriReconst}
A.\ Ori, Phys.\ Rev.\ D {\bf67}, 124010 (2003).

\bibitem{Ori:2000zn}
  A.~Ori and K.~S.~Thorne,
  Phys.\ Rev.\  D {\bf 62}, 124022 (2000).

\bibitem{O'Shaughnessy:2002ez}
  R.~O'Shaughnessy,
  Phys.\ Rev.\  D {\bf 67}, 044004 (2003).

\bibitem{Ottewill:2007mz}
  A.~C.~Ottewill and B.~Wardell,
  Phys.\ Rev.\  D {\bf 77}, 104002 (2008)
  [arXiv:0711.2469 [gr-qc]].

\bibitem{Ottewill:2008uu}
  A.~C.~Ottewill and B.~Wardell,
  Phys.\ Rev.\  D {\bf 79}, 024031 (2009)
  [arXiv:0810.1961 [gr-qc]].

\bibitem{PazosAvalos:2004rp}
  E.~Pazos-Avalos and C.~O.~Lousto,
  Phys.\ Rev.\  D {\bf 72}, 084022 (2005)
  [arXiv:gr-qc/0409065].

\bibitem{Pfenning:2000zf}
  M.~J.~Pfenning and E.~Poisson,
  Phys.\ Rev.\  D {\bf 65}, 084001 (2002)
  [arXiv:gr-qc/0012057].

\bibitem{Poisson:1995vs}
  E.~Poisson,
  Phys.\ Rev.\  D {\bf 52}, 5719 (1995)
  [Addendum-ibid.\  D {\bf 55}, 7980 (1997)]
  [arXiv:gr-qc/9505030].


\bibitem{Poisson:2003nc}
  E.~Poisson,
  Living Rev.\ Rel.\  {\bf 7}, 6 (2004)
  [arXiv:gr-qc/0306052].

\bibitem{Poisson:CQG}
  E.~Poisson,
  Class.\ Quantum Grav.\ {\bf 21}, R153 (2004).

\bibitem{Poisson:2004gg}
  E.~Poisson,
  arXiv:gr-qc/0410127.

\bibitem{Poisson:2004cw}
  E.~Poisson,
  Phys.\ Rev.\  D {\bf 70}, 084044 (2004)
  [arXiv:gr-qc/0407050].

\bibitem{Pound:2005fs}
  A.~Pound, E.~Poisson and B.~G.~Nickel,
  Phys.\ Rev.\  D {\bf 72}, 124001 (2005)
  [arXiv:gr-qc/0509122].

\bibitem{Pound:2007ti}
  A.~Pound and E.~Poisson,
  Phys.\ Rev.\  D {\bf 77}, 044012 (2008)
  [arXiv:0708.3037 [gr-qc]].

\bibitem{Pound:2007th}
  A.~Pound and E.~Poisson,
  Phys.\ Rev.\  D {\bf 77}, 044013 (2008)
  [arXiv:0708.3033 [gr-qc]].

\bibitem{Pretorius:2005gq}
  F.~Pretorius,
  Phys.\ Rev.\ Lett.\  {\bf 95}, 121101 (2005).

\bibitem{Prince:2009uc}
  T.~A.~Prince (for members of the LISA International Science Team),
  {\it The Promise of Low-Frequency Gravitational Wave Astronomy},
  arXiv:0903.0103.

\bibitem{Quinn:2000wa}
  T.~C.~Quinn,
  Phys.\ Rev.\  D {\bf 62}, 064029 (2000)
  [arXiv:gr-qc/0005030].

\bibitem{Quinn:1996am}
  T.~C.~Quinn and R.~M.~Wald,
  Phys.\ Rev.\  D {\bf 56}, 3381 (1997)
  [arXiv:gr-qc/9610053].

\bibitem{RW}
T. Regge and J. A. Wheeler, Phys.\ Rev.\ {\bf 108}, 1063 (1957).

\bibitem{BSprep} N.~Sago and L.~Barack, in preparation; see also Sago's
presentation at the 12th Capra meeting, posted on the meeting's website
\cite{Capra12}.

\bibitem{Sago:2002fe}
  N.~Sago, H.~Nakano and M.~Sasaki,
  Phys.\ Rev.\  D {\bf 67}, 104017 (2003)
  [arXiv:gr-qc/0208060].

\bibitem{Sago:2005gd}
  N.~Sago, T.~Tanaka, W.~Hikida and H.~Nakano,
  Prog.\ Theor.\ Phys.\  {\bf 114}, 509 (2005)
  [arXiv:gr-qc/0506092].

\bibitem{Sago:2005fn}
  N.~Sago, T.~Tanaka, W.~Hikida, K.~Ganz and H.~Nakano,
  Prog.\ Theor.\ Phys.\  {\bf 115}, 873 (2006)
  [arXiv:gr-qc/0511151].

\bibitem{Sago:2008id}
  N.~Sago, L.~Barack and S.~Detweiler,
  Phys.\ Rev.\  D {\bf 78}, 124024 (2008)
  [arXiv:0810.2530 [gr-qc]].

\bibitem{Schutz:2009tz}
  B.~F.~Schutz, J.~Centrella, C.~Cutler and S.~A.~Hughes,
  arXiv:0903.0100 [gr-qc].


\bibitem{Sopuerta:2005gz}
  C.~F.~Sopuerta and P.~Laguna,
  Phys.\ Rev.\  D {\bf 73}, 044028 (2006)
  [arXiv:gr-qc/0512028].

\bibitem{Sopuerta:2005rd}
  C.~F.~Sopuerta, P.~Sun, P.~Laguna and J.~Xu,
  Class.\ Quant.\ Grav.\  {\bf 23}, 251 (2006)
  [arXiv:gr-qc/0507112].

\bibitem{Sundararajan:2008bw}
  P.~A.~Sundararajan,
  Phys.\ Rev.\  D {\bf 77}, 124050 (2008).

\bibitem{Sundararajan:2007jg}
  P.~A.~Sundararajan, G.~Khanna and S.~A.~Hughes,
  Phys.\ Rev.\  D {\bf 76}, 104005 (2007)
  [arXiv:gr-qc/0703028].

\bibitem{Sundararajan:2008zm}
  P.~A.~Sundararajan, G.~Khanna, S.~A.~Hughes and S.~Drasco,
  Phys.\ Rev.\  D {\bf 78}, 024022 (2008)
  [arXiv:0803.0317 [gr-qc]].

\bibitem{Tanaka:2005ue}
  T.~Tanaka,
  Prog.\ Theor.\ Phys.\ Suppl.\  {\bf 163}, 120 (2006)
  [arXiv:gr-qc/0508114].

\bibitem{Teukolsky}
S.\ A.\ Teukolsky, Phys.\ Rev.\ Lett.\ {\bf 29}, 1114 (1972).

\bibitem{TeukolskyPress}
S.\ A.\ Teukolsky and W.~H.~Press, Astrophys.~J.~{\bf 193}, 443 (1974).

\bibitem{Thornburg:2009mw}
  J.~Thornburg,
  arXiv:0909.0036 [gr-qc].

\bibitem{Vega:2007mc}
  I.~Vega and S.~Detweiler,
  Phys.\ Rev.\  D {\bf 77}, 084008 (2008)
  [arXiv:0712.4405 [gr-qc]].
  
\bibitem{Vega:2009qb}
  I.~Vega, P.~Diener, W.~Tichy and S.~Detweiler,
  arXiv:0908.2138 [gr-qc].

\bibitem{WaldReconst}
R. M. Wald, Phys. Rev. Lett. {\bf 41}, 203 (1978).

\bibitem{Wald} R. M. Wald, {\it general relativity} (The University of Chicago
Press, Chicago and London, 1984).

\bibitem{BWprep}  N.~Warburton and L.~Barack, in preparation;
see also Warburton's presentation at the 12th Capra meeting, posted on
the meeting's website \cite{Capra12}.

\bibitem{Wiseman:2000rm}
  A.~G.~Wiseman,
  Phys.\ Rev.\  D {\bf 61}, 084014 (2000)
  [arXiv:gr-qc/0001025].

\bibitem{Zerilli} F. J. Zerilli, J.\ Math.\ Phys.\ {\bf 11}, 2203 (1970);
            Phys.\ Rev.\ D {\bf 2}, 2141 (1970).

\end{thebibliography}
%

\end{document}